\documentclass[prx,aps,twocolumn]{revtex4-1}
\usepackage[utf8]{inputenc}
\usepackage{graphics, bm, psfrag, amsmath, amssymb, epsfig, grffile, float, bbold}
\usepackage{dsfont}
\usepackage{color}
\usepackage{hyperref}
\usepackage[shortlabels]{enumitem}
\usepackage{amsfonts,amsthm,mathtools}
\usepackage{latexsym,graphicx}
\usepackage{amscd}
\usepackage{natbib} 
\usepackage{xcolor}
\usepackage{soul}
\usepackage[capitalize]{cleveref}
\usepackage[percent]{overpic}

\newcommand{\bx}{\mathbf{x}}
\newcommand{\by}{\mathbf{y}}

\newcommand{\ee}{{\rm e}}

\newcommand{\ii}{\mathrm{i}}
 
\newcommand{\bk}{\mathbf{k}}
\newcommand{\bQ}{\mathbf{Q}}

\newcommand{\eps}{\epsilon}

\newcommand{\R}{{\mathbb R}}
\newcommand{\C}{{\mathbb C}}
\newcommand{\Z}{{\mathbb Z}}

\newcommand{\cN}{\mathcal{N}}
\newcommand{\cC}{\mathcal{C}}
\newcommand{\cF}{\mathcal{F}}

\newcommand{\cS}{\mathcal{S}}

\newcommand{\HHF}{H_{\mathrm{HF}}}
\newcommand{\Hqf}{H_{\mathrm{qf}}}
\newcommand{\rhoHF}{\rho_{\mathrm{HF}}} 
\newcommand{\rhoqf}{\rho_{\mathrm{qf}}} 
\newcommand{\ZHF}{Z_{\mathrm{HF}}} 
\newcommand{\Zqf}{Z_{\mathrm{qf}}} 
\newcommand{\FHF}{\mathcal{G}}
\newcommand{\GG}{G}

\newcommand{\pdag}{{\phantom\dag}} 
\newcommand{\Tr}{\mathrm{Tr}}
\newcommand{\tr}{\mathrm{tr}}

\newcommand{\LnT}{\mathrm{Ln}_\beta} 
\newcommand{\fT}{f_\beta} 
\newcommand{\diag}{\mathrm{diag}}
\newcommand{\Epot}{E_{\mathrm{pot}}}

\begin{document}

\title{Update of Hartree--Fock theory for Hubbard-like models}

\author{E. Langmann$^{1}$ and J. Lenells$^{2}$}
\affiliation{$^1$Department of Physics, KTH Royal Institute of Technology, \\ 106 91 Stockholm, Sweden
	\\
	%{\rm orcid: 0000-0001-7481-2245}\\	
	$^{2}$Department of Mathematics, KTH Royal Institute of Technology, \\ 100 44 Stockholm, Sweden\\
	%{\rm orcid: 0000-0001-6191-7769}\\
}
\email{langmann@kth.se}
\email{jlenells@kth.se}

%\date{June 18, 2025} 

\begin{abstract}		
We show that the standard textbook description of (restricted) Hartree--Fock theory for (Fermi) Hubbard-like models is in need of an update, and we present such an update allowing us to correct basic and established results in the condensed matter physics literature that are qualitatively wrong.
Our update amounts to adding a test which reliably checks the thermodynamic stability of solutions of Hartree--Fock equations.
This stability test makes it possible to detect, by simple means and with certainty, regions in phase space where the model exhibits mixed phases where two conventional phases coexist and translation invariance is broken in complicated ways; in such a mixed phase, unconventional physics is to be expected.
Our results show that mixed phases are ubiquitous in Hubbard-like models in arbitrary dimensions.
\end{abstract}

\maketitle

\section{Introduction}\label{sec:intro}  
Hartree--Fock theory is an important method in condensed matter physics, as  Anderson expressed beautifully in his famous textbook as follows:  {\em ``[\ldots] modern many-body theory has mostly just served to show us how, where, and when to use Hartree--Fock  theory and how flexible and useful a technique it can be"} \cite{A1963}. Indeed, the highly successful theories of metals, superconducturs, and magnets are all based on Hartree--Fock theory: it is the starting point for a qualitative understanding by simple means and, guided by it, refined theoretical methods have been developed that make predictions which are often in excellent quantitative agreement with experimental results on real materials  \cite{A1963}. 

For (Fermi) Hubbard-like models, Hartree--Fock theory has a long tradition. 
In 1966, three years after the Hubbard model was introduced \cite{G1963,H1963,K1963}, Penn published a comprehensive paper where he computed the mean-field phase diagram of the three-dimensional (3D) Hubbard model using a method he called {\em self-consistent-field approximation} \cite{P1966}. 
In subsequent work, Penn's method was commonly referred to as  (restricted) Hartree--Fock theory \cite{Comment:HF}, but this is not fully accurate: Penn was more careful than many of his followers in that he also performed a stability analysis of the Hartree--Fock solutions he found. Penn's work was seminal: it contained many important ideas used in numerous subsequent papers since then. One prominent example is Hirsch's computation of the mean-field phase diagram of the 2D Hubbard model in 1985 \cite{H1985}, two years before this model became famous  in the context of high-temperature superconductivity \cite{A1987}. Another example is a recent generalization of the 2D Hubbard model introduced by Das--Leeb--Knolle--Knap (DLKK), which supports altermagnetism that can be studied in cold atom systems \cite{DLKK2024}: the mean-field phase diagram of the DLKK model at half-filling was computed by restricted Hartree--Fock theory, following Hirsch's variant of Penn's method \cite{DLKK2024}. Today, this method is still regarded as state-of-the-art in condensed matter physics; for example, the phase diagrams of the 3D and 2D Hubbard models by Penn \cite[Fig.~15]{P1966} and Hirsch \cite[Fig.~3]{H1985} are reprinted in an influential review on the metal-insulator transition \cite{I1998}; see \cite[Fig.~10]{I1998} and \cite[Fig.~11]{I1998}, respectively. 

There are several indications that Hartree--Fock theory for Hubbard-like models, as applied in well-known papers in the condensed matter literature, is not reliable.
One example is the mean-field phase diagram of the 2D Hubbard model already mentioned: it predicts an anti-ferromagnetic phase in a significant doping range away from half-filling \cite[Fig.~3]{H1985} which, as pointed out by Hirsch \cite[Sec.~VII]{H1985}, is qualitatively wrong. A better understanding of this wrong result came in 1989 when it was discovered that, away from half-filling, the Hartree--Fock free energy of the 2D Hubbard model is minimized by exotic solutions that break translational invariance in complicated ways, such as domain walls \cite{PR1989,ZG1989}; see \cite{M2023,M2024} and references therein for more recent related work. 
Such complicated solutions are interesting since they indicate exotic physics, and they can be studied systematically  using unrestricted Hartree--Fock theory \cite{V1991}. However, unrestricted Hartree--Fock theory can be used only for small system sizes, it requires large computational efforts, and it leads to complicated results that are difficult to use as a starting point for the development of more sophisticated methods; thus, it cannot be a substitute for (restricted) Hartree--Fock theory. Moreover, Hartree--Fock theory has remained an important tool up to this day; for example, it was used recently in the interesting paper by DLKK  \cite{DLKK2024}, as already mentioned. Since Penn's antiferromagnetic solution of Hartree--Fock theory for the Hubbard model in arbitrary dimensions is known to be stable at half-filling \cite{BLS1994}, one would expect that this solution is stable at half-filling for other Hubbard-like models as well. Thus, there was good reason to believe that Penn's method can be used to predict the phase diagram of the half-filled DLKK model \cite{DLKK2024}. However, we recently discovered that some of the Hartree--Fock solutions obtained in \cite{DLKK2024} are qualitatively wrong \cite{LL2025A}: in the DLKK model, mixed phases can occur even at half-filling. This prompted us to revisit Penn's phase diagrams of the 3D Hubbard model \cite{P1966}; we not only found that these phase diagrams are qualitatively wrong as well, but also discovered the underlying mistake. As will be explained, it was an honest mistake; unfortunately, it was never corrected but, instead, repeated in later papers, including \cite{H1985,DLKK2024}. These examples show that Hartree--Fock theory for Hubbard-like models is in need of an update, to make sure that this mistake is not repeated in the future. Hartree--Fock theory is a basic method taught in textbooks which is used to guide studies by more advanced methods --- clearly, it is important that such a method is reliable. 

The aim of this paper is to provide an update of the textbook approach to Hartree--Fock theory for Hubbard-like models that re-establishes it as a trustworthy and versatile tool for this important class of models. 
Our update amounts to adding a stability test which allows us to avoid unstable solutions of the (standard) Hartree--Fock equations.
We also advocate an efficient implementation of Hartree--Fock theory due to Bach, Lieb, and Solovej \cite{BLS1994} (see also \cite{BP1996}); using this implementation, the computational effort required to derive Hartree--Fock equations is minimized, and the stability test is automatically included.

As we show in several examples, the stability test often reveals a mixed phase where two distinct Hartree--Fock solutions are unstable towards phase separation. Finding such a mixed phase does not mean that phase separation actually occurs, but it rather is an indication that states minimizing the Hartree--Fock functional break translation invariance in complicated ways. The details of such a mixed phase can only be understood by more advanced methods (such as unrestricted Hartree--Fock theory \cite{V1991}), but what our updated  Hartree--Fock method can provide, by simple means and with certainty,  is the regions in parameter space where such a mixed phase occurs.
We demonstrate through multiple examples that mixed phases in Hubbard-like models are ubiquitous; in particular, they occur in arbitrary dimensions, and they can occur even at half-filling.

Stability tests were important already to Penn in his groundbreaking work \cite{P1966}. 
However, since he compared energies at fixed density rather than at fixed chemical potential, Penn missed the instability of the antiferromagnetic phase away from half-filling: while it is known today that this antiferromagnetic phase is a robust absolute energy minimum at half-filling for the Hubbard model in arbitrary dimensions \cite{BLS1994}, we show in this paper that {\em it is a local energy maximum away from half-filling} and thus unstable. This was an honest mistake: (i) Penn was a pioneer and, of course, could not have known what we know today, (ii) while his numerical results are impressive (considering the limited computational resources available to him), his ability to test his numerical results was limited as compared to what is possible today, (iii) due to this mistake, he did not have to confront the issue of mixed phases which, probably, would have been shocking at his time, 
(iv) as explained in Section~\ref{sec:bug}, the mistake was subtle. 
However, this mistake was very  unfortunate since it led to phase diagrams that are qualitatively wrong. Since this mistake was never corrected but, instead, repeated for other models \cite{H1985,DLKK2024}, it has led to a bad reputation of restricted Hartree--Fock theory. We strongly believe that this has been a serious handicap in research on Hubbard-like models.
	
While our emphasis in this paper is theory, we stress that the phase diagrams we present are predictions that can be tested in cold atom experiments realizing the conventional 1D, 2D, or 3D Hubbard model in the lab \cite{B2008,S2008,H2015,M2017}: Whenever a mixed phase occurs, one can expect exotic physics. How this exotic physics manifests itself we do not know (this would require further studies beyond  Hartree--Fock theory), but we expect that mixed phases are easily detectable: When changing from a conventional to a mixed phase (which can be done in cold atom experiments by changing the interaction strength $U$, for example), some drastic change of the physical behavior should occur. It would of course be interesting to make such a prediction more specific, but this is beyond the scope of the present paper.

To avoid misunderstandings, we stress that there are papers on Hartree--Fock theory about Hubbard-like models in the literature using the correct stability test, including \cite{LW1997,LW2007,G2015,J2020}, but these papers are not well-known. We also emphasize that we do not claim that pertinent papers we do not mention are wrong --- a systematic check of the literature would be important but is beyond the scope of our paper.

Our plan is as follows. In Section~\ref{sec:summary} we summarize our main results: we introduce notation and give definitions (Section~\ref{sec:notation0}), present a concise summary of our updated version of Hartree--Fock theory (Sections~\ref{sec:HFnutshell} and \ref{sec:rHF}), and present our updated phase diagrams of the Hubbard model in various dimensions (Section~\ref{sec:phasediagram}). 
Section~\ref{sec:method} explains our method: we give a detailed discussion of how mixed phases arise (Section~\ref{sec:mixedAFP} and \ref{sec:otherphases}), examples of mixed phases at half-filling (Section~\ref{sec:halffilled}), and a detailed explanation of the mistake corrected by our update (Section~\ref{sec:bug}). In Sections~\ref{sec:HF} and \ref{sec:BLS}, we give a self-contained account of Hartree--Fock theory, starting with the fundamental principle on which Hartree--Fock theory is based. Our presentation aims to bridge the conventional approach to Hartree--Fock theory in the physics literature with the mathematically precise approach due to Bach, Lieb, and Solovej \cite{BLS1994}. We end with conclusions in Section~\ref{sec:conclusions}.  We also added three appendices \ref{app:Gen}--\ref{app:exchange} where interested readers can find the technical details needed to verify the mathematical claims we make in the paper.

\medskip 

\section{Summary of results}\label{sec:summary} 
In Section~\ref{sec:notation0} we define the models we consider and introduce our notation. Sections~\ref{sec:HFnutshell} and \ref{sec:rHF} summarize the updated Hartree--Fock theory we propose. Our updated phase diagrams are presented in Sections~\ref{sec:phasediagram} and \ref{sec:DLKK}. 

\subsection{Model definitions}\label{sec:notation0}
We set the lattice constant, Planck's constant $\hbar$, and the Boltzmann constant $k_B$ to 1.
We use the symbol $\beta>0$ for the inverse of the temperature, i.e., $1/\beta\downarrow 0$ corresponds to zero temperature. The symbols $t>0$, $U\geq 0$ are reserved for the usual Hubbard parameters, $\mu\in\R$ is the chemical potential, and $\nu\in[-1,1]$ is the doping parameter (defined in \eqref{rho} below). The DLKK model is an extension of the 2D Hubbard model that involves two further real parameters $t'$ and $\delta$ \cite{DLKK2024}.

We now give the definitions of the Hubbard model and the DLKK model for arbitrary dimension $n=1,2,3, \ldots$. We write the $n$D Hubbard Hamiltonian as 
\begin{align}\label{Hdef} 
	H=H_0 + U\sum_{\bx}(n_{\bx,\uparrow}-\tfrac12)(n_{\bx,\downarrow}-\tfrac12), 
\end{align} 
\begin{align}\label{H0def} 
	H_0 = \sum_{\bx,\by}\sum_{\sigma=\uparrow,\downarrow}t_{\bx,\by}c_{\bx,\sigma}^\dag c_{\by,\sigma}\pdag 
	-\mu\sum_{\bx}(n_\bx-1)
\end{align} 
with $c_{\bx,\sigma}^\dag$ and $c_{\bx,\sigma}$ fermion creation and annihilation operators labeled by lattice sites $\bx=(x_1,\ldots,x_n)\in\Z^n$ and a spin index $\sigma\in\{\uparrow,\downarrow\}$,  normalized such that the density operators 
\begin{align}\label{nsigmadef} 
n_{\bx,\sigma}= c_{\bx,\sigma}^\dag c_{\bx,\sigma}^\pdag
\end{align} 
have eigenvalues 0 and 1, and with 
\begin{align}\label{ndef}  
	n_\bx = n_{\bx,\uparrow}+ n_{\bx,\downarrow}	
\end{align} 
the (total) density operator. 
The hopping matrix elements $t_{\bx,\by}$ are $-t$ for nearest-neighbor sites $\bx,\by$ and $0$ otherwise, i.e., the dispersion relation is 
\begin{align}\label{veps0} 
\varepsilon_0(\bk) = -2t\sum_{i=1}^n\cos(k_i) 
\end{align} 
for pseudo-momenta $\bk=(k_1,\ldots,k_n)\in[-\pi,\pi]^n$. 
The DLKK model \cite{DLKK2024} is defined by the same formulas but with the hopping matrix elements $t_{\bx,\by}$ also allowing for next-nearest-neighbor hopping in a staggered pattern; see \cite[Fig.~1(a)]{DLKK2024}. 
Due to this, one obtains two further dispersion relations 
(the model in \cite{DLKK2024} was defined in 2D ($n=2$); our formulas below generalize this definition to an arbitrary number of dimensions)
\begin{equation}\label{veps1veps2} 
	\begin{split} 
	\varepsilon_1(\bk) & = -4t'\sum_{1\leq i<j\leq n} \cos(k_i)\cos(k_j) ,\\
	\varepsilon_2(\bk) & = -4t'\delta \sum_{1\leq i<j\leq n} \sin(k_i)\sin(k_j) 
    \end{split} 
\end{equation} 
such that the non-interacting part of the Hamiltonian in Fourier space can be written as 
\begin{multline}\label{H0}  
H_0 = \sum_{\bk,\sigma}\Big(\big[\varepsilon(\bk)-\mu\big]\big[c_\sigma^\dag(\bk) c_\sigma(\bk)-\tfrac12\big] \\
+ \varepsilon_2(\bk)c_\sigma^\dag(\bk) c_\sigma(\bk+\bQ)
\Big) 	
\end{multline} 	
with $\varepsilon(\bk) = \varepsilon_0(\bk)+\varepsilon_1(\bk)$ and 
\begin{align}\label{bQ} 
\bQ=(\pi,\ldots,\pi) 
\end{align} 
(see Appendix~\ref{app:FT} for further details).
As usual, to make the model  well-defined, we restrict the lattice to a finite hypercube with $L^n$ sites and periodic boundary conditions, i.e., the momenta $\bk=(k_1,\ldots,k_n)$ are constrained by the condition that the components $k_i$ are integer multiples of $2\pi/L$, with 
the even integer $L>0$
a regularization parameter such that the thermodynamic limit corresponds to $L\to +\infty$. For $\delta=0$, the DLKK model reduces to the $t$-$t'$-$U$ model extending the Hubbard model by next-nearest-neighbor hopping of strength $t'$  \cite{LH1987}; the limiting case $t'=0$ corresponds to the Hubbard model. 
%For later reference, we note that 
%\begin{equation} 
%\begin{split} 	
%\frac{\varepsilon(\bk)+\varepsilon(\bk+\bQ)}{2}
%&=\varepsilon_1(\bk),\\
%\frac{\varepsilon(\bk)-\varepsilon(\bk+\bQ)}{2}
%&=\varepsilon_0(\bk),  
%\end{split}
%\end{equation} 
%and $\varepsilon_2(\bk+\bQ)=\varepsilon_2(\bk)$. 

To simplify formulas for particle-hole transformations, we find it convenient to define density relative to half-filling, i.e., we have $n_{\bx,\sigma}-\tfrac12$ rather than $n_{\bx,\sigma}$ in our formulas; due to this, we obtain $c_\sigma^\dag(\bk) c_\sigma(\bk)-\tfrac12$ rather than $c_\sigma^\dag(\bk) c_\sigma(\bk)$ in \eqref{H0}. It is interesting to note that, under a particle-hole transformation, the model parameters change as 
\begin{align}\label{PH} 
(t,t',\delta,U,\mu,\nu)\to (t,-t',\delta,U,-\mu,-\nu);
\end{align} 
thus, for $t' =0$, we have symmetry under particle-hole transformations with half-filling $\nu=0$ corresponding to $\mu=0$, but this is not the case for non-zero $t'$ (see Appendix~\ref{app:PH} for further details).

We also use the spin operator 
\begin{align} 
\vec{s}_{\bx}= (s^x_{\bx},s^y_{\bx},s^z_{\bx}), 
\end{align} 
with the following components defined as usual,
\begin{equation}\label{vecS} 
\begin{split} 	
	s_{\bx}^x & = c_{\bx,\uparrow}^\dag c_{\bx,\downarrow}^\pdag + c_{\bx,\downarrow}^\dag c_{\bx,\uparrow}^\pdag,\\ 
	s_{\bx}^y & = -\ii c_{\bx,\uparrow}^\dag c_{\bx,\downarrow}^\pdag + \ii c_{\bx,\downarrow}^\dag c_{\bx,\uparrow}^\pdag,\\
	s_{\bx}^z &=  n_{\bx,\uparrow}-n_{\bx,\downarrow}. 
\end{split} 
\end{equation} 	
For later reference, we recall the definition of the density of states for the $n$D Hubbard dispersion relation in \eqref{veps0}, 
\begin{equation}\label{DOS} 
	N_n(\eps)= \int_{[-\pi,\pi]^n}\frac{d^n\bk}{(2\pi)^n}\delta(\eps-\varepsilon_0(\bk)); 
\end{equation} 
see Appendix~\ref{app:DOS} for a discussion of useful facts about these functions. 

\subsection{Updated Hartree--Fock theory}\label{sec:HFnutshell}

As is well-known, Hartree--Fock theory is a variational method with mean-fields serving as variational parameters and determined such that the grand canonical potential is minimized; see e.g.\ \cite{BLS1994} (this is discussed in more detail in Section~\ref{sec:foundation}). For the Hubbard-like  models we consider, the variational fields are a real-valued and an $\R^3$-valued field denoted as $d(\bx)$ and $\vec{m}(\bx)=(m^x(\bx),m^y(\bx),m^z(\bx))$, respectively, where $\bx\in\Z^n$ are lattice sites. 
The well-known Hartree--Fock equations determining these fields can be written as 
\begin{equation}\label{HFeqs} 
d(\bx)=\langle n_{\bx}-1\rangle,\quad 
\vec{m}(\bx) = \langle\vec{s}_{\bx}\rangle ,  	
\end{equation} 
which suggests to interpret $d(\bx)$ and $\vec{m}(\bx)$ as local doping and local magnetisation, respectively (here and in the following, {\em doping} is short for {\em fermion density relative to half-filling}).  As usual, the expectation values $\langle\cdots \rangle$  are defined such that \eqref{HFeqs} is a system of non-linear equations allowing to compute the variational fields self-consistently, using the Hartree--Fock Hamiltonian 
\begin{align}\label{HHF}  
\HHF = H_0 + \frac{U}{2}\sum_{\bx}\big( d(\bx)(n_{\bx}-1)-\vec{m}(\bx)\cdot\vec{s}_{\bx} 
\big) 	
\end{align} 	
that can be diagonalized by a Bogoliubov transformation. One important parameter is the (average) {\em doping}, which we denote as $\nu$; it is determined by 
\begin{equation}\label{rho} 
\nu= \frac{\sum_{\bx} \langle n_{\bx}-1\rangle}{\sum_{\bx} 1}
\end{equation} 
where $\sum_{\bx} 1=L^n$ is the number of lattice sites. Note that $-1\leq \nu\leq 1$, and that half-filling corresponds to $\nu=0$. 
Thus, Hartree--Fock theory allows us to determine the variational fields $d(\bx)$, $\vec{m}(\bx)$ at given doping $\nu$. 

Unrestricted Hartree--Fock theory amounts to solving these equations as they stand; for larger systems, this is a formidable variational problem with $4L^n$ variational parameters. It is common to simplify this task by restricting the number of variational parameters by some ansatz; one famous example is the following ansatz going back at least to Penn \cite{P1966}, 
\begin{align}\label{Penn_ansatz} 
	\begin{split} 	
		d(\bx) =&\; d_0 +  (-1)^{\bx}d_1,\\
		\vec{m}(\bx) =&\; m_0\vec{e} +  (-1)^{\bx}m_1\vec{e},
	\end{split} 
\end{align} 		
with four real variational parameters $d_0$, $d_1$, $m_0$, $m_1$,  and $\vec{e}$ an arbitrary unit vector in $\R^3$ (e.g., $\vec{e}=(0,0,1)$); here and in the following, we use the shorthand notation
\begin{align}\label{staggeredfactor} 
(-1)^{\bx} = (-1)^{x_1+\cdots+x_n}
\end{align} 
for $\bx=(x_1,\ldots,x_n)\in \Z^n$ (note that $(-1)^{\bx} = \ee^{\ii\bQ\cdot\bx}$). 
It is important to note that, while this ansatz is our main example, there are many other ans\"atze that have been explored in the literature: the more variational parameters one allows, the more detailed information can be obtained but at the cost of more work.  Our update  applies to all these ans\"atze.

What we discussed up to now is standard. Our update is to add a stability test: as will be shown, the Hartree--Fock equations have solutions which are unstable, and it is important to rule out these solutions to avoid wrong results. We propose to use a known function, $\FHF$, which we call  the {\em Hartree--Fock function} (not to be confused with the Hartree--Fock functional defined in \cite{BLS1994}); it is a function of the variational fields and the chemical potential, $\FHF=\FHF(\boldsymbol{d},\boldsymbol{\vec{m}};\mu)$ (at fixed temperature $1/\beta$), where we write $\boldsymbol{d}$ short for the vector of all $d(\bx)$ and similarly for $\boldsymbol{\vec{m}}$. This function has the following properties.

\begin{enumerate} 
\item The Hartree--Fock equations are given by 
\begin{equation}\label{HFeqs1}  
\frac{\partial\FHF}{\partial d(\bx)}=0,\quad 
\frac{\partial\FHF}{\partial\vec{m}(\bx)}=\vec{0} 	
\end{equation} 	
(we use the notation  $\frac{\partial}{\partial\vec{m}}=(\frac{\partial}{\partial m^x},\frac{\partial}{\partial m^y},\frac{\partial}{\partial m^z})$ and $\vec{0}=(0,0,0)$ to write three similar equations as one).

\item The {\em doping constraint} is given by
\begin{equation}\label{rho1}  
\nu = -\frac{\partial\FHF}{\partial\mu}. 	
\end{equation} 

\item \label{stability} If $\boldsymbol{d}^{*}$, $\boldsymbol{\vec{m}}^{*}$ is a solution of the Hartree--Fock equations, then the {\em free energy (density)} $\cF$ of the state described by this solution is given by 
\begin{equation}\label{cF}
 \cF(\mu) = \FHF(\boldsymbol{d}^{*},\boldsymbol{\vec{m}}^{*};\mu)
\end{equation}
(this result is derived in Section~\ref{sec:variational}).
	
\end{enumerate} 
The specialization of this to restricted Hartree--Fock theory is straightforward, as discussed in Section~\ref{sec:rHF}.

Thus, $\FHF$ is a useful object for practical implementations of Hartree--Fock theory: it not only gives the Hartree--Fock equations and the doping constraint in a simple way, but it also provides a means to test stability: 

\medskip 
\noindent {\bf Stability test:} {\em If you obtain several qualitatively different solutions of the Hartree--Fock equations at a fixed value of $\mu$, compute the free energy $\cF$ for each of these solutions, and take the one giving the smallest free energy density} (the other solutions should be discarded). 

\medskip

This test is essential in restricted Hartree--Fock theory to avoid wrong results. Indeed, in restricted Hartree--Fock theory one typically finds several distinct solutions which are {\em qualitatively} different \cite{Comment:qualitative} and, to avoid mistakes, it is important to know which one to choose. A careful reader might ask: what should I do if I find several solutions with the same free energy $\cF$? The answer to this question can be found in  Section~\ref{sec:method}.

To avoid the mistake mentioned in the introduction, it is important to compare free energies of different Hartree--Fock solutions at the same value of the chemical potential (and {\em not} at the same value of doping); this is similar to BCS theory where the pertinent variational states are superpositions of states with different particle numbers \cite{BCS1957}. 

To be specific, we give an explicit formula for the Hartree--Fock function $\FHF$: for finite $L$, 
\begin{multline*}
\FHF =  
\frac{U}{4L^n}\sum_{\bx}\big(\vec{m}(\bx)^2-d(\bx)^2 \big) 
\\ -\frac1{L^n}\sum_I \frac1\beta\ln\left(2\cosh\left(\frac{\beta E_I}{2}\right)\right)
\end{multline*} 
where $I$ is short for $(\bx,\sigma)$ and $E_I$ are the eigenvalues of a $2L^n\times 2L^n$ hermitian matrix $h=(h_{I,J})$ which can be obtained by writing the  Hartree--Fock Hamiltonian in \eqref{HHF} as 
\begin{align}\label{HHFgen1} 
\HHF = \sum_{I,J} h_{I,J}^\pdag\big( c_I^\dag c_J^\pdag -\tfrac12\delta_{I,J}\big);  	
\end{align}  
see \eqref{h} for an explicit formula for $h_{I,J}$.

We conclude this section by discussing an important  technical point. From our discussion above, one might expect that stable solutions of the Hartree--Fock equations minimize the Hartree--Fock function $\FHF$, but this is not the case. 
In fact, as we will explain, stable solutions $\boldsymbol{d}^{*}$, $\boldsymbol{\vec{m}}^{*}$ of the Hartree--Fock equations are saddle points of $\FHF$ in the following sense, 
\begin{equation}\label{minmax}  
\FHF(\boldsymbol{d}^{*},\boldsymbol{\vec{m}}^{*};\mu)=\min_{\boldsymbol{\vec{m}}}\max_{\boldsymbol{d}}\FHF(\boldsymbol{d},\boldsymbol{\vec{m}};\mu); 
\end{equation} 	
by this notation we mean that, to find stable solutions of Hartree--Fock theory, one should, {\em at fixed value of $\mu$, first maximize $\FHF$ over the set of all $d(\bx)$'s, and then minimize over the set of $\vec{m}(\bx)$'s}. As will be explained, this {\em min-max principle} is useful in practical implementations of Hartree--Fock theory since, by using it, one automatically avoids unphysical solutions of Hartree--Fock equations.

\subsection{Restricted Hartree--Fock theory}\label{sec:rHF}  
The ansatz in \eqref{Penn_ansatz} is interesting since it includes antiferromagnetic, ferromagnetic, ferrimagnetic, and paramagnetic phases \cite{P1966}. 
With this ansatz, one can compute the Hartree--Fock function in the thermodynamic limit exactly by analytic means, for all temperatures $1/\beta> 0$ and even for the DLKK model (as will be discussed, the Hartree--Fock solution of the DLKK model in \cite{DLKK2024} corresponds to a special case of this ansatz). We write our result using the abbreviation
\begin{equation}\label{LT} 
		\LnT(\eps) =		\frac1\beta\ln\left(2\cosh\left(\frac{\beta\eps}{2}\right)\right) 
		\quad (\epsilon\in\R)
\end{equation}
as
\begin{multline}\label{FHFrestr} 
	\FHF(d_0,d_1,m_0,m_1;\mu) = \frac{U}{4}(m_0^2+m_1^2-d_0^2-d_1^2)
	\\ - \frac12\int_{[-\pi,\pi]^n}\frac{d^n\bk}{(2\pi)^n}
	\sum_{r,r'=\pm}\LnT\big( E_{r,r'}(\bk)\big)   	
\end{multline} 
where
\begin{multline}\label{Errp}
	E_{r,r'}(\bk) = \varepsilon_1(\bk) -\mu + \frac{Ud_0}{2}-r\frac{Um_0}{2} \\
	+ r'\sqrt{\varepsilon_0(\bk)^2
		+ \left(\varepsilon_2(\bk)+\frac{Ud_1}{2}- r\frac{Um_1}{2}\right)^2}	
\end{multline} 
for $r,r'=\pm$ are four effective bands, with $\varepsilon_i(\bk)$ for $i=0,1,2$ the tight-binding band relations in \eqref{veps0} and \eqref{veps1veps2}. Note that $\FHF$ in \eqref{FHFrestr} is independent of $\vec{e}$ due to rotation invariance.

For later reference, we note that the special function $\LnT(\epsilon)$ in \eqref{LT} is minus the anti-derivative of 
\begin{align}\label{fT}  
 	\fT(\eps) =-\frac12\tanh\left(\frac{\beta\eps}2\right) = \frac1{\ee^{\beta\eps}+1}-\frac12, 
\end{align} 	
which is the particle-hole symmetric version of the Fermi distribution function; to be precise, 
\begin{align}\label{fTLT} 
	\fT(\epsilon) = -\frac{\partial}{\partial\epsilon}\LnT(\epsilon). 
\end{align} 
We also note that, in the zero-temperature limit $1/\beta\downarrow 0$, $\LnT(\eps)\to \tfrac12|\epsilon|$ and $\fT(\epsilon)\to -\tfrac12\mathrm{sgn}(\epsilon)=\theta(-\epsilon)-\tfrac12$, where $\theta$ is the Heaviside function.

From the explicity known Hartree--Fock function $\FHF$ in \eqref{FHFrestr}, one can obtain the Hartree--Fock equations and the doping constraint in a simple way by differentiation, 
\begin{align}\label{rHFeqs} 
\frac{\partial\FHF}{\partial d_i}=0,\quad 
\frac{\partial\FHF}{\partial m_i}=0\quad (i=0,1),  
\end{align} 
and $\nu=-\frac{\partial\FHF}{\partial\mu}$. 
To make it easy for the reader to see that these are the standard equations obtained in conventional Hartree--Fock theory, we write them out in Section~\ref{sec:staggered}; see \eqref{restr_HFeqs} and \eqref{restr_rho}.
However, we found that the most convenient way to solve these equations is to use the min-max procedure in \eqref{minmax} (this is explained in Section~\ref{sec:mixedAFP}).

We conclude this section with a {\bf warning about a dangerous shortcut} (discussed in greater detail in Section~\ref{sec:mixedAFP}): The Hartree--Fock equations \eqref{restr_HFeqs} and the doping constraint \eqref{restr_rho} suggest that $d_0$ can be identified with the doping $\nu$. By using this identification and introducing the renormalized chemical potential
\begin{align}\label{tmu} 
\tilde\mu=\mu-\frac{U}{2}d_0=\mu-\frac{U}{2}\nu, 
\end{align} 
one can eliminate the variational parameter $d_0$ from the problem (since the $E_{r,r'}$ depend on $d_0$ and $\mu$ only in the combination $\tilde\mu$). 
Doing this is dangerous for two reasons: (i) By taking this shortcut, one ends up comparing free energies of Hartree--Fock solutions at the same value of doping (and not at fixed $\mu$), which can -- and does -- give wrong results. (ii) This simplification assumes that each doping value can be realized, which is often not true: as will be shown, there are doping regimes where no translationally invariant Hartree--Fock solution included in the ansatz \eqref{Penn_ansatz} is thermodynamically stable; to remain open to this possibility, one must avoid this shortcut.
This shortcut was used by DLKK \cite{DLKK2024}, following well-established previous papers \cite{P1966,H1985,LH1987} and, as explained in Section~\ref{sec:method}, it led to wrong results \cite{LL2025A}. We believe this is only one example of many: this shortcut can be used for any restricted Hartree--Fock ansatz. We therefore believe that it would  be important to check and  correct other restricted Hartree--Fock results in the literature as well. 
 
\begin{figure}
	\vspace{0.4cm}
	\begin{center}
		\hspace{-.6cm}
		\begin{overpic}[width=.46\textwidth]{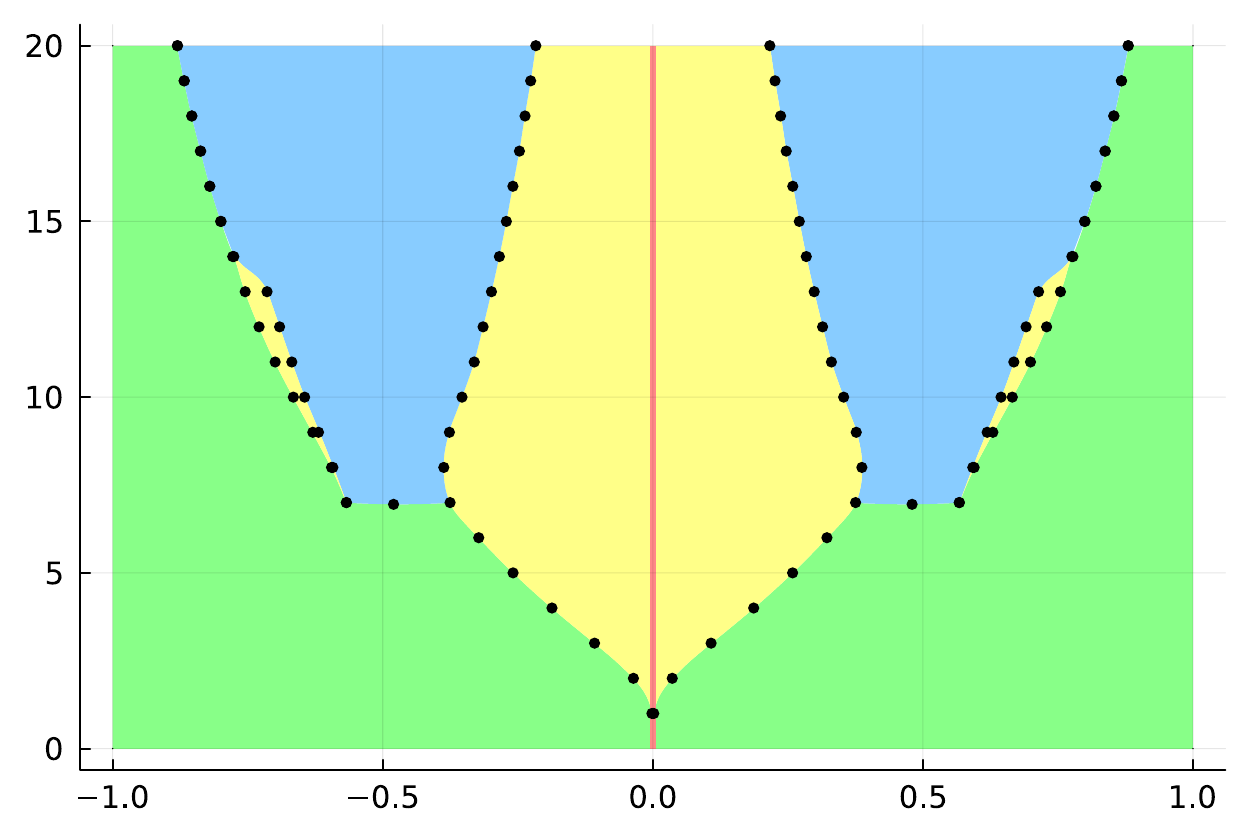 }
			\put(45,66.5){\footnotesize (a) $1/\beta = 0$}
			\put(5.5,66.5){\footnotesize $U/t$}
			\put(99.5,4.5){\footnotesize $\nu$}
			\put(20,15){\footnotesize P}
			\put(80,15){\footnotesize P}
			\put(26,50){\footnotesize F}
			\put(74,50){\footnotesize F}
			\put(39,28){\footnotesize Mixed}
			\put(55.5,28){\footnotesize Mixed}
			\put(46,36){\footnotesize AF}
			\put(49,35){\vector(2,-3){3}}
		\end{overpic}
		
		\vspace{0.5cm}
		\hspace{-.6cm}
		\begin{overpic}[width=.46\textwidth]{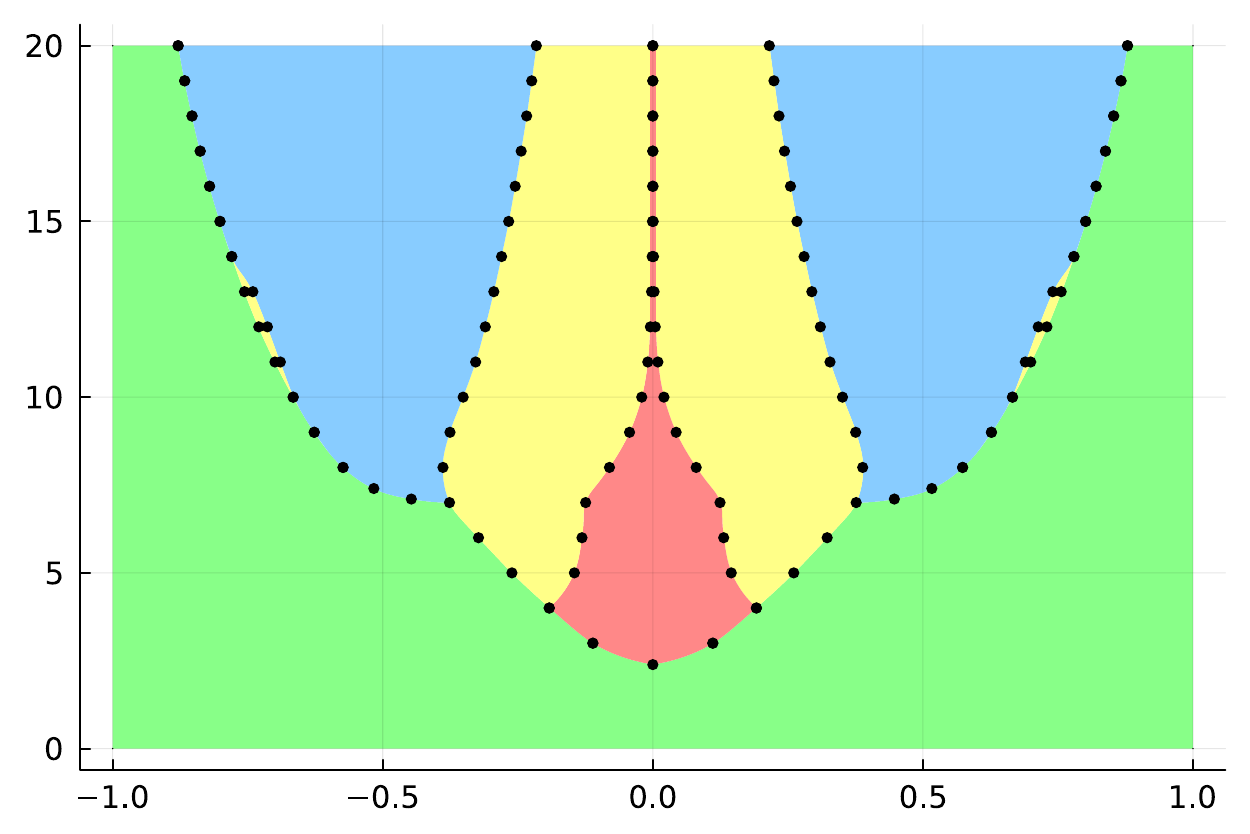 }
			\put(43,66.5){\footnotesize (b) $1/\beta = 0.2t$}
			\put(5.5,66.5){\footnotesize $U/t$}
			\put(99.5,4.5){\footnotesize $\nu$}
			\put(50,20){\footnotesize AF}
			\put(20,15){\footnotesize P}
			\put(80,15){\footnotesize P}
			\put(26,50){\footnotesize F}
			\put(74,50){\footnotesize F}
			\put(39,33){\footnotesize Mixed}
			\put(56,33){\footnotesize Mixed}
		\end{overpic}
		\caption{\label{fig:3D} 
			Mean-field phase diagrams of the 3D Hubbard model for temperatures $1/\beta=0$ (a) and $1/\beta =0.2t$ (b), obtained by Hartree--Fock theory restricted to AF, F and P states. The phases are shown as a function of doping $\nu$ and coupling $U/t$. 
			AF, F, and P regions are red, blue, and green, respectively, and yellow regions indicate mixed phases. 
		}
	\end{center}
	\vspace{-0.4cm}
\end{figure}

\begin{figure}
	\vspace{0.4cm}
	\begin{center}
		\hspace{-.6cm}
		\begin{overpic}[width=.46\textwidth]{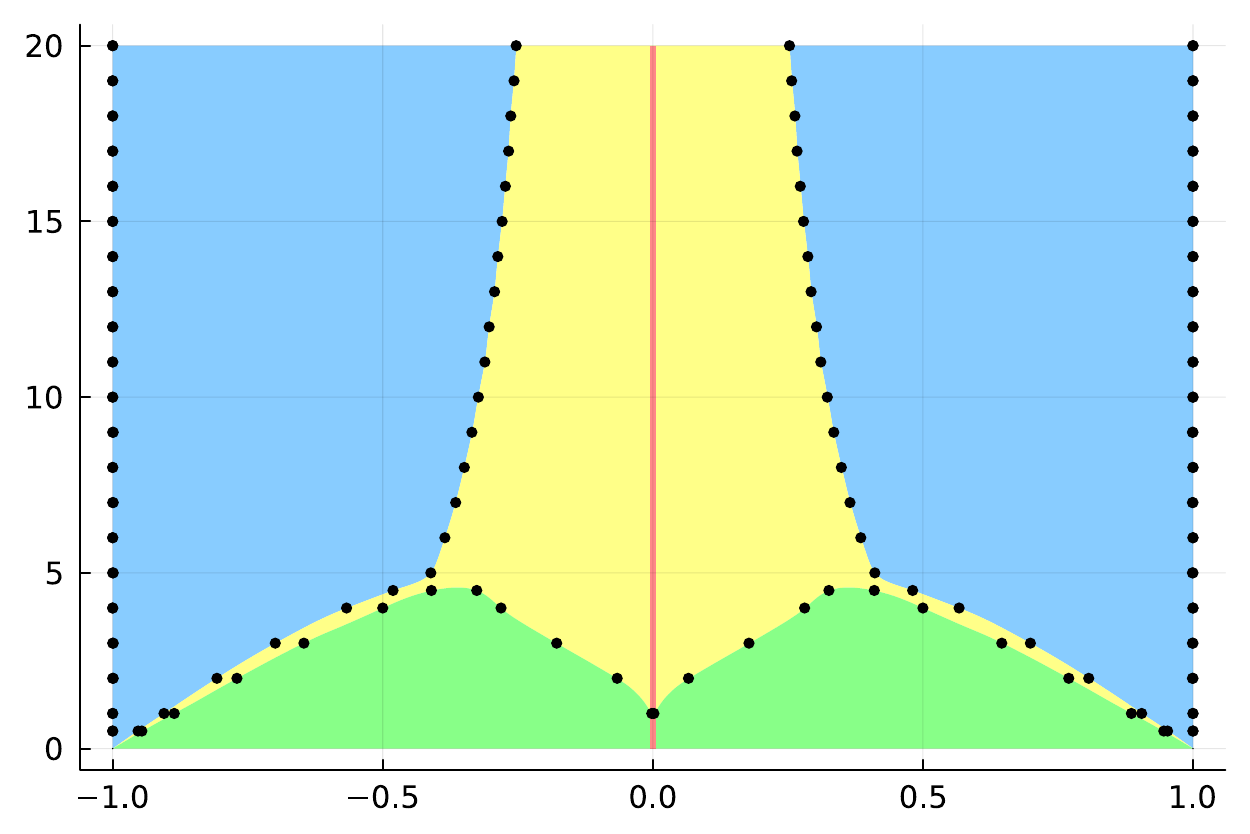 }
			\put(45,66.5){\footnotesize (a) $1/\beta=0$}
			\put(5.5,66.5){\footnotesize $U/t$}
			\put(99.5,4.5){\footnotesize $\nu$}
            \put(30,10){\footnotesize P}
            \put(70,10){\footnotesize P}
            \put(25,45){\footnotesize F}
            \put(78,45){\footnotesize F}			
			\put(39.5,24){\footnotesize Mixed}
			\put(55,24){\footnotesize Mixed}
			\put(46,36){\footnotesize AF}
			\put(49,35){\vector(2,-3){3}}
		\end{overpic}
		
		\vspace{.5cm}
		\hspace{-.6cm}
		\begin{overpic}[width=.46\textwidth]{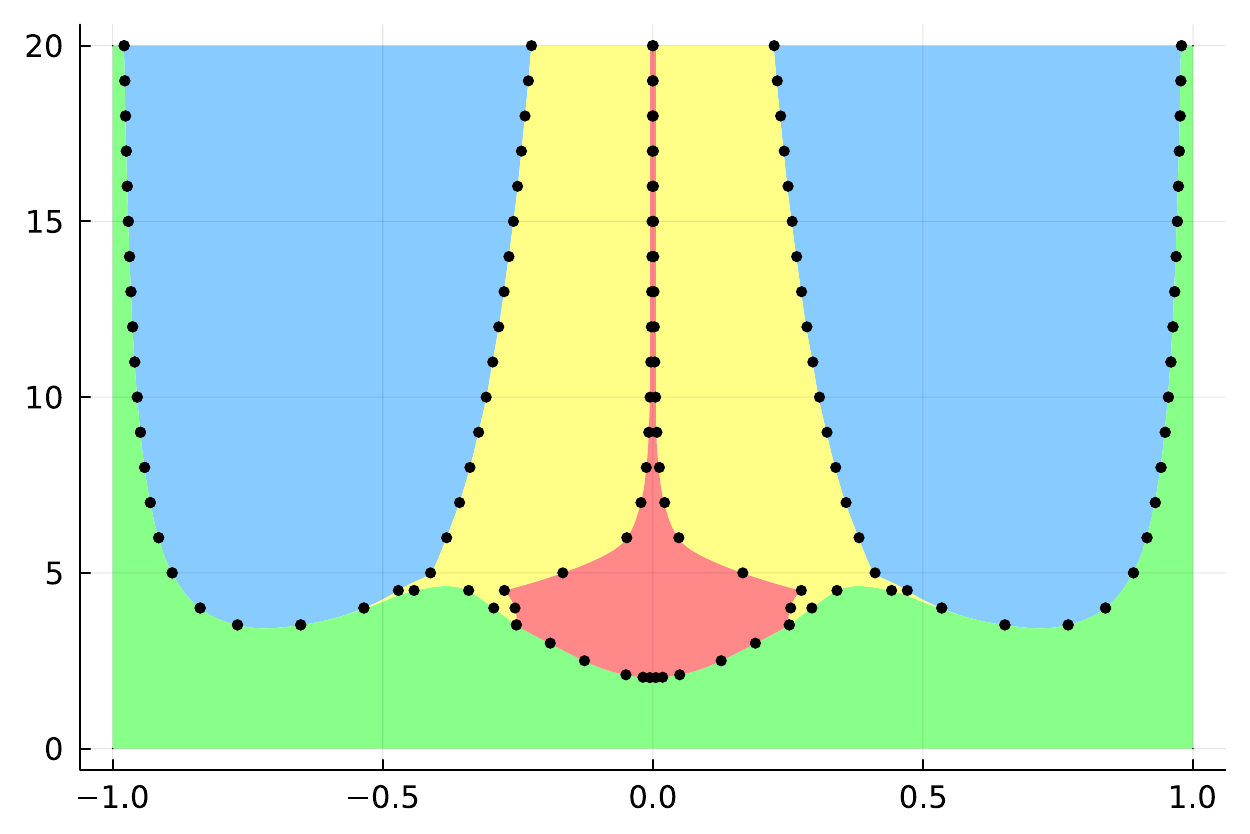 }
			\put(43,66.5){\footnotesize (b) $1/\beta= 0.2t$}
			\put(5.5,66.5){\footnotesize $U/t$}
			\put(99.5,4.5){\footnotesize $\nu$}
            \put(50,17){\footnotesize AF}
            \put(20,10){\footnotesize P}
            \put(80,10){\footnotesize P}
            \put(25,45){\footnotesize F}
            \put(78,45){\footnotesize F}			
			\put(39.5,28){\footnotesize Mixed}
			\put(55,28){\footnotesize Mixed}
		\end{overpic}
		\caption{\label{fig:1D} 
		Same as Fig.~\ref{fig:3D} but for the 1D Hubbard model. 
		}
	\end{center}
	\vspace{-0.4cm}
\end{figure}

\begin{figure}
	\vspace{0.4cm}
	\begin{center}
		\hspace{-.6cm}
		\begin{overpic}[width=.46\textwidth]{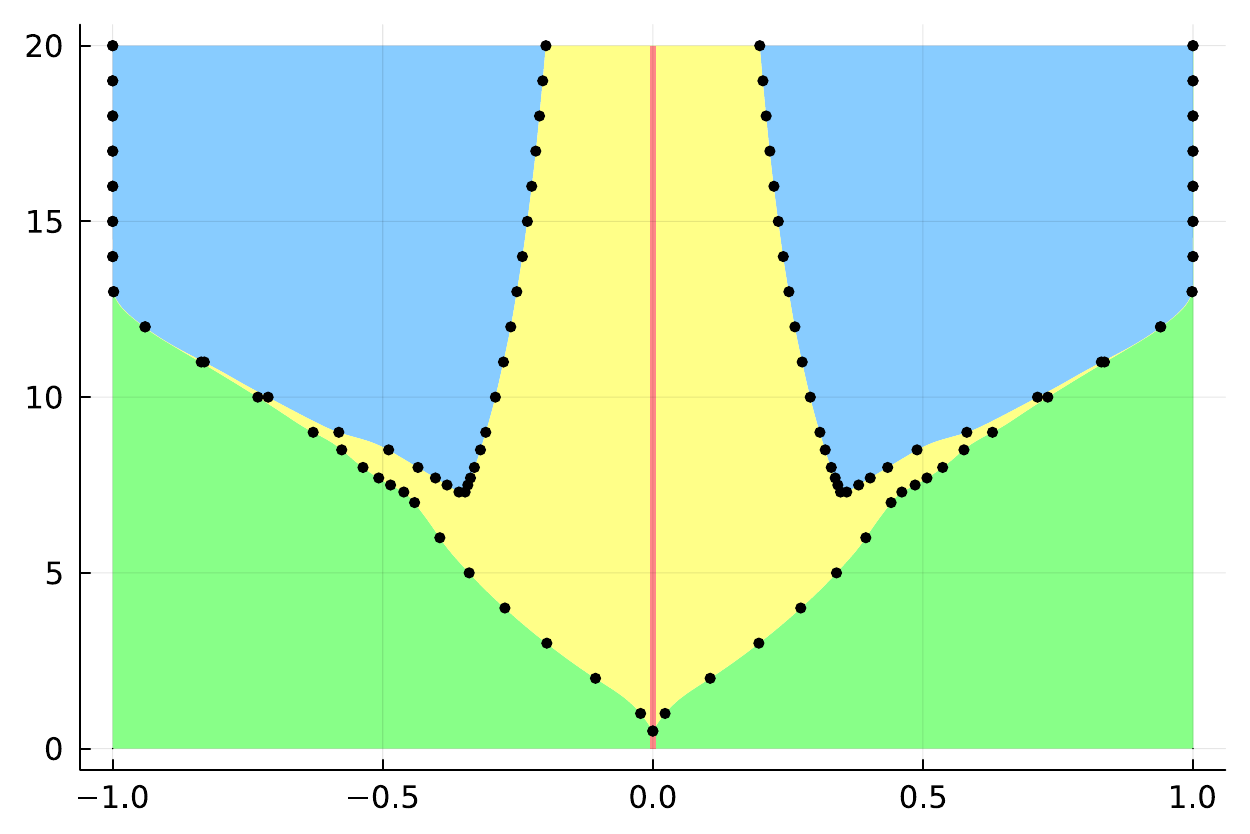 }
			\put(45,66.5){\footnotesize (a) $1/\beta=0$}
			\put(5.5,66.5){\footnotesize $U/t$}
			\put(99.5,4.5){\footnotesize $\nu$}
            \put(20,15){\footnotesize P}
            \put(80,15){\footnotesize P}
            \put(26,45){\footnotesize F}
            \put(77,45){\footnotesize F}				
			\put(39,23){\footnotesize Mixed}
			\put(55.5,23){\footnotesize Mixed}
			\put(46,36){\footnotesize AF}
			\put(49,35){\vector(2,-3){3}}
		\end{overpic}
		
		\vspace{.5cm}
		\hspace{-.6cm}
		\begin{overpic}[width=.46\textwidth]{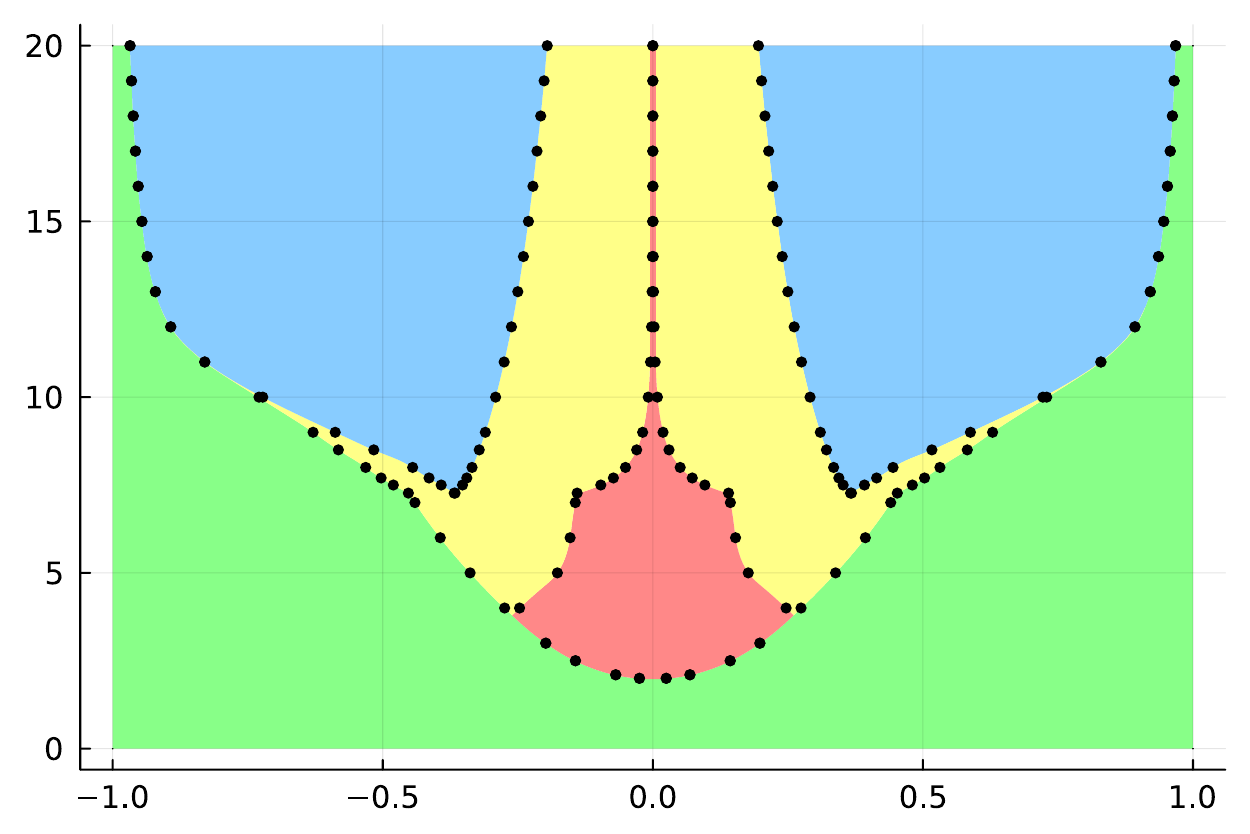 }
			\put(43,66.5){\footnotesize (b) $1/\beta= 0.2t$}
			\put(5.5,66.5){\footnotesize $U/t$}
			\put(99.5,4.5){\footnotesize $\nu$}
            \put(50,19){\footnotesize AF}
            \put(20,15){\footnotesize P}
            \put(80,15){\footnotesize P}
            \put(26,45){\footnotesize F}
            \put(77,45){\footnotesize F}				
			\put(40,31){\footnotesize Mixed}
			\put(54.5,31){\footnotesize Mixed}
		\end{overpic}
		\caption{\label{fig:2D} 
				Same as Fig.~\ref{fig:3D} but for the 2D Hubbard model. 
		}
	\end{center}
	\vspace{-0.4cm}
\end{figure}

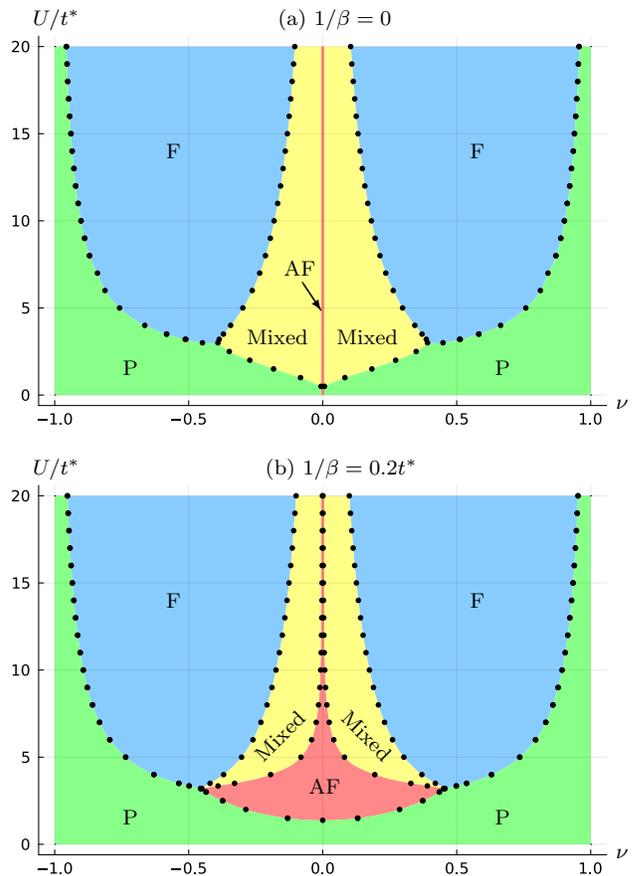
\begin{figure}
	\vspace{0.4cm}
	\begin{center}
		\hspace{-.6cm}
		\begin{overpic}[width=.46\textwidth]{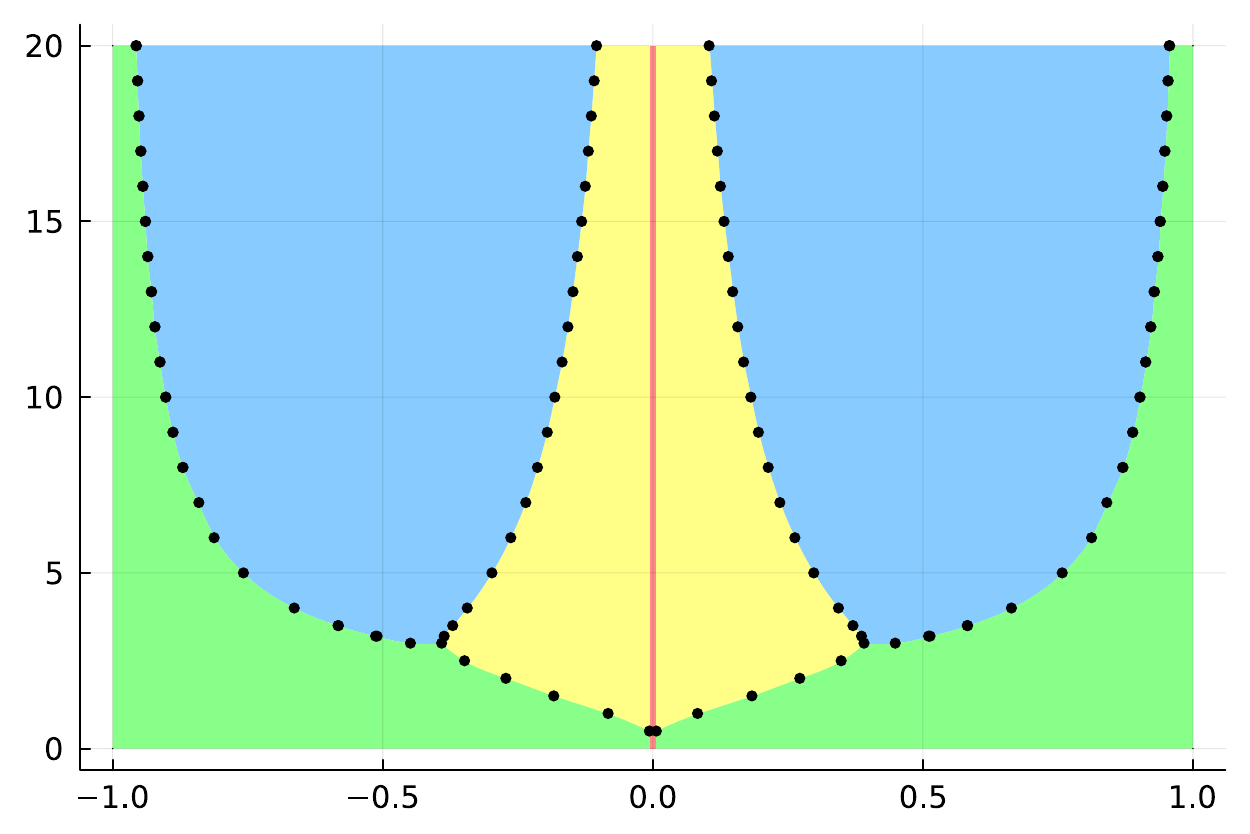 }
			\put(45,66.5){\footnotesize (a) $1/\beta = 0$}
			\put(5.5,66.5){\footnotesize $U/t^*$}
			\put(99.5,4.5){\footnotesize $\nu$}
			\put(20,10){\footnotesize P}
			\put(80,10){\footnotesize P}
			\put(27,45){\footnotesize F}
			\put(76,45){\footnotesize F}	
			\put(40,15){\footnotesize Mixed}
			\put(54.5,15){\footnotesize Mixed}
			\put(46,26){\footnotesize AF}
			\put(49,25){\vector(2,-3){3}}
	\end{overpic}
		
		\vspace{.5cm}
		\hspace{-.6cm}
		\begin{overpic}[width=.46\textwidth]{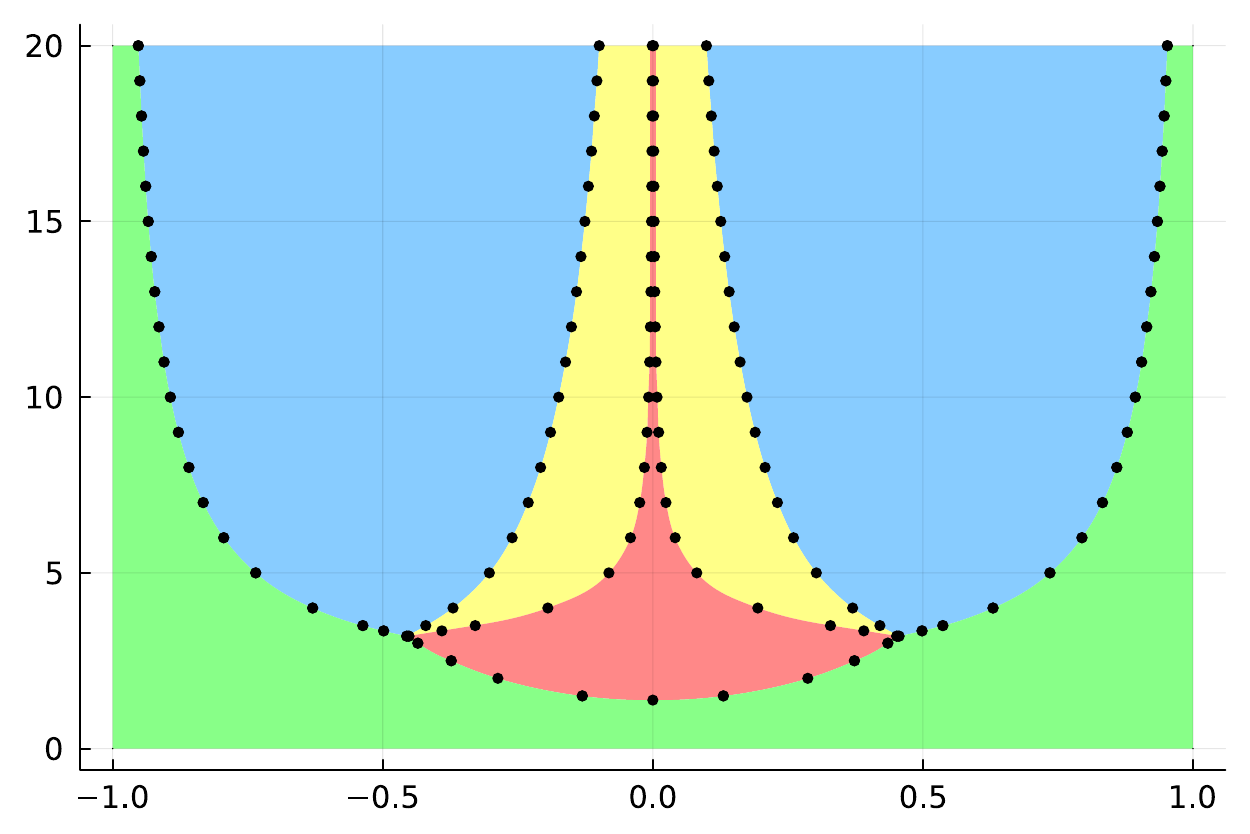 }
			\put(43,66.5){\footnotesize (b) $1/\beta= 0.2t^*$}
			\put(5.5,66.5){\footnotesize $U/t^*$}
			\put(99.5,4.5){\footnotesize $\nu$}
			\put(50,15){\footnotesize AF}
			\put(20,10){\footnotesize P}
			\put(80,10){\footnotesize P}
			\put(27,45){\footnotesize F}
			\put(76,45){\footnotesize F}	
			\put(41,20){\footnotesize {\rotatebox{45}{Mixed}}}
			\put(55,27){\footnotesize {\rotatebox{-45}{Mixed}}}
		\end{overpic}
		\caption{\label{fig:inftyD} 
		Same as Fig.~\ref{fig:3D} but for the Hubbard model in infinite dimensions, scaling $ t=t^*/\sqrt{2n}$ as $n\to\infty$ \cite{MV1989}.	
		}
	\end{center}
	\vspace{-0.4cm}
\end{figure}

\subsection{Updated phase diagrams of the Hubbard model}\label{sec:phasediagram} 
The ansatz in \eqref{Penn_ansatz} includes three important phases which, as is well-known since the pioneering papers on the Hubbard model \cite{G1963,H1963,K1963,P1966}, are relevant for Hubbard-like models: the anti-ferromagnetic (AF) phase, the ferromagnetic (F) phase, and the paramagnetic (P) phase; these phases  correspond to the following special cases of \eqref{Penn_ansatz},   
\begin{equation}\label{MF}  
	\begin{split} 
\text{AF}:&\quad (d_0,d_1,m_0,m_1) = (d_{\mathrm{AF}},0,0,m_{\mathrm{AF}})\\	\text{F}:&\quad (d_0,d_1,m_0,m_1) = (d_{\mathrm{F}},0,m_{\mathrm{F}},0), \\
\text{P}:&\quad (d_0,d_1,m_0,m_1) = (d_{\mathrm{P}},0,0,0),
	\end{split} 
\end{equation}
with $m_{\mathrm{F}}> 0$ and $m_{\mathrm{AF}}> 0$; note that the paramagnetic phase is a limit of the ferromagnetic and the anti-ferromagnetic phases (the limiting cases are $m_{\mathrm{F}}=0$ and $m_{\mathrm{AF}}=0$, respectively). 

We note in passing that, for the Hubbard model at half-filling, the restriction of Hartree--Fock theory to AF states can be done without loss of generality: it is known by mathematical proof that, for the Hubbard model on $\Z^n$, the absolute minimizer of unrestricted Hartree--Fock theory for $\nu=0$ is an AF state \cite{BLS1994}.
There exist similar mathematical results for $U/t\to \infty$ and/or $|\nu|\uparrow 1$ concerning F states  \cite{BLT2006}, but these results are weaker.

We use the name {\em mean-field theory} for Hartree--Fock theory restricted to AF, F, and P states in \eqref{MF}.

The benefit of using our updated Hartree--Fock theory is that it not only makes it possible to detect pure phases but also mixed phases where two distinct phases co-exist in a phase-separated state \cite{LW1997,LW2007}; see Section~\ref{sec:method} for more detailed explanations. Thus, the phase diagrams we obtain are richer than the established ones in the literature: in addition to conventional AF, F, and P phase regions, we also find mixed phase regions. 

Fig. \ref{fig:3D}(a) is a main result in this paper; as discussed in the introduction, it corrects a well-known phase diagram by Penn \cite[Fig.~9]{P1966} (the phase boundaries in Fig. \ref{fig:3D}(a) were obtained by interpolating between points, indicated by dots, which we computed numerically from scratch; a similar remark applies to all other phase diagrams in this paper). We find that, for $1/\beta=0$, the AF phase exists only strictly at half-filling, in agreement with what is known today by other methods. Instead of Penn's doped AF region, there are large mixed phases in the phase diagram (marked by yellow) away from half-filling; these mixed phases are between the AF and P phase for smaller values of $U$ (including $U/t=5$), and between AF and F for larger values of $U$ (including $U/t=10$ and $U/t=20$). For intermediate values of $U$ (including $U/t=10$), we also find (small) mixed regions  between F and P. 

To demonstrate the effect of finite temperature, we computed this phase diagram also at $1/\beta=0.2t$; our result is in Fig.~\ref{fig:3D}(b). The phase diagram at $1/\beta=0.2t$ is qualitatively similar to the one at $1/\beta=0$, and, in particular, there are still mixed regions. The main qualitative differences are as follows:  (i) while AF exists at half-filling down to $U\downarrow 0$ for $1/\beta=0$, there is a finite $U_c>0$ (depending on $1/\beta$) below which AF is absent if $1/\beta>0$, (ii) for $1/\beta>0$, there is a region away from half-filling where AF exists. We note in passing that the inverse of the function $U_c(T)$, $T=1/\beta$, is the N\'eel temperature $T_N(U)$; see \cite{LL2024} for recent analytic results on $T_N(U)$ for the 3D Hubbard model. 

The corresponding mean-field phase diagrams of the Hubbard model for  dimensions $n=1$, $n=2$, and $n=\infty$ are given in Figs.~\ref{fig:1D}, \ref{fig:2D}, and \ref{fig:inftyD}, respectively; in the latter case, we used the well-known scaling $t=t^*/\sqrt{2n}$ allowing the limit $n\to\infty$ to be non-trivial \cite{W1983,MV1989}. The  mean-field phase diagram for the 2D Hubbard model at zero temperature corrects the well-established one by Hirsch \cite[Fig.~3]{H1985}.  It appeared already in \cite[Fig.~1]{LW2007}, but Fig.~\ref{fig:2D} is an improvement in that the phase boundaries in the regions $|\nu|\to 1$ are more accurate (\cite[Fig.~1]{LW2007} was computed for a Hubbard model on a finite square lattice with $60^2$ sites, while Fig.~\eqref{fig:2D}(a) is for the infinite square lattice -- the free energy difference between the F and P phases is tiny and sensitive to finite size effects for small $1-|\nu|$, which led to large uncertainties in the phase boundaries in this region in \cite[Fig.~1]{LW2007}). 

To conclude, we mention that we performed extensive numerical computations of mean-field phase diagrams using the more general ansatz in \eqref{Penn_ansatz} (with four variational parameters), but we never found a Hartree--Fock solution different from the ones included in \eqref{MF}. 
This disagrees with results of Penn \cite{P1966} who found ferrimagnetic solutions (where both $m_0$ and $m_1$ are non-zero at the same time), but this is no contradiction since Penn used a different stability criterion. Thus, we have good reason to believe that the restriction from \eqref{Penn_ansatz} to \eqref{MF} is done without loss of generality; it would be interesting to prove this by mathematical arguments, or to find counterexamples.

\section{Mixed phases}\label{sec:method}
While the existence of mixed phases in the phase diagram of the 2D Hubbard model ($n=2$) was pointed out before \cite{LW1997,LW2007}, it has received little attention in the physics literature. For this reason, and to keep this paper self-contained, we review a method proving the existence of mixed phases \cite{LW1997,LW2007}, taking the 3D Hubbard model ($n=3$) as example. We choose this example to emphasize that mixed phases are not particular to 2D but exist in any dimension. 

\begin{figure}
	\vspace{0.4cm}
	\begin{center}
		\hspace{-.6cm}
		\begin{overpic}[width=.46\textwidth]{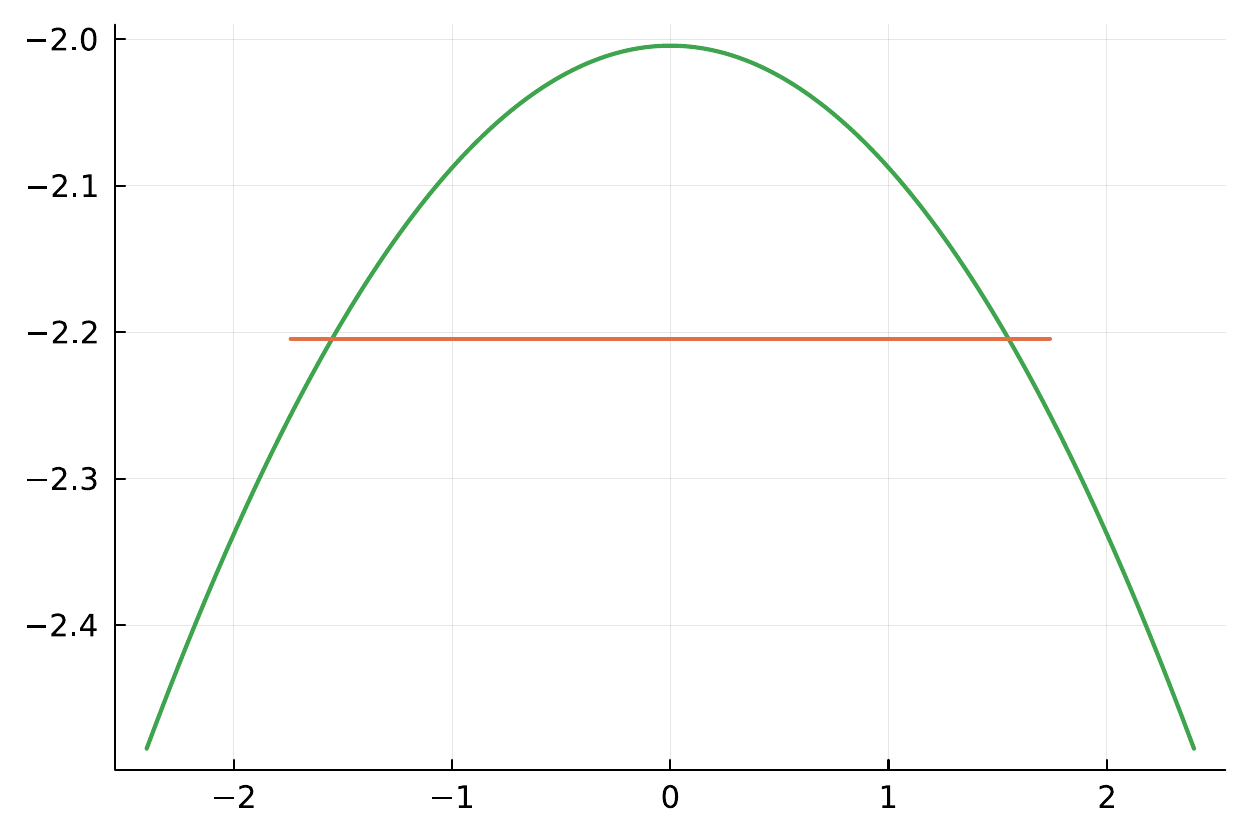 }
			\put(45,66.5){\footnotesize (a) $U = 5$}
			\put(8,66.5){\footnotesize $\cF$}
			\put(100,4){\footnotesize $\mu$}
		\end{overpic}
		
		\vspace{.5cm}
		\hspace{-.6cm}
		\begin{overpic}[width=.46\textwidth]{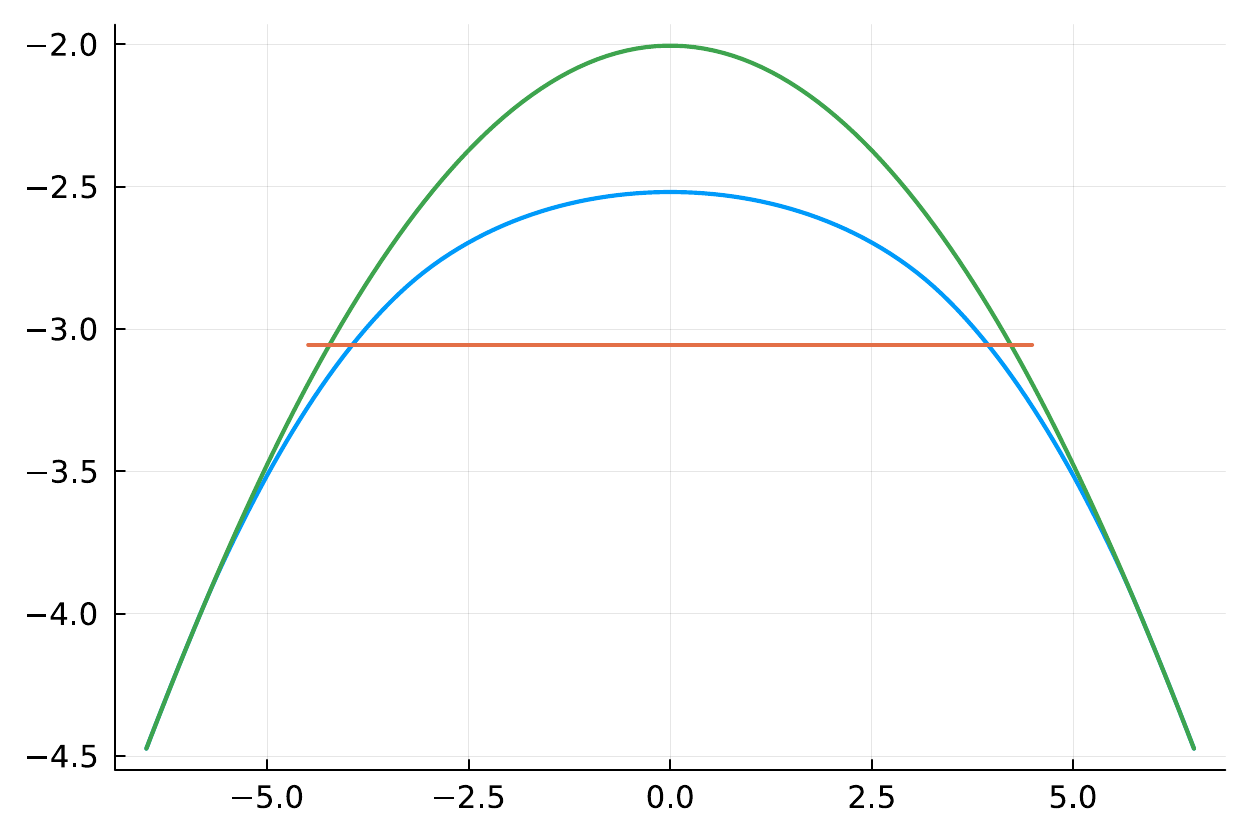 }
			\put(45,66.5){\footnotesize (b) $U = 10$}
			\put(8,66.5){\footnotesize $\cF$} 
			\put(100,4){\footnotesize $\mu$}
		\end{overpic}
		
		\vspace{.5cm}
		\hspace{-.6cm}
		\begin{overpic}[width=.46\textwidth]{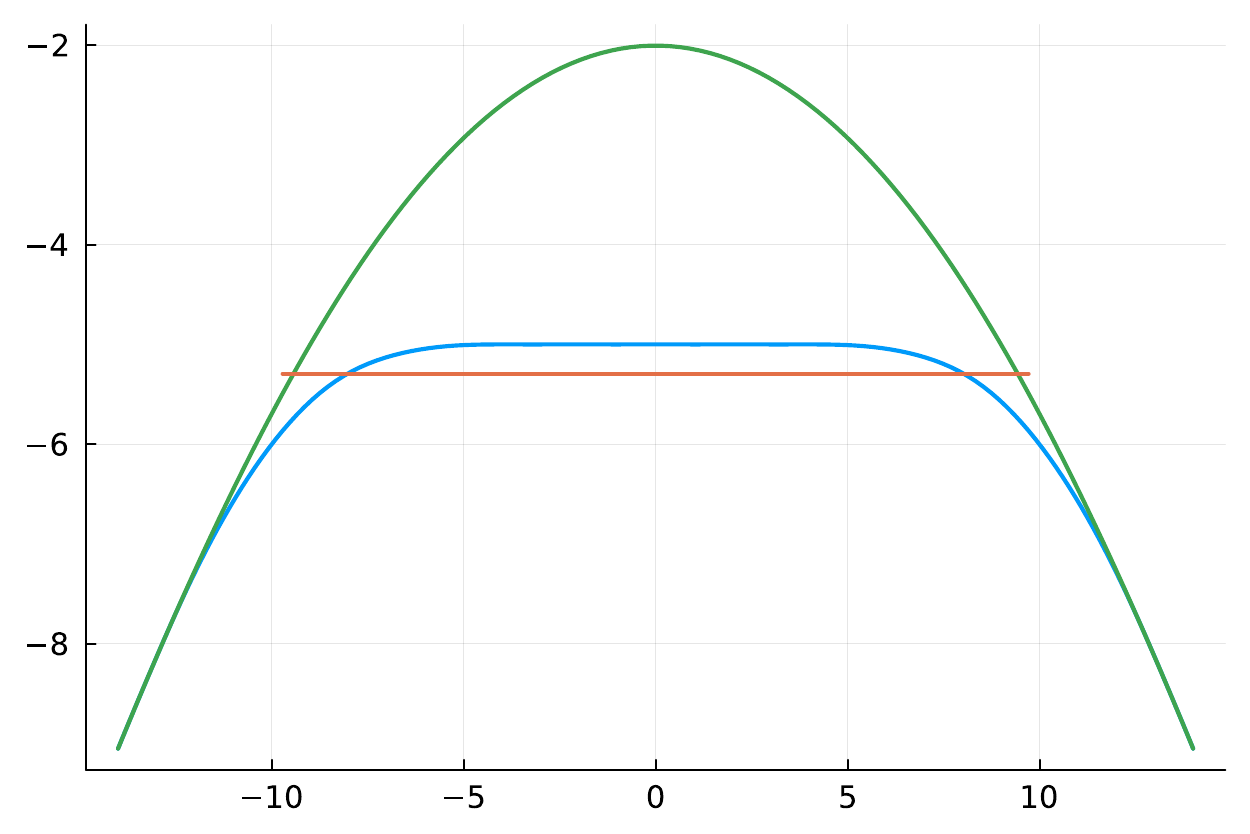 }
			\put(45,66.5){\footnotesize (c) $U = 20$}
			\put(6,66.5){\footnotesize $\cF$} 
			\put(100,4){\footnotesize $\mu$}
		\end{overpic}
		\caption{\label{fig2} 
			The Hartree--Fock free energy $\cF=\cF_{\mathrm{X}}$ (X=AF, F, P) for the 3D Hubbard model at zero temperature as a function of the chemical potential $\mu$, and for different phases X=AF (red curves), X=F (blue curves), and X=P (green curves); we set $t=1$.
			\\
			(a) $U=5$ with X=AF and X=P. The AF curve exists for $|\mu| \lesssim 1.74$. The F curve does not exist.
			\\
			(b) $U=10$ with X=AF, X=F, and X=P. 
			The AF curve exists for $|\mu| \lesssim 4.49$. The F and P curves intersect at $|\mu|\approx 5.73$ with slighly different slopes, as can be seen by zooming in; see  Fig.~\ref{fig:D=3,U=10}. 
			\\
			(c) $U=20$ with X=AF, X=F, and X=P. 
			The AF curve exists for $|\mu| \lesssim 9.71$. Within the numerical accuracy we can reach, the F and P curves have the same slope where they merge. 			
		}
	\end{center}
	\vspace{-0.4cm}
\end{figure}

To simplify formulas, we choose units such that $t=1$ in this section. Moreover, to simplify our discussion, we restrict ourselves to doping values in the range $0\leq \nu\leq 1$, which  corresponds to $\mu\geq 0$; corresponding results for negative doping values can be obtained in a simple way using particle--hole symmetry.

\subsection{The AF-P mixed phase in the Hubbard model}\label{sec:mixedAFP}  
As a representative example allowing us to focus on the key points, we discuss Hartree--Fock theory for the 3D Hubbard model at zero temperature restricted to AF states, 
\begin{align}\label{ansatz_AF}  
	d(\bx) = d_0,\quad \vec{m}(\bx) = (-1)^{\bx}m_1\vec{e}. 
\end{align} 
In this case, the Hartree--Fock function is the special case $n=3$ of 
\begin{multline}\label{FHF_AF} 
	\FHF(d_0,m_1;\mu)=\frac{U}{4}(m_1^2-d_0^2)
	\\ -\int_{[-\pi,\pi]^n} \frac{d^n\bk}{(2\pi)^n}
	\Big(\LnT(E_+(\bk)) + \LnT(E_-(\bk)) \Big),
\end{multline}  
where
\begin{align}\label{E_AF}  
	E_\pm(\bk) = \frac{U}{2}d_0-\mu \pm\sqrt{\varepsilon_0(\bk)^2 + \left( \frac{U}{2}m_1 \right)^2} 
\end{align} 
are the well-known AF band relations with the tight-binding band relation $\varepsilon_0(\bk)$ given in \eqref{veps0}.

We note in passing that, to obtain the plots for the Hubbard model presented in this paper, we computed the $\bk$-integral in \eqref{FHF_AF} etc.\ using the density of states in \eqref{DOS}; see \eqref{kintegralwithDOS} and Appendix~\ref{app:DOS} for further details.

\begin{figure}
	\vspace{0.4cm}
	\begin{center}
		\hspace{-.6cm}
		\begin{overpic}[width=.46\textwidth]{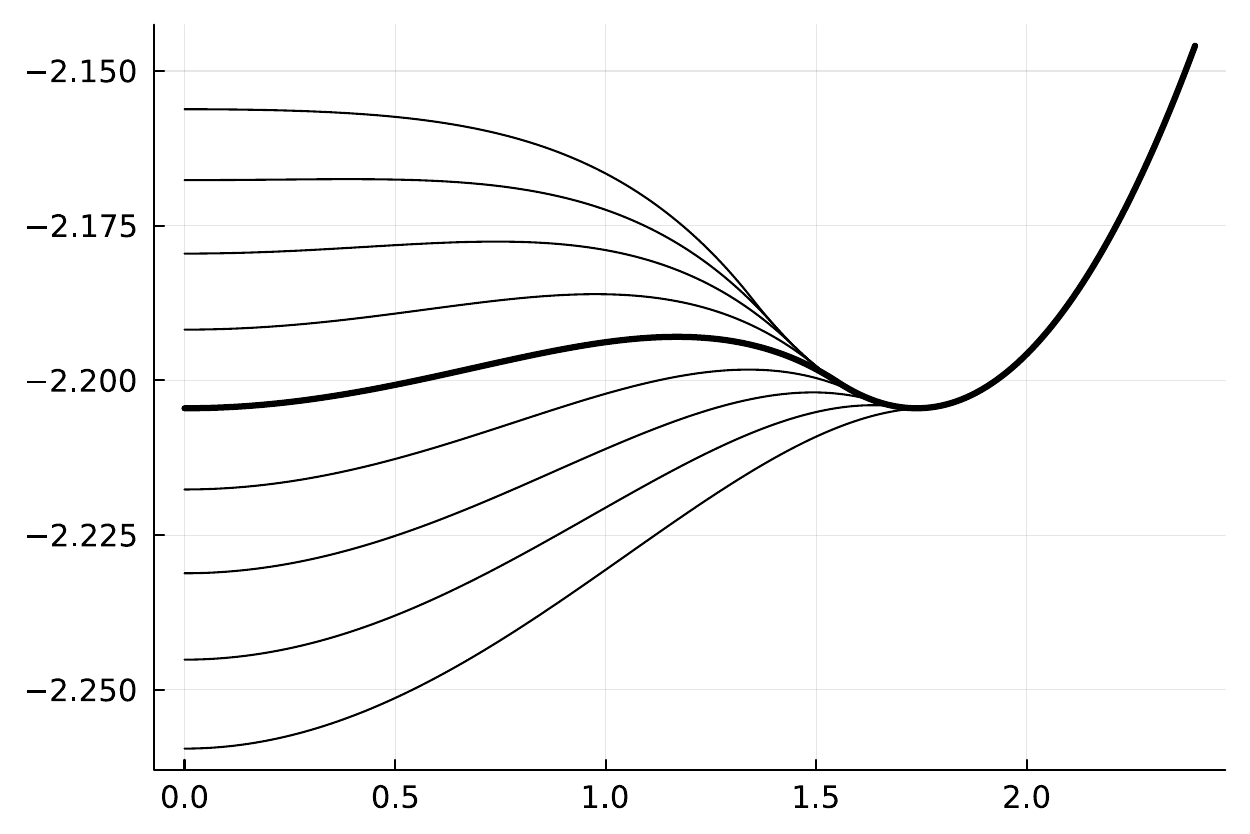 }
			\put(5,66.5){\footnotesize $\max_{d_0} \FHF$}
			\put(99.3,4.2){\footnotesize $\frac{U}2 m_1$}
			\put(14.5,59.5){\footnotesize $n=-4$}
			\put(14.5,53.5){\footnotesize $n=-3$}
			\put(14.5,47.7){\footnotesize $n=-2$}
			\put(14.5,41.9){\footnotesize $n=-1$}
			\put(14.5,35.6){\footnotesize $n=0$}
			\put(14.5,29){\footnotesize $n=1$}
			\put(14.5,22.6){\footnotesize $n=2$}
			\put(14.5,15.5){\footnotesize $n=3$}
			\put(14.5,8.5){\footnotesize $n=4$}
		\end{overpic}
		\caption{\label{fig:D=3,U=5} 
Details on how the mixed phase (yellow region) between AF and P in Fig.~\ref{fig:3D}(a) at $U=5$ is obtained: $\max_{d_0}\FHF$ with $\FHF=\FHF(d_0,m_1;\mu)$ defined in \eqref{FHF_AF} as a function of the AF order parameter $\Delta=\tfrac{U}2m_1$ of the 3D Hubbard model at $U=5$ and zero temperature for different values $\mu = \mu_c+ 0.05n$ ($n=-4, \dots, 4$), with $\mu_c \approx 1.550$ the value at which the free energies of the P and AF solutions coincide (we set $t=1$).
		}
	\end{center}
	\vspace{-0.4cm}
\end{figure}

A convenient and safe way to solve the Hartree--Fock equations is to use the min-max procedure in \eqref{minmax} which, for our ansatz in \eqref{ansatz_AF}, reduces to 
\begin{align}\label{minmaxAF}  
\FHF(d_0^{*},m_1^{*};\mu) =\min_{m_1}\max_{d_0}\FHF(d_0,m_1;\mu). 
\end{align} 
To find the relevant Hartree--Fock solutions $(d_0^{*},m_1^{*})$, one can also solve the standard  Hartree--Fock equations $\frac{\partial\FHF}{\partial d_0}=0$, $\frac{\partial\FHF}{\partial m_1}=0$ given by 
\begin{equation}\label{AF_HFeqs} 
	\begin{split}  
		d_0 = & \int \frac{d^n\bk}{(2\pi)^n}
		\Big(\fT(E_+(\bk)) + \fT(E_-(\bk)) \Big) ,\\
		m_1 = -& 
		\frac{U}{2}\int \frac{d^n\bk}{(2\pi)^n}
		\frac{\fT(E_+(\bk)) - \fT(E_-(\bk))}{\sqrt{\varepsilon_0(\bk)^2 + \left( \frac{U}{2}m_1 \right)^2}}m_1, 
	\end{split} 
\end{equation} 	
respectively, with $f_\beta(\eps)$ in \eqref{fT} and $\int$ short for $\int_{[-\pi,\pi]^n}$. For later reference, we also write the doping constraint in \eqref{rho1}, 
\begin{align}\label{rho_AF}  
\nu = \int \frac{d^n\bk}{(2\pi)^n}
\Big(\fT(E_+(\bk)) + \fT(E_-(\bk)) \Big). 
\end{align} 
As already mentioned, even though the first equation in \eqref{AF_HFeqs} and \eqref{rho_AF} show that $d_0$ and $\nu$ have the same value at a solution of the Hartree--Fock equations, this does {\em not} mean that the variational parameter $d_0$ can be eliminated from the problem by introducing a renormalized chemical potential as in \eqref{tmu}: as explained in Section~\ref{sec:bug}, this is a dangerous shortcut that has led to incorrect results in the literature.

A {\em stable} solution $(d_0^{*},m_1^{*})=(d_{\mathrm{AF}},m_{\mathrm{AF}})$ of the Hartree--Fock equations \eqref{AF_HFeqs} such that $m_{\mathrm{AF}}>0$ describes an AF phase, 
and a solution $(d_0^{*},m_1^{*})=(d_{\mathrm{P}},0)$  describes a P phase. We emphasize the importance of the qualifier {\em stable} in the previous sentence: as will be shown, the Hartree--Fock equations can have two AF solutions, $(d_0^{(i)},m_1^{(i)})$ with $m_1^{(i)}>0$ for $i=1,2$ and,  in such a case, we take as $(d_{\mathrm{AF}},m_{\mathrm{AF}})$ the solution with the lowest free energy $\FHF(d_0^{(i)},m_1^{(i)};\mu)$, i.e., $\FHF(d_{\mathrm{AF}},m_{\mathrm{AF}};\mu) = \min_{i}\FHF(d_0^{(i)},m_1^{(i)};\mu)$.
With this prescription, the free energies of the AF and P states are given by (if both exist) 
\begin{align}\label{cF_AFP}  
	\cF_{\mathrm{AF}}(\mu) = \FHF(d_{\mathrm{AF}},m_{\mathrm{AF}};\mu),
	\quad \cF_{\mathrm{P}}(\mu) = \FHF(d_{\mathrm{P}},0;\mu),
\end{align} 	
and the Hartree--Fock free energy for states as in \eqref{ansatz_AF} is 
\begin{align}\label{FHFd0*m1*} 
	\FHF(d_0^*,m_1^*;\mu)= \min(\cF_{\mathrm{AF}}(\mu),\cF_{\mathrm{P}}(\mu)); 
\end{align} 
it is the latter free energy which determines doping as
\begin{align}\label{rho_AFpartial} 
	\nu = -\frac{\partial\FHF(d_0^{*},m_1^{*};\mu)}{\partial\mu} .
\end{align}

We mention a technical point which careful readers might wonder about otherwise. In general, a solution $(d_0^{*},m_1^{*})$ depends on $\mu$ and, for this reason, it seems important that the derivative in \eqref{rho_AFpartial} is a partial one. 
Fortunately, this is not the case: in \eqref{rho_AFpartial}, it does not matter if one uses the partial $\mu$-derivative (as written) or the total $\mu$-derivative (which can be more convenient); this is true since the $d_0$- and $m_1$-derivatives of $\FHF$ vanish at solutions of the Hartree--Fock equations. Thus, one can insert a solution $(d_0^{*},m_1^{*})$ in $\FHF$ before or after the $\mu$-differentiation to obtain $\nu$: it gives the same result.

\subsubsection{Free energy plots}\label{sec:Fplots}
We discuss Figs.~\ref{fig2}(a) and \ref{fig:D=3,U=5}
to explain how the different phases in Fig.~\ref{fig:3D}(a) at $U=5$ and $1/\beta=0$ are obtained. The red and green curves in Fig.~\ref{fig2}(a) show the AF and P free energies (defined in \eqref{cF_AFP}) as functions of the chemical potential $\mu$, respectively, and Fig.~\ref{fig:D=3,U=5} shows in more detail how Fig.~\ref{fig2}(a) arises. For this example where $U=5$, we can ignore the F solution (since it does not exist).

From Fig.~\ref{fig2}(a) we see that the AF solution exists for $0\leq \mu \lesssim 1.739$, the P solution exists for all $\mu\geq 0$, and the AF and P free energies coincide at $\mu=\mu_c\approx 1.550$; moreover, for $0\leq \mu<\mu_c$ the AF free energy is lower, and for $\mu>\mu_c$ the P free energy is lower. Thus, the Hartree--Fock free energy in \eqref{FHFd0*m1*} is a continuous function of $\mu$: it follows the red curve between $\mu=0$ and $\mu=\mu_c$ and it follows the green curve for $\mu\geq\mu_c$. Since the doping is the negative slope of this function (by \eqref{rho_AFpartial}), the doping is only a piecewise continuous function of $\mu$ with a jump at $\mu=\mu_c$ from  
\begin{align}\label{rho_AFmax} 
	\nu^{\mathrm{max}}_{\mathrm{AF}} = \left. -\frac{\partial \cF_{\mathrm{AF}}(\mu)}{\partial\mu}\right|_{\mu=\mu_c}
\end{align} 	
to 
\begin{align}\label{rho_Pmin}  
	\nu^{\mathrm{min}}_{\mathrm{P}} = \left. -\frac{\partial \cF_{\mathrm{P}}(\mu)}{\partial\mu}\right|_{\mu=\mu_c}. 
\end{align} 
In our example, $\nu^{\mathrm{max}}_{\mathrm{AF}}=0$ and $\nu^{\mathrm{min}}_{\mathrm{P}}\approx 0.258$ (at zero temperature and for the Hubbard model, one always finds  $\nu^{\mathrm{max}}_{\mathrm{AF}}=0$, but $\nu^{\mathrm{max}}_{\mathrm{AF}}$ can be non-zero in other cases). 
Thus, we find an AF phase for $0\leq \nu\leq \nu^{\mathrm{max}}_{\mathrm{AF}}$ and a P phase for $\nu^{\mathrm{min}}_{\mathrm{P}}\leq\nu\leq 1$, but for 
\begin{align}
\label{rho_mixed} 
\nu^{\mathrm{max}}_{\mathrm{AF}}<\nu< \nu^{\mathrm{min}}_{\mathrm{P}}
\end{align} 
we find that neither the AF nor the P phase is thermodynamically stable. As explained in Section~\ref{sec:details}, in this unstable doping regime \eqref{rho_mixed}, our ansatz \eqref{ansatz_AF} describes a mixed state where macroscopic AF and P regions coexist; we call this a {\em mixed phase}. 

To summarize: {\em The graphs in Fig.~\ref{fig2}(a) prove that, for $n=3$, $U=5$, $1/\beta=0$, and within the ansatz in \eqref{ansatz_AF}, the best possible approximation to the unrestricted Hartree--Fock state at $\nu=0$ is the AF state, for $0<\nu<\nu^{\mathrm{min}}_{\mathrm{P}}\approx 0.258$ it is a mixed state where macroscopically large AF and P regions coexist, and for $\nu^{\mathrm{min}}_{\mathrm{P}}\leq \nu\leq 1$  it is the P-state;} see  Fig.~\ref{fig:3D}(a) at $U=5$ where the AF, mixed, and P phases are colored in red, yellow, and green, respectively. 

Fig.~\ref{fig:D=3,U=5} shows in greater detail how the mixed phase comes about: it shows 
\begin{align*}  
	\max_{d_0}\FHF=\max_{d_0}\FHF(d_0,m_1;\mu) 	
\end{align*} 	
as a function of the AF order parameter $\Delta=\frac{U}{2}m_1$ for different values of $\mu$, namely $\mu=\mu_c$ (bold  curve) and $\mu=\mu_c \pm 0.05n$  for $n=1,\ldots,4$. 
One way to obtain these curves is to compute $d_0=d^{*}_0(m_1,\mu)$ by solving the Hartree--Fock equation for $d_0$ in \eqref{AF_HFeqs} (this solution exists and is unique, and it is easy to compute it numerically), which gives 
\begin{align*} 
\max_{d_0}\FHF = \FHF(d^{*}_0(m_1,\mu),m_1;\mu).
\end{align*} 
The function $\max_{d_0}\FHF$ of $\Delta=\frac{U}{2}m_1$ is interesting since its absolute minimum at fixed $\mu$ determines the best variational state within the ansatz \eqref{ansatz_AF}; see \eqref{minmaxAF}. 

In Fig.~\ref{fig:D=3,U=5} one clearly sees how the AF minimum at $m_1=m_{\mathrm{AF}}>0$ for $0\leq \mu<\mu_c$ jumps to the P minimum at $m_1=0$ for $\mu>\mu_c$, and that this jump at $\mu=\mu_c$ leads to a non-trivial doping region where neither the AF nor the P phase is stable: the P-minimum of $\max_{d_0}\FHF $ becomes less than its AF minimum at $\mu=\mu_c$ because $\max_{d_0}\FHF$ decreases faster with $\mu$ at $m_1=0$ than at $m_1=m_{\mathrm{AF}}>0$, which by \eqref{rho_AFpartial} implies
\begin{align*} 
\nu^{\mathrm{min}}_{\mathrm{P}}
= \left. -\frac{\partial \cF_{\mathrm{P}}(\mu)}{\partial\mu}\right|_{\mu=\mu_c}
> \left. -\frac{\partial \cF_{\mathrm{AF}}(\mu)}{\partial\mu}\right|_{\mu=\mu_c}
= \nu^{\mathrm{max}}_{\mathrm{AF}}.
\end{align*} 
It is also interesting to note that the P solution at $m_1=0$ is a local free energy maximum for $0\leq \mu\lesssim 1.380$. This instability of the P solution was noted already by Penn \cite{P1966}; however, since $1.380<\mu_c$,  this local instability of the P solution does not concern us: it occurs at $\mu$-values $0\leq \mu\lesssim 1.380$ where the AF solution is stable.

We mention a simple physics interpretation of why the AF solution is robust in the sense that it is the absolute free energy minimum in a significant $\mu$-regime but, despite of this, it is only stable at  half-filling $\nu=0$ \cite{LW1997}: as is well-known, the AF bands in \eqref{E_AF} describe a  system with a gap  $\Delta=\frac{U}2m_1$. At half-filling $\mu=0$, the AF state with $m_1=m_{\mathrm{AF}}>0$ is stable, and the chemical potential is in the middle of the AF gap. As we increase the chemical potential within the AF gap, $0\leq \mu< \frac{U}{2}m_{\mathrm{AF}}$, the  AF free energy cannot change. However, as we increase $\mu$, the P free energy at $m_1=0$ decreases (since the P state is not gapped, its doping increases with increasing $\mu$), and at  $\mu=\mu_c<\frac{U}{2}m_{\mathrm{AF}}$ it becomes lower than the AF free energy. Thus, if we restrict ourselves to translationally invariant states as in \eqref{ansatz_AF}, we jump from a half-filled AF insulator at $\mu=\mu_c-0^+$ to a significantly doped metal for  $\mu=\mu_c+0^+$.

\subsubsection{Details on mixed phases}\label{sec:details}  
We now explain that, at the value $\mu=\mu_c$ of the chemical potential where the AF and P free energies coincide, 
one can reach doping values in the unstable range \eqref{rho_mixed} by variational states where a fraction $w$ is in the AF state and a fraction $1-w$ in the P state ($0<w<1$); there are many such states, and any of them is described by the Hartree--Fock function  
\begin{align}\label{FHFmixed} 
\FHF_{\mathrm{mixed}} = w\FHF(d_0,m_1;\mu) +(1-w)  \FHF(\tilde d_0,0;\mu) 	
\end{align} 
at $\mu=\mu_c$, for $d_0=d_{\mathrm{AF}}$, $m_1=m_{\mathrm{AF}}$ and $\tilde d_0=d_{\mathrm{P}}$. 
By \eqref{rho}, the corresponding doping is minus the $\mu$-derivative of this function at $\mu=\mu_c$, i.e., 
\begin{align*}
	\nu = w\nu^{\mathrm{max}}_{\mathrm{AF}} +  (1-w)\nu^{\mathrm{min}}_{\mathrm{P}}, 
\end{align*}
using \eqref{rho_AFmax}--\eqref{rho_Pmin}. 
This allows us to determine the parameter $w$ in terms of the doping: 
\begin{align} 
	w = \frac{\nu^{\mathrm{min}}_{\mathrm{P}} - \nu}{\nu^{\mathrm{min}}_{\mathrm{P}} - \nu^{\mathrm{max}}_{\mathrm{AF}}}
\end{align} 
for $\nu$ as in \eqref{rho_mixed}. Thus, we refer to the unstable doping range in \eqref{rho_mixed} as a mixed phase. 

In a nutshell: the Hartree--Fock variational states describing the mixed phase are phase separated states. More specifically, one can construct such  states as follows. 
\begin{enumerate} 
\item Consider the hypercube $[0,1]^n$, i.e., the set of  all $\mathbf{s}=(s_1,\ldots,s_n)$ such that  $0\leq s_i\leq 1$ for $i=1,\ldots,n$ (this is the unit interval, square, and cube for $n=1$, $2$, and $3$, respectively). For fixed $w$ in the range $0\leq w\leq 1$, choose any subset $R_w$ of $[0,1]^n$ that has volume $w$. 
(One simple example for $R_w$ would be the set of all $\mathbf{s}\in[0,1]^n$ such that $0\leq s_1\leq w$.) 
\item For the model on the finite lattice with $L^n$ lattice  sites $\bx=(x_1,\ldots,x_n)$ where $x_i=1,\ldots,L$ (and periodic boundary conditions), we consider the following variational field configurations, 
\begin{align*} 
(d(\bx),\vec{m}(\bx))
= \begin{cases} (d_0,m_1\vec{e}(-1)^{\bx})& (\bx/L\in R_w) \\ 
(\tilde d_0,\vec{0}) & (\bx/L\notin R_w) \end{cases} 		
\end{align*} 	
where $d_0$, $m_1$, and $\tilde d_0$ are real variational parameters, and $\vec{e}\in\R^3$ is an arbitrary unit vector (note that $\bx/L\in[0,1]^n$ for all lattice sites $\bx$). 
\end{enumerate} 
Clearly, there are very many such variational states (since there are many possible choices for $R_w$; note that $R_w$ need not be simply connected --- e.g.\ for $n=1$, one can choose $R_w$ as the union of an arbitrary number of disjoint intervals of total length $w$, and for $n>1$ the boundaries of $R_w$ can be curved). 

For any such variational state, $N_1$ sites are in the AF phase and $N_2=L^n-N_1$ sites are in the P phase, with $N_1/L^n\to w$ in the limit $L\to\infty$. For such an ansatz, the Hartree--Fock function converges to $\FHF_{\mathrm{mixed}}$ in \eqref{FHFmixed} in the limit $L\to\infty$, independently of the choice of $R_w$ and $\vec{e}$: there are certainly contributions from the phase boundaries to the Hartree--Fock function, but such contributions scale like $L^{n-1}/L^n=1/L$ for large $L$ and, for this reason, phase boundary contributions to $\FHF$ vanish in the limit $L\to \infty$. 

Thus, one can obtain $\FHF_{\mathrm{mixed}}$ in \eqref{FHFmixed} from a variational ansatz depending on the variational parameters $p=d_0$, $m_1$, $\tilde d_0$, and $w$. 
For each of these variational parameters $p$, one obtains a corresponding Hartree--Fock equation $\frac{\partial\FHF_{\mathrm{mixed}}}{\partial p}=0$. 
Thus, for fixed $w$, the parameters $d_0=d_{\mathrm{AF}}$, $m_1=m_{\mathrm{AF}}$, $\tilde d_0=d_{\mathrm{P}}$ are fixed by the same equation that fix them in the limiting cases $w=0$ and $w=1$ (this is true since $\frac{\partial \FHF_{\mathrm{mixed}}}{\partial d_0}=0$, $\frac{\partial \FHF_{\mathrm{mixed}}}{\partial m_1}=0$ are the Hartree--Fock equations in \eqref{AF_HFeqs}, and $\frac{\partial \FHF_{\mathrm{mixed}}}{\partial \tilde d_0}=0$ are these very equations in the special case $m_1=0$ with  $d_0$ renamed to $\tilde d_0$). Moreover, for $1\leq \mu<\mu_c$ and $\mu>\mu_c$ the Hartree--Fock free energy is minimized by $w=1$ (AF state) and $w=0$ (P state), respectively. 
However, at $\mu=\mu_c$,  the Hartree--Fock equation 
\begin{align*}
\frac{\partial\FHF_{\mathrm{mixed}}}{\partial w}
= \FHF(d_{\mathrm{AF}},m_{\mathrm{AF}};\mu) -   \FHF(d_{\mathrm{P}},0;\mu) 	=0	 
\end{align*} 	
is fulfilled, allowing for solutions where $0<w<1$ which can reach doping values in \eqref{rho_mixed}, as explained. 

To summarize: {\em By a simple ansatz for the Hartree--Fock variational fields as in \eqref{ansatz_AF}, we include automatically, and without additional computational effort, a plethora of phase separated states where two such simple states coexist in macroscopic regions, and it can happen that such a phase separated state is a better approximation to an unrestricted Hartree--Fock state than any of the simple states described by this ansatz.} 

\begin{figure}
	\vspace{0.4cm}
	\begin{center}
		\hspace{-.6cm}
		\begin{overpic}[width=.46\textwidth]{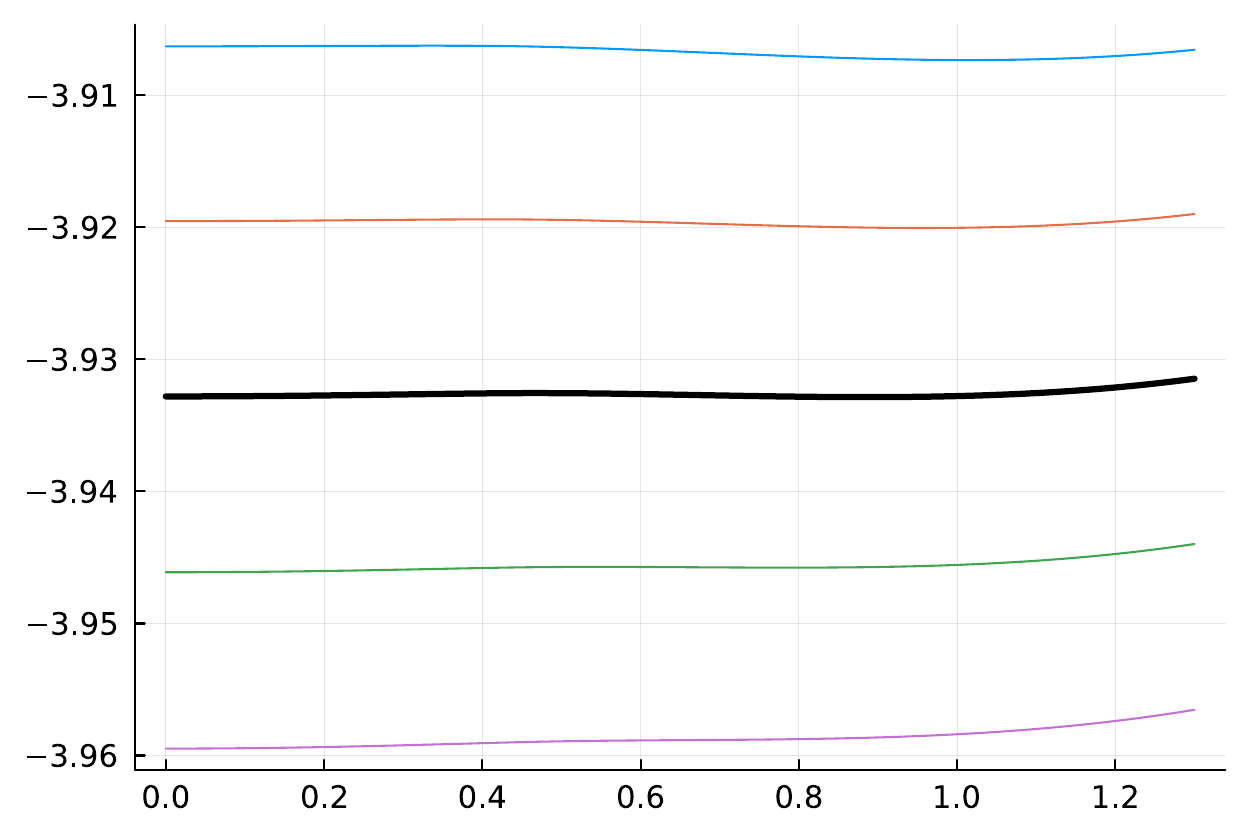 }
			\put(55,66.5){\footnotesize (a)}
			\put(6,66.5){\footnotesize $\max_{d_0} \FHF$}
			\put(100,4){\footnotesize $\frac{U}{2}m_0$}
			\put(14.5,60){\footnotesize $n=-2$}
			\put(14.5,50){\footnotesize $n=-1$}
			\put(14.5,36){\footnotesize $n=0$}
			\put(14.5,22){\footnotesize $n=1$}
			\put(14.5,8){\footnotesize $n=2$}
		\end{overpic}
		\\
		\vspace{.4cm}
		\hspace{-.6cm}
		\begin{overpic}[width=.46\textwidth]{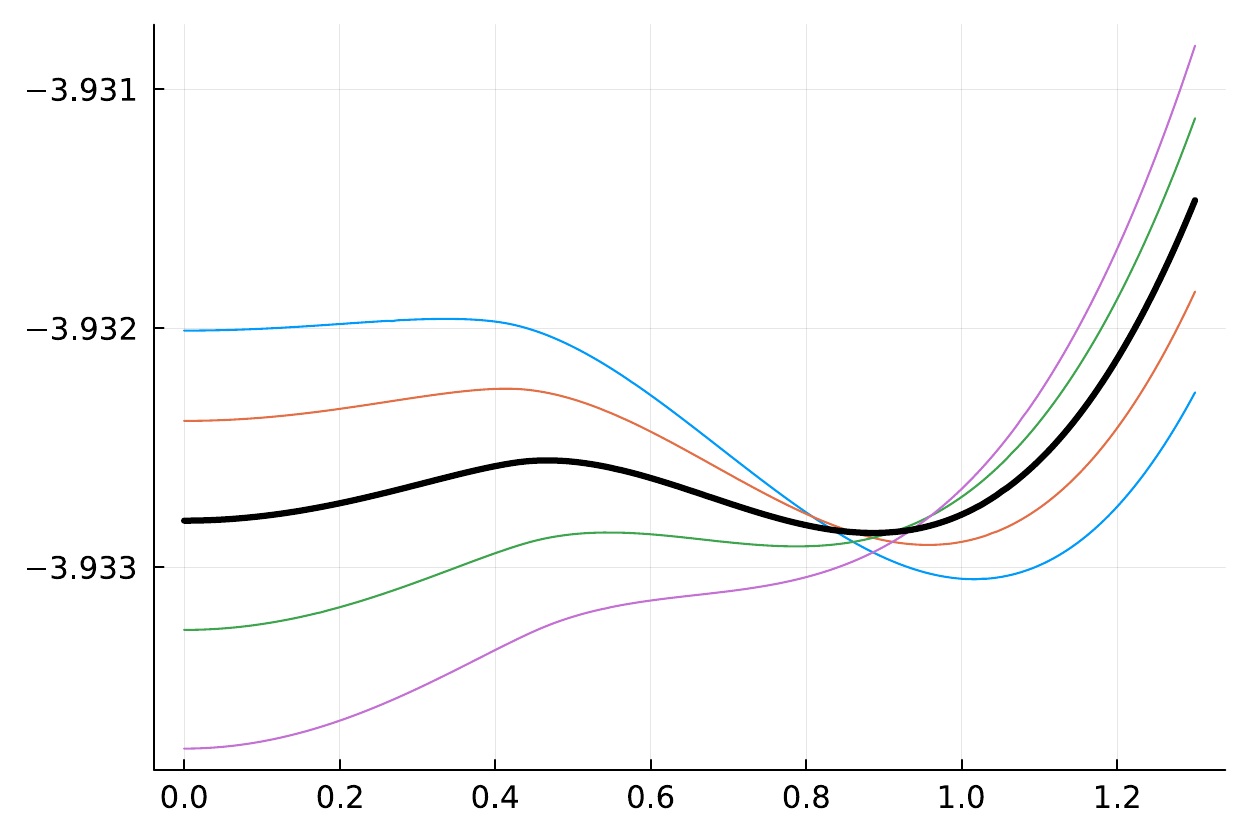 }
			\put(55,66.5){\footnotesize (b)}
			\put(6,66.5){\footnotesize $\max_{d_0}\FHF + 0.02n\nu_{\mathrm{F}}^{\mathrm{max}}$}
            \put(100,4){\footnotesize $\frac{U}{2}m_0$}
			\put(14.5,41.5){\footnotesize $n=-2$}
			\put(14.5,34.5){\footnotesize $n=-1$}
			\put(14.5,27){\footnotesize $n=0$}
			\put(14.5,18){\footnotesize $n=1$}
			\put(14.5,8.5){\footnotesize $n=2$}
		\end{overpic}
		\caption{\label{fig:D=3,U=10}
			Details on how the mixed phase (yellow region) between F and P in Fig.~\ref{fig:3D}(a) at $U=10$ is obtained: (a) $\max_{d_0}\FHF$ with $\FHF=\FHF(d_0,m_0;\mu)$ given by \eqref{FHF_F}--\eqref{E_F} as a function of the F order parameter $\Delta_{\mathrm{F}}=\tfrac{U}2m_0$ of the 3D Hubbard model at $U=10$ and zero temperature for different values $\mu = \mu_c+ 0.02n$ ($n=-2, \dots, 2$), with $\mu_c \approx 3.9328$ the value at which the free energies of the P and F solutions coincide (we set $t=1$).
			(b) Same as (a), but we have subtracted off the downward motion of the curves by plotting
			$\max_{d_0} \mathcal{F}(d_0, m_0; \mu) + 0.02n \nu_{\mathrm{F}}^{\mathrm{max}}$ where $\nu_{\mathrm{F}}^{\mathrm{max}}\approx 0.643$ is the doping of the F curve for $\mu = \mu_{c}$. The figure shows that the P solution is moving downwards at speed $\nu_{\mathrm{P}}^{\mathrm{min}} > \nu_{\mathrm{F}}^{\mathrm{max}}$, where $\nu_{\mathrm{P}}^{\mathrm{min}}\approx 0.665$ is the doping of the P curve for $\mu = \mu_{c}$. 
			This shows that a mixed phase exists for $\nu_{\mathrm{F}}^{\mathrm{max}}<\nu<\nu_{\mathrm{P}}^{\mathrm{min}}$. 
		}
	\end{center}
	\vspace{-0.4cm}
\end{figure}

The example discussed in this section is representative. Mixed phases are ubiquitous in Hubbard-like models, and while details differ, the mechanism leading to mixed phases is always the same: {\em The free energy vs.\ $\mu$ curves of two qualitatively different Hartree--Fock solutions intersect at some $\mu=\mu_c$, leading to a discontinuity of the $\mu$-derivative of the free energy; this discontinuity corresponds to a jump of doping and a mixed phase at $\mu=\mu_c$.}

\subsection{Other mixed phases in the Hubbard model}\label{sec:otherphases} 
For larger values of $U$, Hartree--Fock theory for the 3D Hubbard model also has F solutions described by the ansatz 
\begin{align}\label{ansatz_F}  
	d(\bx) = d_0,\quad \vec{m}(\bx) = m_0\vec{e}. 
\end{align}   
For this ansatz, the Hartree--Fock function reduces to the special case $n=3$ of  
\begin{multline}\label{FHF_F} 
	\FHF(d_0,m_0;\mu)=\frac{U}{4}(m_0^2-d_0^2)
	\\ -\int_{[-\pi,\pi]^n} \frac{d^n\bk}{(2\pi)^n}
	\Big(\LnT(E_+(\bk)) + \LnT(E_-(\bk)) \Big) 
\end{multline}  
with the well-known F band relations   
\begin{align}\label{E_F}  
	E_\pm(\bk) = \frac{U}{2}d_0-\mu + \varepsilon_0(\bk) \pm  \frac{U}{2}m_0
\end{align} 
and $\varepsilon_0(\bk)$ in \eqref{veps0}.
(To keep our notation simple, we use the same symbols $\FHF$ and $E_\pm$ here as for the AF ansatz \eqref{ansatz_AF} discussed in Section~\ref{sec:mixedAFP}; we hope that no confusion arises from this.)

Again, it is convenient to solve the Hartree--Fock equations by the min-max procedure \eqref{minmax},
\begin{align}\label{minmaxF} 
	\FHF(d_0^{*},m_0^{*};\mu) =\min_{m_0}\max_{d_0}\FHF(d_0,m_0;\mu), 
\end{align} 
and a stable solution $(d_0^{*},m_0^{*})=(d_{\mathrm{F}},m_{\mathrm{F}})$ with $m_{\mathrm{F}}>0$ describes an F phase with the F free energy 
\begin{align}\label{cFF}  
\cF_{\mathrm{F}}(\mu) = \FHF(d_{\mathrm{F}},m_{\mathrm{F}};\mu). 
\end{align} 
Again, the qualifier {\em stable} is important in the previous sentence: we often find that the Hartree--Fock equations have several F solution $(d_0^{(i)},m_0^{(i)})$ with $m_0^{(i)}>0$ for $i=1,2,\ldots$ and, in such a case, we take the one with the lowest free energy,  $\FHF(d_{\mathrm{F}},m_{\mathrm{F}};\mu) = \min_i \FHF(d_0^{(i)},m_0^{(i)};\mu)$.
In general, we have AF, P, and F solutions, and the Hartree--Fock free energy is 
\begin{align*} 
\min\big(\cF_{\mathrm{AF}}(\mu),\cF_{\mathrm{F}}(\mu),\cF_{\mathrm{P}}(\mu) \big), 
\end{align*} 
with $\cF_{\mathrm{X}}$ for X=AF, P defined in Section~\ref{sec:mixedAFP}. 
In Fig.~\ref{fig2}(b) we plot the free energy $\cF_{\mathrm{X}}(\mu)$, for X=AF (red), F (blue), and P (green), as a function of $\mu$ for the 3D Hubbard model at $U=10$ and $1/\beta=0$. Again, the AF free energy is lowest at $\mu=0$, and it remains lowest in some $\mu$-interval until it intersects with the F free energy, leading to a mixed phase between the AF phase at half-filling and the F phase; since this is very similar to the mixed phase between the AF and P phase at $U=5$ discussed in detail in Section~\ref{sec:mixedAFP}, we do  not discuss this further. 

A new feature at $U=10$ is that, by increasing $\mu$ further, the P free energy descreases faster than the F free energy, and at a value $\mu=\mu_c\approx 5.73$ they intersect. From Fig.~\ref{fig2}(b) it is difficult to see if the slopes of the P and F curves at $\mu=\mu_c$ are the same or not but, by zooming in as in Fig.~\ref{fig:D=3,U=10}, one can see that these slopes, 
\begin{align*} 
	\nu^{\mathrm{min}}_{\mathrm{P}}
	= \left. -\frac{\partial \cF_{\mathrm{P}}(\mu)}{\partial\mu}\right|_{\mu=\mu_c}
	,\quad \nu^{\mathrm{max}}_{\mathrm{F}}=  \left. -\frac{\partial \cF_{\mathrm{F}}(\mu)}{\partial\mu}\right|_{\mu=\mu_c}, 	
\end{align*} 
are different: they are $\nu^{\mathrm{min}}_{\mathrm{P}}\approx 0.665$, $\nu^{\mathrm{max}}_{\mathrm{F}}\approx 0.643$, and thus there is a finite doping region $0.643\lesssim \nu\lesssim 0.665$ with a mixed phase between F and P; see Fig.~\ref{fig:3D}(a) for $U=10$. (Note that we can be sure that $\nu^{\mathrm{min}}_{\mathrm{P}}\geq \nu^{\mathrm{min}}_{\mathrm{F}}$ since otherwise the two curves cannot intersect; what is non-trivial here is that  $\nu^{\mathrm{min}}_{\mathrm{P}}> \nu^{\mathrm{min}}_{\mathrm{F}}$.)

Fig.~\ref{fig2}(c) shows the free energy curves for the AF, F, and P states as functions of $\mu$ for the 3D Hubbard model at $U=20$ and $1/\beta=0$. The curves are similar to the ones in Fig.~\ref{fig2}(b) for $U=10$, except that the transition
from F to P seems to be smooth: within the numeric accuracy we can reach,  $\nu_{\mathrm{F}}^{\mathrm{max}}$ and $\nu_{\mathrm{P}}^{\mathrm{min}}$ are the same, i.e., the transition from F to P in this case seems to be a standard second order phase transition; however, we cannot rule out the possibility that a (tiny) mixed phase between the F and P phases will be detected in more accurate future computations.  

We note in passing that, for $U=20$ (but not for $U=10$), there is a finite $\mu$-interval where $\cF_{\mathrm{F}}$ is constant. This is due to the fact that, in this $\mu$-regime, the F bands in \eqref{E_F} are separated, and the chemical potential can be changed within the F gap without changing doping or free energy. However, this is of no concern since the AF solution is stable in this $\mu$-regime.   

The other mean-field phase diagrams for the Hubbard model were obtained in the same way. 

\subsection{Mixed phases in half-filled Hubbard-like models}\label{sec:halffilled}  
\begin{figure}
	\vspace{0.4cm}
	\begin{center}
		\hspace{-1cm}
		\begin{overpic}[width=.46\textwidth]{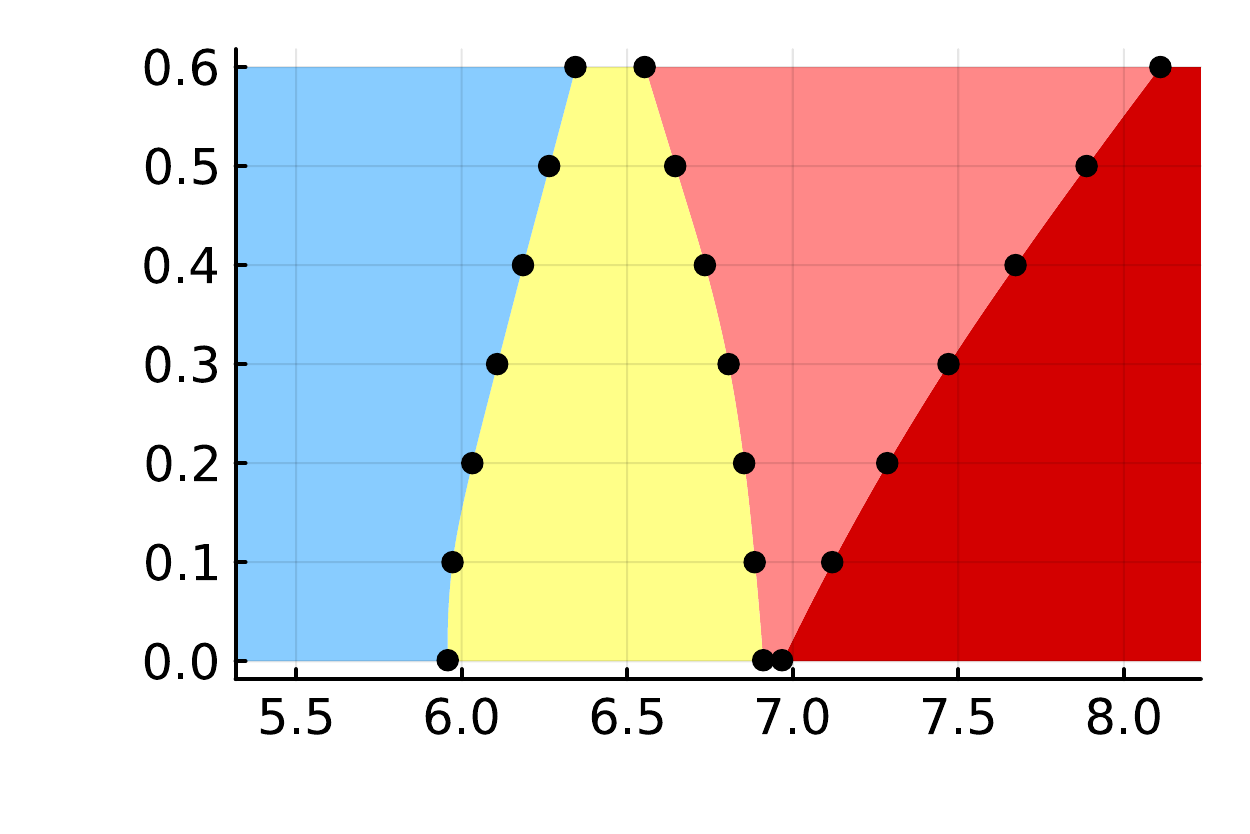 }
			\put(40,66.5){\footnotesize (a) $t'/t = 1, \delta = 0.2$}
			\put(15,65.5){\footnotesize $1/\beta t$}
			\put(98.5,11){\footnotesize $U/t$}
			\put(28,40){\footnotesize F}
			\put(44,32.5){\footnotesize Mixed}
			\put(63,40){\footnotesize AMM}
			\put(81,32.5){\footnotesize AMI}
		\end{overpic}
		\\
		\hspace{-1cm}				
		\begin{overpic}[width=.46\textwidth]{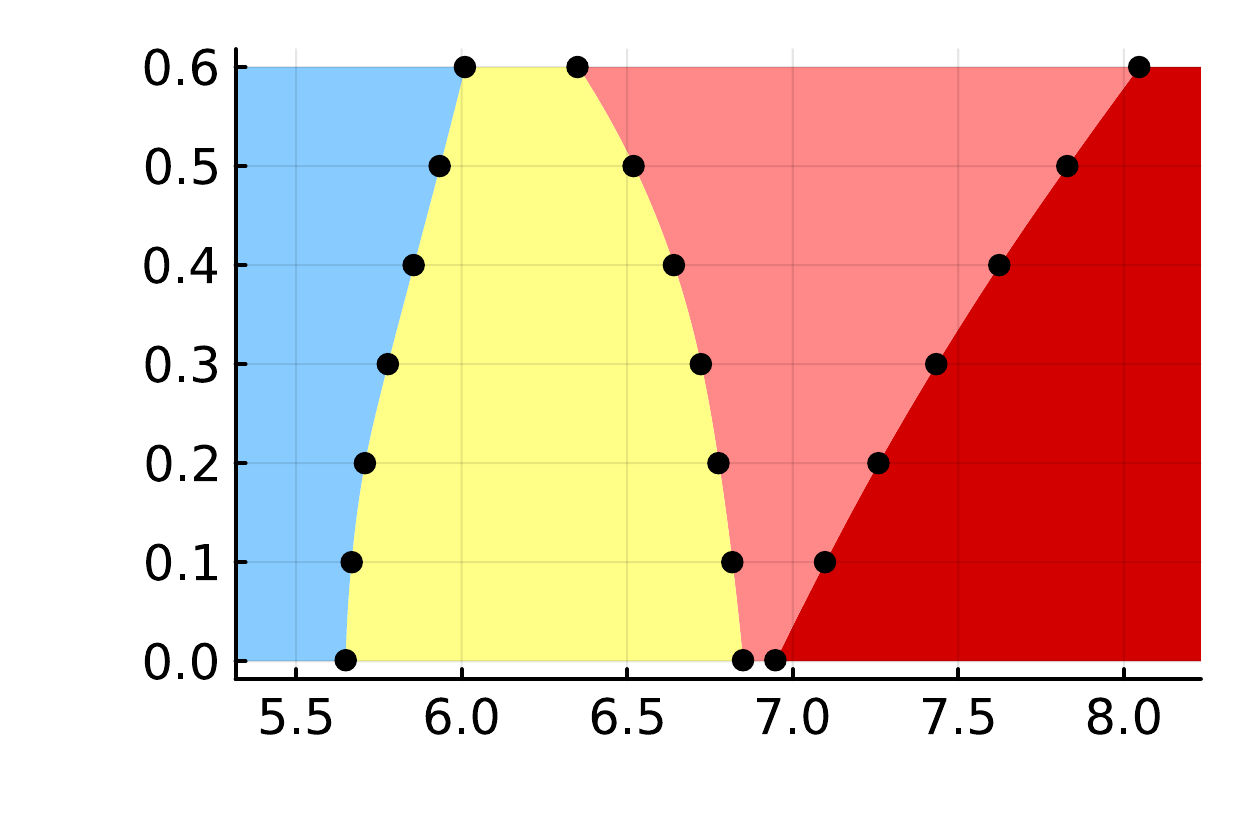 }
			\put(40,66){\footnotesize (b) $t'/t = 1, \delta = 0$}
			\put(15,65.5){\footnotesize $1/\beta t$}
			\put(98.5,11){\footnotesize $U/t$}
			\put(24,40){\footnotesize F}
			\put(38.8,32.5){\footnotesize Mixed}
			\put(62,40){\footnotesize AFM}
			\put(81,32.5){\footnotesize AFI}
		\end{overpic}
		\caption{\label{fig:DLKK} 
			(a) Phase diagram of the 2D DLKK model at half-filling: phases as function of the interaction strength $U/t$ and temperature $1/\beta t$ for $t'/t=1$, $\delta=0.2$. The different phases are  F (ferromagnetic, blue), Mixed (yellow), AMM (altermagnetic metallic, red), AMI (altermagnetic insulator, dark red). 
			\\
			(b) Same as (a) but for $\delta=0$ corresponding to the 2D $t$-$t'$-$U$ model. Similar to (a), except that there is AF instead of AM for $\delta=0$.  		
		}
	\end{center}
	\vspace{-0.4cm}
		\vspace{0.4cm}
	\begin{center}
		\hspace{-.6cm}
		\begin{overpic}[width=.46\textwidth]{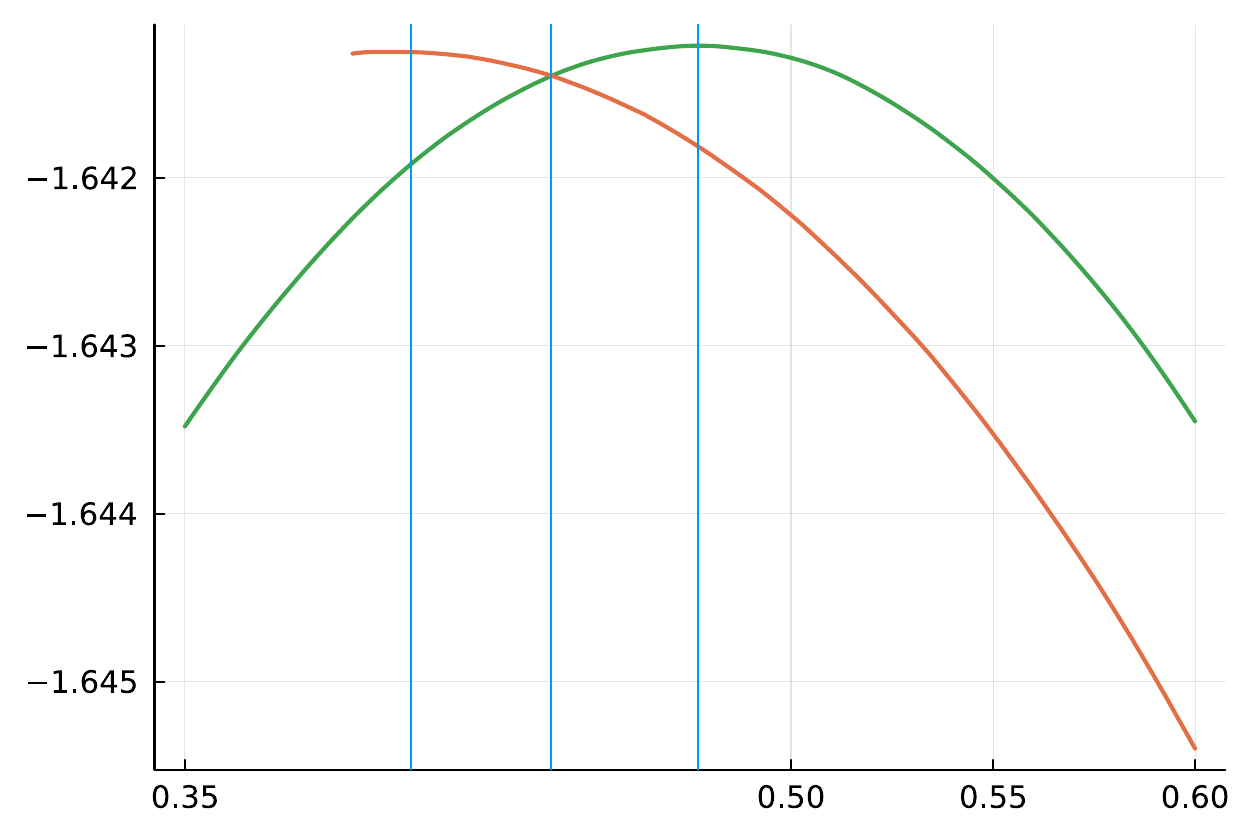 }
			\put(11,66.5){\footnotesize $\cF$}
			\put(100,4.5){\footnotesize $\mu$}
			\put(54,2.5){\footnotesize $\mu_{\mathrm{P}}$}
			\put(43,2.5){\footnotesize $\mu_c$}
			\put(30,2.5){\footnotesize $\mu_{\mathrm{AF}}$}
		\end{overpic}
		\caption{\label{fig8} 
			Mixed phase in the half-filled $t$-$t'$-$U$ model in 2D for $t' = 0.2$, $U = 2.065$, $1/\beta=0$ ($t=1$): Hartree--Fock free energy $\cF=\cF_{\mathrm{X}}$ (X=AF, P) vs.\  $\mu$ for different phases X=AF (red curve) and X=P (green curve). The half-filled AF and P solutions at $\mu=\mu_{\mathrm{AF}}$ and $\mu=\mu_{\mathrm{P}}$ are both unstable, and the stable solution at half-filling occurs at $\mu=\mu_c$ is mixed. 
		}
	\end{center}
	\vspace{-0.4cm}
\end{figure}

The AF solution is very robust for the half-filled Hubbard model and, for this reason, one can safely  ignore the possibility of mixed phases for the half-filled Hubbard model. However, for extended Hubbard models, this is not true: we now present examples that demonstrate that mixed phases can occur even at half-filling provided next-nearest-neighbor hopping is allowed.

The 2D DLKK model at half-filling was proposed as a system exhibiting altermagnetism (AM) which can be studied in cold atom systems \cite{DLKK2024}. More specifically, it was shown that Hartree--Fock theory with the ansatz \ref{ansatz_AF} has solutions with $m_1=m_{\mathrm{AM}}>0$, and it was proposed that these solutions describe an AM state which can be either metallic (AMM) or insulating (AMI), depending on whether the chemical potential intersects an effective band or not \cite{DLKK2024}. We checked and refined these results using our updated Hartree--Fock method where, in addition, we also allowed for states as in \eqref{ansatz_F} with $m_0=m_{{\mathrm F}}>0$ which, for lack of a better name, we call F states. We found that the half-filled DLKK model not only has P, AMI, and AMM phases but, in addition,  also F and mixed phases in significant parts of the phase diagram \cite{LL2025A}. 

As an example, we show in Fig.~\ref{fig:DLKK}(a) the mean-field phase diagram we obtained for the parameters $t'/t=1$ and $\delta=0.2$; shown are the phases as a function of coupling $U/t$ and temperature $1/\beta t$. 
One can see a large mixed phase (yellow region) between the F and AMM phases which exists in a large temperature range. To investigate how important the DLKK parameter $\delta$ is, we also computed the corresponding phase diagram for $\delta=0$; see Fig.~\ref{fig:DLKK}(b). To our surprise, we found that the phase diagrams for $\delta=0.2$ and $\delta=0$ are qualitatively very similar: mixed phases occur even for the half-filled $t$-$t'$-$U$ model in 2D. 

We conclude with an example that mixed phases at half-filling and $\delta=0$ can occur even at much smaller values of $t'/t$: the curves in Fig.~\ref{fig8} prove that, for the 2D $t$-$t'$-$U$ model at parameters $t' = 0.2$, $U = 2.065$, $1/\beta=0$ (in units where $t=1$),  the stable state at half-filling is mixed. More specifically, the curves show the AF and P free energies as functions of the chemical potential $\mu$ (red and green curves, respectively). Since half-filling corresponds to a horizontal tangent, one can easily see  that the AF curve is at half-filling for $\mu=\mu_{\mathrm{AF}}\approx 0.41$ where the P curve has lower free energy, and the P curve is at half-filling for $\mu=\mu_{\mathrm{P}}\approx 0.48$ where the AF curve has lower free energy. Thus, neither the AF and P phase is stable at half-filling: the stable phase at half-filling is mixed, and it is realized at $\mu=\mu_c\approx 0.44$ where the AF and P curves intersect. 
For completeness, we mention that the AF and P dopings in Fig.~\ref{fig8} at $\mu_c$ are $\nu_{\mathrm{AF}}^{{\rm min}}\approx 0.0076$ and $\nu_{\mathrm{P}}^{{\rm max}}\approx -0.011$, respectively, which reveals a mixed phase in the doping regime $-0.011\lesssim\nu\lesssim 0.0076$. This mixed phase has interesting implications for the Mott transition in the 2D $t$-$t'$-$U$ Hubbard model --- we plan to come back to  this in future work.

\subsection{Root of mistake fixed by update}\label{sec:bug} 
In this section, we go to the root of the mistake fixed by our update. We explain how Penn obtained the doped AF region in the phase diagram of the 3D Hubbard model at zero temperature \cite{P1966}, what went wrong, and how his method became established in well-known papers \cite{H1985,LH1987}, leading to incorrect results up to this day \cite{DLKK2024}.

\subsubsection{3D Hubbard model}\label{sec:Penn}  
We use the same example as in Section~\ref{sec:mixedAFP}: the 3D Hubbard model at $U/t=5$ and $1/\beta=0$. To find a solution of the Hartree--Fock equations \eqref{AF_HFeqs} at any doping value $\nu$, 
we use the shortcut described at the end of Section~\ref{sec:rHF}: we use \eqref{rho_AF} and the first equation in \eqref{AF_HFeqs} to set $d_0=\nu$.  This allows us to ignore the first equation in \eqref{AF_HFeqs}: from \eqref{rho_AF}, we get  $\tilde\mu$ in \eqref{tmu}, and by inserting this in the second equation in \eqref{AF_HFeqs} we can find $m_1=m_{\mathrm{AF}}\geq 0$; if $m_{\mathrm{AF}}>0$ and $\nu\neq 0$,  we call this the {\em unstable AF solution}. As shown below, this unstable AF solution exists in a significant doping regime away from half-filling. Penn used this unstable AF solution to conclude that the 3D Hubbard model at zero temperature has a significant AF region away from half-filling \cite[Fig.~15]{P1966}. 
   
The plots of $\max_{d_0}\FHF$ in Fig.~\ref{fig:D=3,U=5} give a better understanding of how the unstable AF solution arises: as discussed in Section~\ref{sec:mixedAFP}, for $\mu\gtrsim 1.380$, the P solution at $m_1=0$ is a local minimum of this function, and the stable AF minimum at $m_1=m_{\mathrm{AF}}>0$ exists for $\mu\lesssim 1.739$; thus, clearly, for $1.380\lesssim\mu\lesssim 1.739$, the function $\max_{d_0}\FHF$ of $m_1$ has a local maximum at a value $m_1$ between $0$ and $m_{\mathrm{AF}}$, and this maximum changes with $\mu$ so as to allow for a finite doping range: {\em the unstable AF solution corresponds to this free energy maximum}. 

Fig.~\ref{fig2withinset} is an extension of Fig.~\ref{fig2}(a) where, in addition to the free energies of the stable AF (solid red curve) and P (green curve) solutions discussed already in Section~\ref{sec:mixedAFP}, we also include the free energy of the doped AF solution (red dotted curves).
Clearly, any doping in the rangle $-1\leq \nu\leq 1$ can be reached either by the P or the doped AF curve (since the slopes to these curves take all values between $-1$ and $1$): this explains why the shortcut-method always gives a solution which either is P or AF. However, it is clear from Fig.~\ref{fig2withinset} that this unstable AF solution cannot be realized: the stable AF solution and the P solution both have lower free energy. Thus, the doped AF solution must be discarded.   

It is interesting to note that, if one compares the free energies of the P and unstable AF solutions {\em at fixed doping $\nu$} (rather than fixed $\mu$), one finds that the unstable AF solution has lower free energy than the P solution whenever it exists; this can be seen in Fig.~\ref{fig2withinset}. Actually, Penn did not have a chemical potential in his equations, and he restricts to zero temperature \cite{P1966}. Thus, the mistake was that energies were compared at fixed densities rather than at fixed $\mu$.

By looking at Fig.~\ref{fig2withinset}, one can see that the unstable AF solution behaves in a strange way: at $\mu\approx 1.739$ (where the stable AF solution ends) the doping is zero; then the doping increases as $\mu$ is decreased, following the dashed red curve until it merges with the green curve at $\mu\approx 1.380$; finally, changing direction by increasing $\mu$ and following the green curve from $\mu\approx 1.380$, the doping increases even further; thus, if one takes this solution, doping is not a monotonically increasing function of $\mu$, which is weird. This strange behavior is a strong indication that something is wrong with the unstable AF solution. 

\begin{figure}
	\vspace{0.4cm}
	\begin{center}
		\hspace{-.6cm}
		\begin{overpic}[width=.46\textwidth]{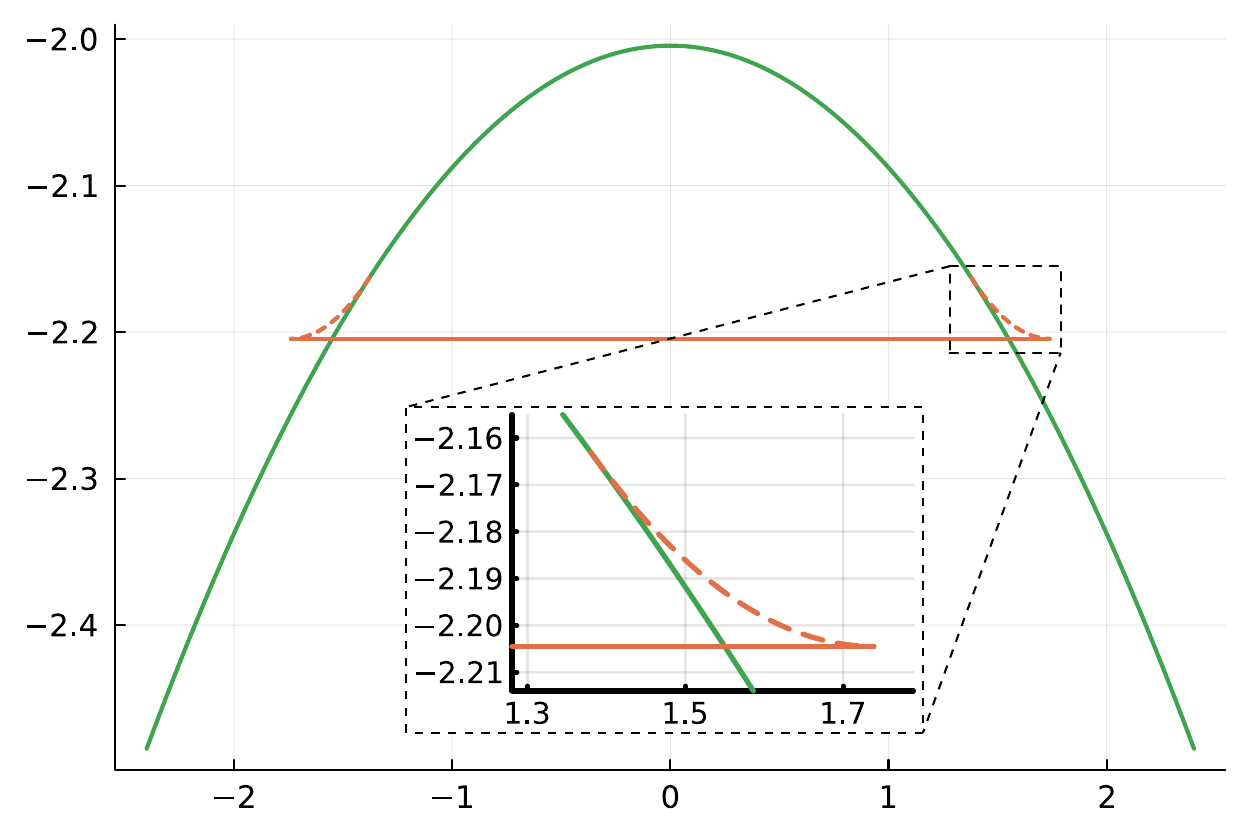 }
			\put(45,66.5){\footnotesize $U = 5$}
			\put(8,66.5){\footnotesize $\cF$}
			\put(100,4){\footnotesize $\mu$}
		\end{overpic}
		\caption{\label{fig2withinset} 
			Free energies of the AF and P Hartree--Fock solutions for the 3D Hubbard model at $U=5$ and zero temperature ($t=1$): Same as Fig.~\ref{fig2}(a) but with the unstable AF solution (red dashed curves) added. The unstable AF solution exists for $1.380 \lesssim |\mu| \lesssim 1.739$; it describes states away from half-filling, but it always has larger free energy than both the stable AF solution (solid red curve) and the P solution (green curve) and thus is unstable.  Inset: Enlargement of the unstable AF solution. 
		}
	\end{center}
	\vspace{-0.4cm}
\end{figure}
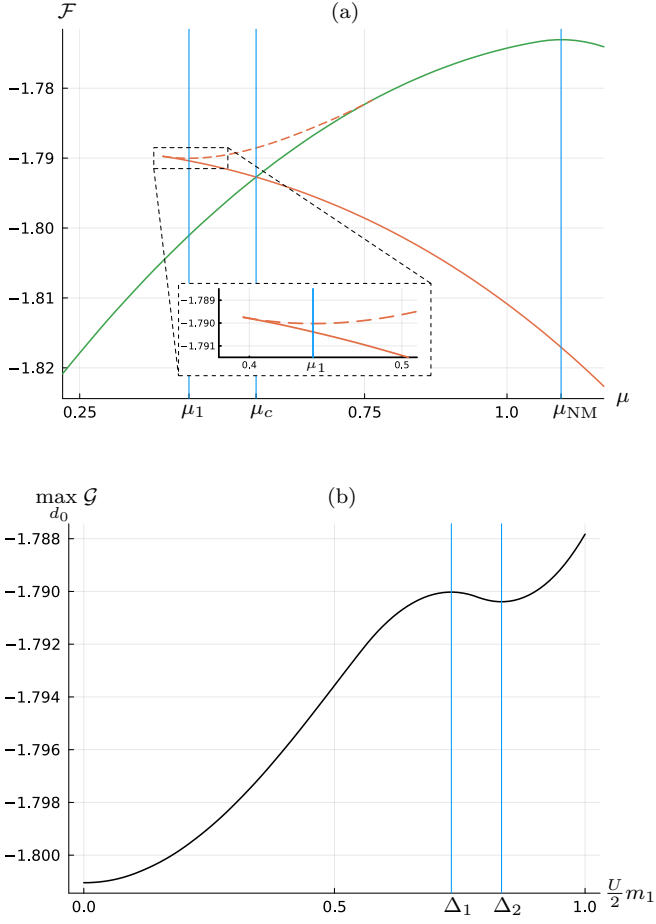
\begin{figure}
	\vspace{0.4cm}
	\begin{center}
		\hspace{-.6cm}
		\begin{overpic}[width=.46\textwidth]{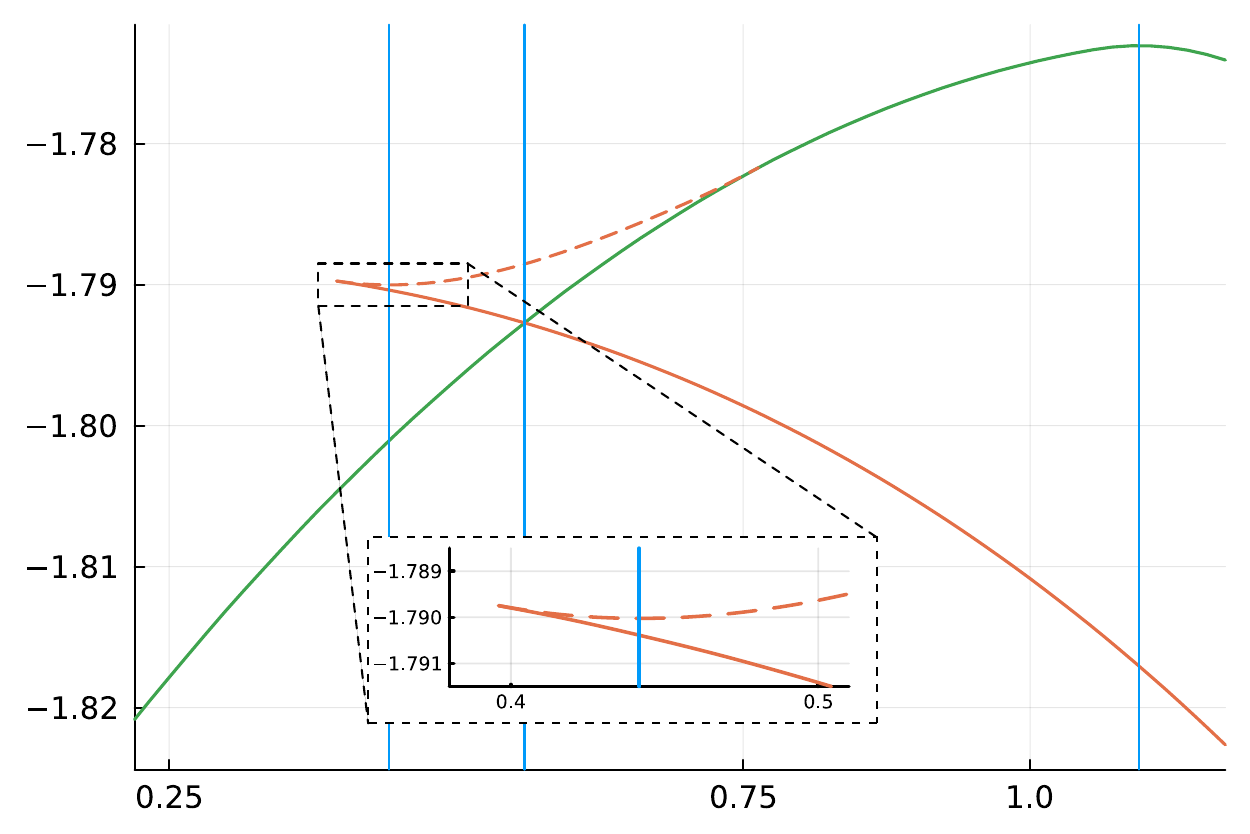 }
			\put(54,66.5){\footnotesize (a)}
			\put(10,66.5){\footnotesize $\cF$}
			\put(100,4.5){\footnotesize $\mu$}
			\put(89.8,2.6){\footnotesize $\mu_{\mathrm{NM}}$}
			\put(29.9,2.6){\footnotesize $\mu_1$}
			\put(41,2.6){\footnotesize $\mu_c$}
			\put(50,10){\tiny $\mu_1$}
		\end{overpic}
	\end{center}
	\vspace{0.4cm}
	\begin{center}
		\hspace{-.7cm}
		\begin{overpic}[width=.46\textwidth]{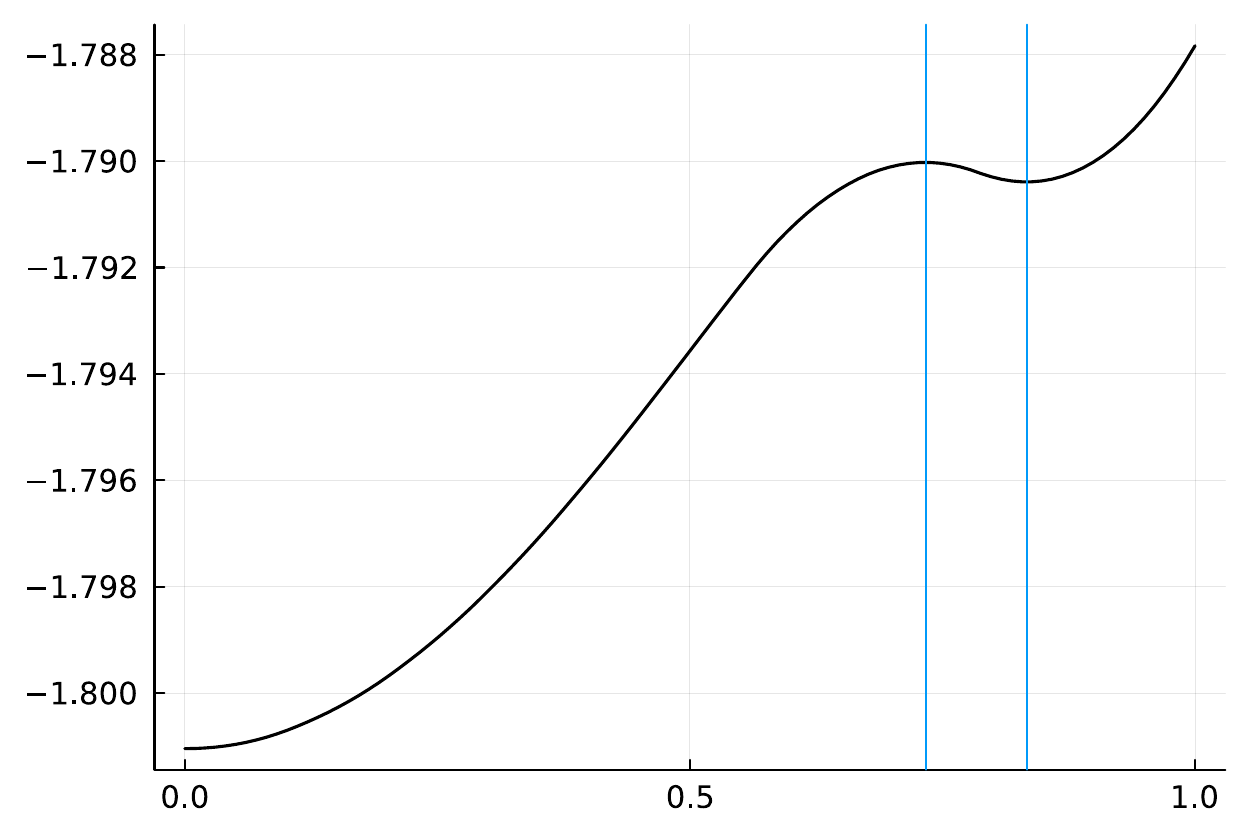 }
			\put(54,68){\footnotesize (b)}
			\put(7,68){\footnotesize $\underset{d_0}{\max \;} \FHF$}
			\put(98.7,4.2){\footnotesize $\frac{U}{2}m_1$}
			\put(72.8,2){\footnotesize $\Delta_1$}
			\put(80.8,2){\footnotesize $\Delta_2$}
		\end{overpic}
		\caption{\label{figDLKKunstable} Hartree--Fock solutions for the 2D DLKK model at the same values of the parameters as in \cite[Fig.~1(b)]{DLKK2024}: $U=3.5$, $t’=0.3$, $\delta=0.9$, $1/\beta=0$ ($t=1$). (a) Hartree--Fock free energy $\cF$ as a function of $\mu$ for the stable altermagnetic (AM, red line), non-magnetic (NM, green line), and unstable AM (dashed red line) solutions. The unstable AM solution exists for $0.396\lesssim\mu\lesssim 0.761$ and corresponds to  half-filling at $\mu=\mu_1\approx 0.442$, but it has larger free energy than both the stable AM solution and the NM solution. There is a half-filled NM solution at $\mu=\mu_{\mathrm NM}$ which is unstable as well. The stable half-filled solution is at $\mu=\mu_c\approx 0.560$ and describes a mixed state. Inset: Enlargement of unstable AM solution. 
			(b) Details on how the unstable AM solution in (a) arises: $\max_{d_0}\FHF$
			with $\FHF=\FHF(d_0,0,0,m_1;\mu)$ given by \eqref{FHFrestr}--\eqref{Errp} as a function of $\Delta=\frac{U}{2}m_1$ for fixed $\mu=\mu_1\approx 0.442$; $\Delta_1\approx 0.733$ corresponds to the unstable AM solution (red dashed line) in (a), while $\Delta_2 \approx 0.833$ corresponds to the locally stable AM solution which describes a doped AM state; the stable Hartree--Fock solution for $\mu=\mu_1$ is at $\Delta=0$ and describes a doped NM state.}
	\end{center}
	\vspace{-0.4cm}
\end{figure}

\subsubsection{2D Hubbard model}\label{sec:Hirsch}  
As discussed in Section~\ref{sec:intro}, the established mean-field phase diagram of the 2D Hubbard model at zero temperature is due to Hirsch \cite[Fig.~3]{H1985}. As Hirsch observed, his phase diagram is  qualitatively similar to Penn's phase diagram of the 3D Hubbard model \cite[Fig.~15]{P1966}. In particular, Hirsch obtained an AF phase in a significant doping regime away from half-filling. However, as shown in Fig.~\ref{fig:2D}, the 2D Hubbard model at zero temperature has an AF phase only at half-filling. We found that, while the stable AF solution in the 2D Hubbard model at $1/\beta=0$ can never be doped, there is an unstable AF solution, similarly as in 3D; it was this unstable AF solution Hirsch used to construct his phase diagram. 

Hirsch writes that he obtained his phase diagram by {\em ``choosing the solution that gives the lowest energy (if more than one solution exists)''} \cite{H1985}. Thus, he chose the solution giving the lowest energy at fixed value of doping $\nu$ (and not at fixed value of $\mu$), following Penn \cite{P1966}. 

We also mention mean-field phase diagrams of the 2D $t$-$t'$-$U$ model obtained by Lin and Hirsch using the same method \cite{LH1987} which are qualitatively wrong by the same mistake \cite{LW2007}. 
 
\subsubsection{2D DLKK model at half-filling}\label{sec:DLKK} 
Our final example is the Hartree--Fock solution of the 2D DLKK model at parameters $U/t=3.5$, $t'/t=0.3$, $\delta=0.9$, and $1/\beta=0$ with the ansatz \eqref{ansatz_AF}. This example was used in recent work to exhibit the spin-resolved band structure of the altermagnetic state at zero temperature in the 2D DLKK model at half-filling; see \cite[Fig.~1(b)]{DLKK2024}. 
Following \cite{DLKK2024}, we refer to solutions with $m_1>0$ and $m_1=0$ as altermagnetic (AM) and non-magnetic (NM) states, respectively. 
The mean-field results in \cite{DLKK2024} were obtained using the method in \cite{H1985} and, by the same mistake going back to \cite{P1966}, some of the results are wrong \cite{LL2025A}.   

To demonstrate  the similarities of important details in different models, we present in Fig.~\ref{figDLKKunstable} (a) and (b) plots for the DLKK model that correspond to Figs.~\ref{fig2withinset} and \ref{fig:D=3,U=5} for the Hubbard model, respectively. From these plots, it is clear that the stable solution of the 2D DLKK model at half-filling for these parameters describes a mixed state, and that the AM solution used in \cite[Fig.~1(b)]{DLKK2024} is unstable and thus should be discarded. Unfortunately, for a mixed state, there is no well-defined band structure: another example should be found. 
We emphasize that the authors obtained their result using the well-established method of Hirsch \cite{H1985} going back to Penn \cite{P1966}.
  
\section{Hartree--Fock theory}\label{sec:HF}
We present a self-contained discussion of Hartree--Fock theory. In particular, we recall the  fundamental principle underlying Hartree--Fock theory, and we give a short heuristic derivation of the Hartree--Fock equations where we try to bridge the conventional derivation and the one based on the Hartree--Fock function $\FHF$. We also describe the stability test included in our updated version of the Hartree--Fock method.

\subsection{Derivation of Hartree--Fock theory}

\subsubsection{Foundation}\label{sec:foundation}  
As is well-known, Hartree--Fock theory is a variational method where the Gibbs state 
\begin{equation}\label{rhoG} 
	\rho_{\mathrm{G}} = \frac{\ee^{-\beta H}}{\Tr(\ee^{-\beta H})}	
\end{equation} 
for the Hamiltonian $H$ in \eqref{Hdef} is approximated by the best possible  quasi-free state. We recall that a {\em state} (also called a {\em density matrix}), $\rho$, is a self-adjoint operator on the fermion Fock space which has eigenvalues in the interval $[0,1]$ and which satisfies $\Tr(\rho)=1$. Furthermore, a state $\rho$ is {\em quasi-free} if expectation values defined as $\langle A\rangle=\Tr(\rho A)$ satisfy the Wick theorem, which we use in the following form (the first equality is implied by \eqref{CAR})
\begin{multline}\label{Wick}  
	\langle c^\dag_I c^\pdag_J c^\dag_K c^\pdag_L \rangle =
	\langle c^\dag_I c^\dag_K c^\pdag_L c^\pdag_J \rangle + 
	\delta_{J,K}\langle c^\dag_I c^\pdag_L \rangle 
	\\=
	\langle c^\dag_I c^\pdag_J\rangle\langle c^\dag_K c^\pdag_L \rangle
	- \langle c^\dag_I c^\pdag_L\rangle \langle c^\dag_K c^\pdag_J \rangle 
    +\delta_{J,K}\langle c^\dag_I c^\pdag_L \rangle,
\end{multline} 
where $I$ is short for $(\bx,\sigma)$ etc; see \cite[Sec. 2]{BLS1994} for further details on Wick's theorem (there are of course many more relations like the one in \eqref{Wick}).
Here and in the following, we simplify our discussion by assuming $\langle c_Ic_J\rangle=0$, i.e., we rule out the possibility of superconducting states; it is known by mathematical proof that, for the Hubbard model, this simplification can be done without loss of generality \cite{BLS1994}.

To quantify how well an arbitrary state $\rho$ approximates the Gibbs state in \eqref{rhoG}, one can use the {\em grand canonical potential} $\Omega$ defined by
\begin{equation}\label{Omegadef} 
	\Omega(\rho) = \Tr(\rho H) + \frac1\beta\Tr(\rho\ln(\rho)), 
\end{equation} 
with $\Tr$ the trace in the fermion Fock space.
This can be understood as follows: for fixed Hamiltonian $H$ and temperature $1/\beta>0$, the Gibbs state in \eqref{rhoG} is the absulute miniumum of $\Omega$ in the set of {\em all} states $\rho$ (since this fundamental result in thermodynamics is important for us, we include a proof in Appendix~\ref{app:Gibbs}). 
Hartree--Fock theory amounts to finding the minimum of $\Omega$ among all quasi-free states.

To summarize: {\em if $\cS$ is the set of all states and $\cS_{\mathrm{qf}}$ the subset of $\cS$ containing all quasi-free states, then the Gibbs state can be found by solving the variational problem 
\begin{equation}\label{OmegarhoG}
\Omega(\rho_{\mathrm{G}})=\min_{\rho\in\cS}\Omega(\rho), 
\end{equation}
while Hartree--Fock theory is the approximation to the latter exact solution obtained by restricting the search of the absolute minimum to the subset $\cS_{\mathrm{qf}}$ of $\cS$:}  
\begin{equation}\label{OmegarhoHF}
\Omega(\rhoHF^{*})=\min_{\rho\in\cS_{\mathrm{qf}}}\Omega(\rho). 
\end{equation}
From this point of view, restricted Hartree--Fock theory is natural: it is an approximation to Hartree--Fock theory where the search for an absolute minimum of $\Omega$ is further restricted to some convenient subset $\cS_{\mathrm{qf,rst}}$ of $\cS_{\mathrm{qf}}$: 
\begin{equation}\label{OmegarhoHFrst}
\Omega(\rho_{\mathrm{HF,rst}}^{*})=\min_{\rho\in\cS_{\mathrm{qf,rst}}}\Omega(\rho). 
\end{equation}
The art in Hartree--Fock theory is to choose $\cS_{\mathrm{qf,rst}}$ well.
Whereas the Gibbs state $\rho_{\mathrm{G}}$ is uniquely determined by \eqref{OmegarhoG}, the optimization problems \eqref{OmegarhoHF} and \eqref{OmegarhoHFrst} may have multiple solutions, so the states $\rhoHF^*$ and $\rho_{\mathrm{HF,rst}}^*$ are in general not uniquely determined. In Appendix~\ref{app:remarks}, we shortly discuss how abstract mathematical notions introduced in Appendix~\ref{app:Gibbs} shed light on the physical interpretation of restricted Hartree--Fock theory and, in particular, mixed states.

The grand canonical potential $\Omega$ is also known as the Landau free energy. We therefore refer to $\Omega(\rho)/L^n$ as the {\em free energy (density)} of a state $\rho$.

\subsubsection{Preliminary results}\label{sec:formulas}  
We collect well-known formulas and results we need. 

First, we recall that, for $\beta<\infty$, any relevant quasi-free state for the lattice fermion models we consider can be represented as  
\begin{equation}\label{rhoqf} 
	\rhoqf = \frac{\ee^{-\beta \Hqf}}{\Zqf} ,\quad 
	\Zqf = \Tr(\ee^{-\beta \Hqf}),
\end{equation} 
where $\Hqf$ is a quadratic Hamiltonian, i.e., it is of the form 
\begin{equation}\label{Hqf}  
\Hqf = \sum_{\bx,\sigma,\by,\sigma'} h_{\bx,\sigma;\by,\sigma'}\big( c^\dag_{\bx,\sigma}c^\pdag_{\by,\sigma'}
	-\tfrac12\delta_{\bx,\by}\delta_{\sigma,\sigma'}\big)
\end{equation} 
with $h=(h_{\bx,\sigma;\by,\sigma'})$ a hermitian $\cN\times\cN$ matrix of size $\cN=2L^n$ (the constant $-\tfrac12\sum_{\bx,\sigma}h_{\bx,\sigma;\bx\sigma}=-\tfrac12\tr(h)$ is added for convenience). 
(It is folklore in physics that any quasi-free state is either of the form \eqref{rhoqf} or can be obtained from such a state as a limiting case --- one important such limiting case is the zero temperature limit $1/\beta\downarrow 0$; for the convenience of the reader, we give a pedestrian derivation of this result in Appendix~\ref{app:qf}, based on results in Appendix~\ref{app:QF}.)
The matrix $h$ has the physical interpretation of a one-particle Hamiltonian.
Note that $\HHF$ in \eqref{HHF} is a special case of $\Hqf$ in \eqref{Hqf} --- as will be discussed, it is non-trivial but true that one can restrict from the general quasi-free states in \eqref{rhoqf}--\eqref{Hqf} to  
\begin{equation}\label{rhoHF} 
	\rhoHF = \frac{\ee^{-\beta \HHF}}{\ZHF} ,\quad 
	\ZHF = \Tr(\ee^{-\beta \HHF}),
\end{equation} 	
with $\HHF$ in \eqref{HHF} without loss of generality.

Second, we give a useful formula for the grand canonical potential restricted to quasi-free states: Defining 
\begin{align}\label{Aexp}  
	\langle A\rangle=\Tr(\rhoqf A),
\end{align} 	
we compute, using \eqref{Omegadef} and \eqref{rhoqf}, 
	\begin{equation}\label{OmegaWHF}
	\begin{split}  
			\Omega(\rhoqf) &= \langle H + \tfrac{1}{\beta}\ln(\rhoqf)\rangle \\
			&= \langle H\rangle -\langle\Hqf\rangle - \tfrac{1}{\beta}\ln(\Zqf).
		\end{split} 
\end{equation} 

Third, we recall the computation of the energy expectation value $\langle H\rangle$ using the Wick theorem \eqref{Wick}: the key step in this computation is
\begin{multline*} 
	\langle n^\pdag_{\bx,\uparrow}n^\pdag_{\bx,\downarrow}\rangle 
	=
	\langle c^\dag_{\bx,\uparrow}c^\pdag_{\bx,\uparrow}\rangle\langle c^\dag_{\bx,\downarrow}c^\pdag_{\bx,\downarrow}   \rangle - \langle c^\dag_{\bx,\uparrow}c^\pdag_{\bx,\downarrow}\rangle\langle c^\dag_{\bx,\downarrow}c^\pdag_{\bx,\uparrow}   \rangle \\ =  
	\langle n^\pdag_{\bx,\uparrow}\rangle \langle n^\pdag_{\bx,\downarrow}\rangle
	- \langle s_{\bx}^+\rangle  
	\langle s_{\bx}^-\rangle = 
	\frac14\big(\langle n_\bx\rangle^2-\langle\vec{s}_\bx\rangle^2\big) 
\end{multline*} 
where $s_{\bx}^+ := c^\dag_{\bx,\uparrow}c^\pdag_{\bx,\downarrow}=\tfrac12(s_{\bx}^x+\ii s_{\bx}^y)$, $s_{\bx}^- = (s_{\bx}^+)^\dag$, and $\langle\vec{s}_\bx\rangle$ is short for $(\langle s_{\bx}^x\rangle,\langle s_{\bx}^y\rangle,\langle s_{\bx}^z\rangle )$.  
Using this, one finds by straightforward computations 
\begin{equation}\label{expectationH} 
	\langle H\rangle  = \langle H_0\rangle + \frac{U}{4}\sum_{\bx}
	\big( \langle n_{\bx}-1\rangle^2 - \langle\vec{s}_\bx\rangle^2 \big).  
\end{equation} 

Finally, we recall an important formula allowing us to compute the Hartree--Fock partition function $\ZHF$ in \eqref{rhoqf}--\eqref{Hqf}. We use a simplified notation where $I$ is short for $(\bx,\sigma)$ etc.\ to write the Hartree--Fock Hamiltonian in \eqref{Hqf} as in \eqref{HHFgen1}  
where $h=(h_{I,J})$ is the one-particle Hamiltonian. It is well-known that 
\begin{equation}\label{TlnZqf} 
\frac1\beta \ln(\Zqf) = \tr\big(\LnT(h)\big)
\end{equation} 
with $\LnT(E)$ the function defined in \eqref{LT} and $\tr$ the one-particle Hilbert space trace (which in our case is the trace of $2L^{n} \times 2L^{n}$ matrices); 
note that, by the spectral theorem for self-adjoint matrices,
\begin{align} 
\tr\big( \LnT(h)\big) = 	\sum_I \LnT(E_I), 
\end{align} 
where $E_I$ are the eigenvalues of the one-particle Hamiltonian $h$ (to keep this paper self-contained, we include a derivation of this well-known result in Appendix~\ref{app:QF}). 

\subsubsection{Heuristic derivation of Hartree--Fock theory}\label{sec:heuristic} 
We give a heuristic argument, similar to one commonly used in the physics literature, allowing us to derive Hartree--Fock equations in a few lines. We also introduce the Hartree--Fock function $\FHF$ in a pragmatic way; the conceptual significance of $\FHF$ is explained in Section~\ref{sec:variational} below.

The common heuristic argument in the physics literature leads to  Hartree--Fock theory restricted to states where spin rotation invariance can be broken only in the $z$-direction; see e.g.\ \cite{DLKK2024}. We give a generalization of this argument without this restriction. To prepare for this generalization, we note that 
$n_{\bx,\uparrow,\downarrow}-\tfrac12 = \tfrac12(n_{\bx}-1\pm s_{\bx}^z)$,
which makes manifest that the formula defining the Hubbard Hamiltonian in  \eqref{Hdef} hides an important symmetry, namely spin rotation invariance: in \eqref{Hdef}, the unit vector $\vec{e}_z=(0,0,1)$ in the  $z$-direction plays a special role (since $s_{\bx}^z=\vec{e}_z\cdot\vec{s}_\bx$), but this is only apparent since, using $n_{\bx,\sigma}^2=n_{\bx,\sigma}$, one can write  
$$
(n_{\bx,\uparrow}-\tfrac12) (n_{\bx,\downarrow}-\tfrac12) =
\tfrac12(n_\bx-1)^2 - \tfrac{1}{4}, 
$$
which makes the spin rotation invariance of the model manifest. 
This observation is useful since it makes clear that we can replace $\vec{e}_z$ in \eqref{Hdef} by an arbitrary unit vector $\vec{e}_\bx=(e^x_{\bx},e^y_{\bx},e^z_{\bx})$ at each site $\bx$, i.e., we can write the Hamiltonian $H$ in \eqref{Hdef} as 
\begin{equation} \label{HH0Usumnunu}
	H = H_0 + U\sum_{\bx} \nu_{\bx,\uparrow}\nu_{\bx,\downarrow}
\end{equation} 
where 
$$
\nu_{\bx,\uparrow,\downarrow} = \tfrac12(n_{\bx}-1\pm \vec{e}_{\bx}\cdot\vec{s}_{\bx}), 
$$
with arbitrary $\vec{e}_{\bx}\in\R^3$ satisfying $\vec{e}_{\bx}^{\,2}=1$ (readers not satisfied with our symmetry argument can find a direct proof of this result in Appendix~\ref{app:nunu}). 

We are now ready for the short heuristic derivation of Hartree--Fock theory. 
We use the mean-field approximation, which amounts to expanding  the Hubbard interaction in density fluctuations $\delta \nu_{\bx,\sigma} = \nu_{\bx,\sigma} - \langle \nu_{\bx,\sigma}\rangle$ and keeping only the linear terms: 
\begin{equation*}
	\nu_{\bx,\uparrow}\nu_{\bx,\downarrow}
	\approx
	\langle \nu_{\bx,\uparrow}\rangle \nu_{\bx,\downarrow}  + \nu_{\bx,\uparrow}\langle \nu_{\bx,\downarrow}\rangle -\langle \nu_{\bx,\uparrow}\rangle\langle \nu_{\bx,\downarrow}\rangle.
\end{equation*} 
Inserting this into the Hubbard Hamiltonian $H$ above, we obtain 
$
H\approx\HHF -\Epot 
$
with the linearized Hubbard Hamiltonian (we use the notation $\HHF$ instead of $\Hqf$ to emphasize that this is a particular special case of a quasi-free Hamiltonian which, as we will show, will lead to the one in \eqref{HHF})
\begin{multline*}\label{HHF1} 
	\HHF  = H_0 + U\sum_{\bx}\big(\langle \nu_{\bx,\uparrow}\rangle \nu_{\bx,\downarrow}  + \nu_{\bx,\uparrow}\langle \nu_{\bx,\downarrow}\rangle\big) \\
	= H_0 + \frac{U}{2}\sum_{\bx}\big(\langle n_{\bx}-1\rangle (n_{\bx}-1)   - \langle \vec{e}_\bx\cdot\vec{s}_\bx\rangle\vec{e}_{\bx}\cdot\vec{s}_{\bx} \big), 
\end{multline*} 
and the following constant which can be interpreted as the potential energy coming from the interactions,  
\begin{equation*} 
	\Epot = U\sum_{\bx} \langle \nu_{\bx,\uparrow}\rangle\langle \nu_{\bx,\downarrow}\rangle\\
	=  \frac{U}{4}\sum_{\bx} \big( \langle n_{\bx}-1\rangle^2 - \langle \vec{e}_{\bx}\cdot \vec{s}_{\bx}\rangle^2
	\big).  
\end{equation*}
In the formulas above, the directions $\vec{e}_{\bx}$ are arbitrary. We now observe that these directions can be fixed by the requirement that the mean-field approximation above preserves the expectation value, i.e., $\langle H\rangle = \langle \HHF -\Epot\rangle$. Indeed, by a simple computation, 
\begin{equation*} 
\langle \HHF-\Epot\rangle = \langle H_0\rangle + \Epot, 
\end{equation*} 
which coincides with the result given in \eqref{expectationH} provided    $\langle\vec{e}_{\bx}\cdot\vec{s}_{\bx}\rangle^2=
\langle\vec{s}_{\bx}\rangle^2$; the latter is true if and only if $\vec{e}_{\bx}$ is parallel with $\langle\vec{s}_{\bx}\rangle$, and this implies
\begin{align*}
	\HHF  
	= H_0 + \frac{U}{2}\sum_{\bx}\big(\langle n_{\bx}-1\rangle (n_{\bx}-1)   - \langle\vec{s}_{\bx}\rangle\cdot\vec{s}_{\bx} \big).  
\end{align*}
This linearized Hamiltonian gives conventional Hartree--Fock theory: if we write it as in \eqref{HHF} using the equations in \eqref{HFeqs}, we can interpret $\langle A\rangle$ for $A=n_{\bx}-1$ and $A= \vec{s}_{\bx}$ as expectation values in \eqref{Aexp} with $\rhoHF$ in  \eqref{rhoqf} and the Hartree--Fock Hamiltonian in \eqref{HHF}. With that interpretation, \eqref{HFeqs} are Hartree--Fock equations: a set of non-linear equations allowing us to determine the mean-fields $d(\bx)$ and $\vec{m}(\bx)$.

For later reference, we note that $\HHF$ in \eqref{HHF} is of the form \eqref{Hqf} with the one-particle Hamiltonian $h=(h_{\bx,\sigma;\by,\sigma'})$ given by 
\begin{multline}\label{h} 
	h_{\bx,\sigma;\by,\sigma'}
	= \big(t_{\bx,\by}-\mu\delta_{\bx,\by}\big)\delta_{\sigma,\sigma'}
	\\ + \frac {U}{2}\delta_{\bx,\by}\big(d(\bx)\delta_{\sigma,\sigma'} - \vec{m}(\bx)\cdot\vec{\sigma}_{\sigma,\sigma'}\big),
\end{multline} 
where $\vec{\sigma}=(\sigma^x,\sigma^y,\sigma^z)$ are the usual Pauli sigma matrices (this can be seen using $\vec{s}_{\bx} = \sum_{\sigma,\sigma'} c_{\bx,\sigma}^\dag\vec{\sigma}_{\sigma,\sigma'}^\pdag c_{\bx,\sigma'}^\pdag$ and  $-\tfrac12\sum_{\bx,\sigma}h_{\bx,\sigma;\bx,\sigma}=\sum_{\bx}(\mu-(U/2)d(\bx))$). Thus, the heuristic argument above suggests a significant simplification: since a hermitian $\cN\times\cN$-matrix $h$ is determined by $\cN^2$ real numbers, \eqref{rhoqf}--\eqref{Hqf} is an ansatz with $\cN^2$ variational parameters where $\cN=2L^n$, while the specialization in \eqref{HHF} reduces this to the $4L^n=2\cN$ variational parameters $(d(\bx),\vec{m}(\bx))$. 
Our argument above suggests that this simplification can be done without loss of generality; a mathematical proof that this is indeed the case is reviewed in Section~\ref{sec:BLS}.

We proceed by deriving a more convenient form of the Hartree--Fock equations. We note that the expectation values in \eqref{HFeqs} can be computed as follows: since $\frac{\partial}{\partial d(\bx)}\HHF=(U/2)(n_{\bx}-1)$ for $\HHF$ in \eqref{HHF}, we have
\begin{multline}\label{ni-1}
	\frac{U}{2}\langle n_{\bx}-1\rangle = 
	\frac{\Tr\big(\ee^{-\beta \HHF}(U/2)(n_{\bx}-1)\big)}{\Tr\big(\ee^{-\beta \HHF}\big)}
	\\= 
	-\frac1\beta \frac{\partial}{\partial d(\bx)}\ln(\Tr(\ee^{-\beta \HHF}))\\= -\frac1\beta\frac{\partial}{\partial d(\bx)}\ln(\ZHF) 
\end{multline} 
with $\ZHF$ in \eqref{rhoHF}, and similarly 
\begin{equation*} 
	\frac{U}{2} \langle \vec{s}_{\bx}\rangle = \frac1\beta\frac{\partial}{\partial\vec{m}(\bx)}\ln(\ZHF)  	
\end{equation*} 
(a non-trivial mathematical fact which we used to derive these formulas is explained and proved in Appendix~\ref{app:Id}).
Thus, the function defined as 
\begin{equation}\label{cGdef}  
\GG  =  \frac{U}{4}\sum_{\bx}\big(\vec{m}(\bx)^2-d(\bx)^2\big)  - \frac1\beta\ln(\ZHF) 
\end{equation} 
satisfies 
\begin{equation}\label{cGprime} 
	\begin{split} 
	\frac{\partial \GG}{\partial d(\bx)} = & -\frac{U}{2}d(\bx)+\frac{U}{2}\langle n_{\bx}-1\rangle,\\ 
	\frac{\partial \GG}{\partial\vec{m}(\bx)}= &\; \frac{U}{2}\vec{m}(\bx)-\frac{U}{2}\langle \vec{s}_{\bx} \rangle, 
	\end{split} 
\end{equation} 
i.e., the Hartree--Fock equations in \eqref{HFeqs} are equivalent to $\frac{\partial}{\partial d(\bx)}\GG=0$, $\frac{\partial}{\partial\vec{m}(\bx)}\GG=\vec{0}$. 
Moreover, since $\sum_{\bx} \langle n_{\bx}-1\rangle = \frac{1}{\beta}\frac{\partial}{\partial\mu}\ln(\ZHF)$ (this follows from $\sum_{\bx}(n_{\bx}-1)=-\frac{\partial}{\partial\mu}\HHF$), we can write the density constraint in \eqref{rho} as 
\begin{equation}\label{rho2}
	\nu = -\frac1{L^n}\frac{\partial \GG}{\partial\mu}. 	
\end{equation} 
The results above motivate us to define the Hartree--Fock function as $\FHF=\GG/L^n$;  using \eqref{TlnZqf} and the definition in \eqref{cGdef}, it is given by 
\begin{equation}\label{FHF}  
	\boxed{ \FHF = \frac{1}{L^n}\Big(\frac{U}{4}\sum_{\bx}\big( \vec{m}(\bx)^2-d(\bx)^2\big) -
		\tr\big(\LnT(h\big)  \Big) } 
\end{equation}  
with $h=h(\boldsymbol{d},\boldsymbol{\vec{m}};\mu)$ the one-particle Hamiltonian in \eqref{h}. As shown above, this function gives the Hartree--Fock equations and the density constraint as \eqref{HFeqs1} and \eqref{rho1}, respectively. 

\subsubsection{Stability test}\label{sec:variational}
We now show that $\FHF$ at solutions of the Hartree--Fock equation is equal to the free energy density, and we explain in more detail why the stability test is important. 

The derivation of the Hartree--Fock equations in the previous section is only  heuristic. A complete derivation should prove that the mean-fields $d(\bx)$, $\vec{m}(\bx)$ satisfying the Hartree--Fock equations minimize the grand canonical potential $\Omega$, as discussed in Section~\ref{sec:foundation}. Such a proof, due to  Bach, Lieb, and Solovej \cite{BLS1994}, is reviewed in Section~\ref{sec:BLS}; this proof is important since it not only justifies the Hartree--Fock equations but also provides a criterion for stability. In this section, we present a simplified argument making this result plausible.

We compute the grand canonical potential for a quasi-free state given by \eqref{rhoqf} for some Hartree--Fock Hamiltonian $\HHF$ of the form \eqref{HHF}. By \eqref{HHF} and \eqref{expectationH},
\begin{multline*} 
	\langle H\rangle -\langle\HHF\rangle = 
	\frac{U}{4}\sum_{\bx} \big( \langle n_{\bx}-1\rangle^2-\langle\vec{s}_{\bx}\rangle^2 \big) \\
	- \frac{U}{2}\sum_{\bx} \big( d(\bx)\langle n_{\bx}-1\rangle -\vec{m}(\bx)\cdot \langle \vec{s}_{\bx}\rangle \big)
\end{multline*}
which, by completing squares, can be written as 
\begin{equation*} 
	\langle H\rangle -\langle\HHF\rangle = \frac{U}{4}\sum_{\bx} \big( \vec{m}(\bx)^2-d(\bx)^2\big) + \cC
\end{equation*}
where we introduce 
\begin{equation}\label{cC}  
\cC = \frac{U}{4}\sum_{\bx}\Big( \big[d(\bx) - \langle n_{\bx}-1\rangle\big]^2 -\big[\vec{m}(\bx)-\langle\vec{s}_{\bx}\rangle\big]^2
\Big) .
\end{equation} 
Inserting this into \eqref{OmegaWHF} and using \eqref{TlnZqf}, we obtain
\begin{equation} 
\Omega = \frac{U}{4}\sum_{\bx}\big( \vec{m}(\bx)- d(\bx)^2\big) -\tr\big(\LnT(h)\big)  + \cC 	
\end{equation} 
with $h=h(\boldsymbol{d},\boldsymbol{\vec{m}};\mu)$ the one-particle Hamiltonian in \eqref{h}. This gives the grand canonical potential as a function of the variational fields and the chemical potential: $\Omega=\Omega(\boldsymbol{d},\boldsymbol{\vec{m}};\mu)$. 

The variational formulation of Hartree--Fock theory is that the variational fields should be determined such that they minimize $\Omega$, i.e., 
the requirement 
\begin{equation}\label{Omegamin}
	\Omega(\boldsymbol{d}^{*},\boldsymbol{\vec{m}}^{*};\mu)=\min_{\boldsymbol{\vec{m}},\boldsymbol{d}}\Omega(\boldsymbol{d},\boldsymbol{\vec{m}};\mu) 	
\end{equation} 
determines the stable solutions $\boldsymbol{d}^{*}$, $\boldsymbol{\vec{m}}^{*}$of Hartree--Fock theory.

By comparing with \eqref{cGdef}, we find 
\begin{equation*} 
\Omega=\GG+\cC,  
\end{equation*} 
and using \eqref{cGprime} and \eqref{rho2} we obtain 
\begin{equation*} 
	\begin{split} 
		\frac{\partial \Omega}{\partial d(\bx)} &= -\frac{U}{2}d(\bx)+\frac{U}{2}\langle n_{\bx}-1\rangle + \frac{\partial\cC}{\partial d(\bx)},\\ 
		\frac{\partial \Omega}{\partial\vec{m}(\bx)} &=
		\frac{U}{2}\vec{m}(\bx)-\frac{U}{2}\langle \vec{s}_{\bx} \rangle +  \frac{\partial\cC}{\partial\vec{m}(\bx)}, 
	\end{split} 
\end{equation*} 
and 
\begin{equation*} 
\nu = -\frac1{L^n}\frac{\partial\Omega}{\partial\mu}	
+ \frac1{L^n}\frac{\partial\cC}{\partial\mu}, 
\end{equation*} 
respectively. 

We note that, using \eqref{cGprime}, $\cC$ in \eqref{cC} can be written as 
\begin{equation*} 
	\cC = \frac{1}{U}\sum_{\bx} \left(
	\Big( \frac{\partial \GG}{\partial d(\bx)}\Big)^2 - \Big(\frac{\partial \GG}{\partial\vec{m}(\bx)}\Big)^2 \right) 
\end{equation*} 
where $\frac{\partial\GG}{\partial d(\bx)}=0$, $\frac{\partial\GG}{\partial\vec{m}(\bx)}=\vec{0}$ are equivalent to the Hartree--Fock equations in \eqref{HFeqs}: if the Hartree--Fock equations are fullfilled, not only $\cC$ but also all its first-order derivatives are zero. This shows that the Hartree--Fock equations in \eqref{HFeqs} imply 
\begin{equation}\label{Omegaprime} 
	\frac{\partial \Omega}{\partial d(\bx)}=0,\quad 
	\frac{\partial \Omega}{\partial\vec{m}(\bx)} = \vec{0}, 
\end{equation} 
\begin{equation} 
\nu = -\frac1{L^n}\frac{\partial\Omega}{\partial\mu}	, 
\end{equation} 
and $\Omega=\GG$: {\em at solutions of the Hartree--Fock equations, $\Omega$ is stationary with respect to variations of the variational fields and equal to $\GG$}. 

Since $\Omega$ at its absolute minimum is stationary and equal to the free energy of the system, we showed that a solution of the Hartree--Fock equation providing an absolute miniumum of $\Omega$ determines the free energy density as $\cF=\Omega/L^n=\GG/L^n=\FHF$. 

However, as we all know from calculus, \eqref{Omegaprime} is not only satisfied at the absolute minimum of $\Omega$ but also at maxima, saddle points, and local minima of $\Omega$. Since only solutions of the Hartree--Fock equations which are absolute minima of $\Omega$ correspond to stable states, it is clear that solving the Hartree--Fock equations without checking the stability of the solutions can lead to wrong results. We therefore update Hartree--Fock theory by emphasizing that the stability test in Section~\ref{sec:HFnutshell} is important. 

\subsection{Restricted Hartree--Fock theory}\label{sec:update} 

\subsubsection{Strategy} 
We discussed the conceptual foundation of Hartree--Fock theory in Section~\ref{sec:foundation}, denoting as $\cS_{\mathrm{qf,rst}}$ some set of quasi-free states restricted by some ansatz. 
There are of course many possible such sets $\cS_{\mathrm{qf,rst}}$; the art is to find a good compromise between two contradicting requirements: (i) the set should be large enough to contain relevant/interesting states, (ii) it should be as small as possible to limit computational effort. Fortunately, there are many results in the literature that can serve as a good guide to finding useful such sets; see e.g.\ \cite{P1966,ZG1989,PR1989,V1991,M2023,M2024}. 

%We discuss examples and give derivations of results presented in Section~\ref{sec:rHF}.

\subsubsection{Staggered states}\label{sec:staggered}
We derive a generalization of the results presented in Section~\ref{sec:rHF}. 

The set of all Hartree--Fock states which are invariant under translations by two sites can be defined by the following ansatz for the variational fields, 
\begin{align}\label{Penn_ansatz1} 
	\begin{split} 	
		d(\bx) =& \; d_0 +  (-1)^{\bx}d_1,\\
		\vec{m}(\bx) =&\; m_0\vec{e}_0 +  (-1)^{\bx}m_1\vec{e}_1,
	\end{split} 
\end{align} 
with real parameters $d_i$, $m_i$ and unit vectors $\vec{e}_i\in\R^3$ for $i=0,1$. Due to rotation invariance of the model, the Hartree--Fock function only depends on the angle $\theta$ between the unit vectors $\vec{e}_0$ and $\vec{e}_1$ and, for this reason, we can choose, without loss of generality, $\vec{e}_0=(0,0,1)$ and $\vec{e}_1=(\sin(\theta),0,\cos(\theta))$; thus
\begin{equation*} 
\vec{e}_0\cdot\vec{\sigma}_{\sigma,\sigma'}
= \sigma\delta_{\sigma,\sigma'},\quad 
\vec{e}_1\cdot\vec{\sigma}_{\sigma,\sigma'}
= \cos(\theta)\sigma\delta_{\sigma,\sigma'}
+ \sin(\theta)\delta_{\sigma,-\sigma'}
\end{equation*} 	
where we identify $\uparrow$ and $\downarrow$ with $+1$ and $-1$, respectively. Note that, for $\theta=0$, \eqref{Penn_ansatz1} reduces to the ansatz in \eqref{Penn_ansatz}.

Using this, \eqref{H0}, and the abbreviations
\begin{align}\label{phiB}  
	\phi_i=\frac{U}{2}d_i,\quad B_i=-\frac{U}{2}m_i\quad (i=0,1),  	
\end{align} 	
we can write the Hartree--Fock Hamiltonian in \eqref{HHF} as
\begin{multline*}
	\HHF = \sum_{\bk,\sigma}\Big(\big[\varepsilon(\bk)-\mu+\phi_0+B_0\sigma\big]c_\sigma^\dag(\bk) c_{\sigma}(\bk) \\
	+ \big[\varepsilon_2(\bk)+\phi_1 
	+ B_1\cos(\theta) \sigma \big]  c_{\sigma}^\dag(\bk) c_\sigma(\bk+\bQ)\\
	+ B_1\sin(\theta)c_{\sigma}^\dag(\bk) c_{-\sigma}(\bk+\bQ)
	\Big) -E_0, 
\end{multline*} 	
where the constant $E_0$ (coming from $-\tfrac12\delta_{I,J}$ in \eqref{HHFgen1}) 
is of no concern for us in this computation (note that $c_\sigma(\bk+\bQ)$ can be identified with $c_\sigma(\bk-\bQ)$ and $c_\sigma(\bk\pm 2\bQ)$ with $c_\sigma(\bk)$; we also use $\varepsilon_2(\bk+\bQ)=\varepsilon_2(\bk)$). To diagonalize this Hamiltonian, we introduce a 4-vector of fermion operators 
\begin{align*}
	C(\bk) = \big(c_\uparrow^\dag(\bk),c_{\uparrow}^\dag(\bk+\bQ),c_\downarrow^\dag(\bk),c_{\downarrow}^\dag(\bk+\bQ)\big)^{\dagger},
\end{align*}
which allows us to write  
	 \begin{align*} 
 	\HHF = \sum_{\bk\in\Lambda^*_{L,\text{half}}} C(\bk)^\dag h(\bk)C(\bk)-E_0 , 
 \end{align*} 
where $\Lambda_{L,\text{half}}^*$ is a set  including exactly one of the points $\mathbf{k}$ and $\mathbf{k} + \mathbf{Q}$ for each $\mathbf{k} \in \Lambda_L^*$ (e.g., one can restrict from $\Lambda_L^*$ to $\Lambda^*_{L,\text{half}}$ by imposing the condition $-\pi\leq k_1<0$ on $\bk=(k_1,\ldots,k_n)$);
by simple computations, we find the  following hermitian matrix 
\begin{widetext} 
\begin{equation*} 
	h(\bk) = \left(\begin{array}{cccc}
		\varepsilon(\bk)-\mu+ \phi_0 + B_0 & \varepsilon_2(\bk)+ \phi_1+B_1\cos(\theta) & 0 & B_1\sin(\theta) \\
		\varepsilon_2(\bk)+\phi_1+B_1\cos(\theta) & \varepsilon(\bk+\bQ)-\mu+\phi_0 + B_0 & B_1\sin(\theta) &  0 \\
		0 & B_1\sin(\theta) & \varepsilon(\bk)-\mu+ \phi_0 - B_0 & \varepsilon_2(\bk)+ \phi_1-B_1\cos(\theta)    \\
	    B_1\sin(\theta) & 0 & \varepsilon_2(\bk)+\phi_1-B_1\cos(\theta) & \varepsilon(\bk+\bQ)-\mu+\phi_0 -B_0 
	\end{array}\right). 	
\end{equation*}
The eigenvalues $E_I$ of $h$ in \eqref{h} can be identified with the eigenvalues of this matrix $h(\bk)$, $\bk\in \Lambda^*_{L,\text{half}}$. Thus, 
\begin{align*} 
	\frac1{L^n}\sum_I\LnT(E_I) = \frac1{L^n}\sum_{\bk\in\Lambda^*_{L,\text{half}}}\sum_{r,r'=\pm} \LnT(E_{r,r'}(\bk)) 
	= \frac12\frac1{L^n}\sum_{\bk}\sum_{r,r'=\pm} \LnT(E_{r,r'}(\bk)) , 
\end{align*} 	
using $E_{r,r'}(\bk)=E_{r,r'}(\bk+\bQ)$. 
Moreover, by inserting \eqref{Penn_ansatz1} and using that $L$ is even, we compute 
\begin{multline*} 
	\frac{U}{4L^n}\sum_{\bx}(\vec{m}(\bx)^2-d(\bx)^2)
	= \frac{U}{4L^n}\sum_{\bx}\big(m_0^2+m_1^2+2m_0m_1\cos(\theta)(-1)^{\bx}
	-d_0^2-d_1^2-2d_0d_1(-1)^{\bx}\big)
	= \frac{U}{4}(m_0^2+m_1^2-d_0^2-d_1^2). 
\end{multline*} 
Thus, \eqref{FHF} yields 
\begin{align*} 
\FHF = 	\frac{U}{4}(m_0^2+m_1^2-d_0^2-d_1^2)
- \frac12\frac1{L^n}\sum_{\bk}\sum_{r,r'=\pm} \LnT(E_{r,r'}(\bk)), 
\end{align*} 	 	 
which becomes \eqref{FHFrestr} in the thermodynamic limit $L\to\infty$. 

For $\theta=0$, the matrix $h(\bk)$ has block-diagonal form, and we obtain
\begin{align*} 
E_{r,r'}(\bk) = \frac{\varepsilon(\bk)+\varepsilon(\bk+\bQ)}{2}+\phi_{0}+rB_{0}
+ r'\sqrt{\left( \frac{\varepsilon(\bk)-\varepsilon(\bk+\bQ)}{2}\right)^2 +\left(\varepsilon_2(\bk)+\phi_1+rB_1\right)^2} , 
\end{align*} 
which is \eqref{Errp}; to see this, recall \eqref{phiB} and $\varepsilon(\bk)=\varepsilon_0(\bk)+\varepsilon_1(\bk)$ where $\varepsilon_0(\bk+\bQ)=-\varepsilon_0(\bk)$ and  $\varepsilon_1(\bk+\bQ)=\varepsilon_1(\bk)$. 

As explained in Section \ref{sec:summary}, from these results one can obtain the Hartree--Fock equations and the doping constraint by differentiation; see \eqref{rHFeqs} and \eqref{rho1}. For the convenience of the reader, we write down the equations thus obtained: using \eqref{fTLT}, we obtain, for $\theta = 0$,
\begin{equation}\label{restr_HFeqs}
	\begin{split}
		d_0 &=  \frac12\int_{[-\pi,\pi]^n}\frac{d^n\bk}{(2\pi)^n}  \sum_{r,r'=\pm}\fT\big( E_{r,r'}(\bk)\big) ,\\	 	 
		m_0 &=  \frac12\int_{[-\pi,\pi]^n}\frac{d^n\bk}{(2\pi)^n} \sum_{r,r'=\pm}r \fT\big( E_{r,r'}(\bk)\big)  ,\\	 	 
		d_1 &=  \frac12\int_{[-\pi,\pi]^n}\frac{d^n\bk}{(2\pi)^n} \sum_{r,r'=\pm}r'\fT\big( E_{r,r'}(\bk)\big) \frac{\varepsilon_2(\bk)+\frac{Ud_1}{2}-r\frac{Um_1}{2}}{\sqrt{\varepsilon_0(\bk)^2
				+ \left(\varepsilon_2(\bk)+\frac{Ud_1}{2}- r\frac{Um_1}{2}\right)^2}}  ,\\	 	 
		m_1 &=  \frac12\int_{[-\pi,\pi]^n}\frac{d^n\bk}{(2\pi)^n} \sum_{r,r'=\pm}rr'\fT\big( E_{r,r'}(\bk)\big) \frac{\varepsilon_2(\bk)+\frac{Ud_1}{2}-r\frac{Um_1}{2}}{\sqrt{\varepsilon_0(\bk)^2
				+ \left(\varepsilon_2(\bk)+\frac{Ud_1}{2}- r\frac{Um_1}{2}\right)^2}} ,
	\end{split} 
\end{equation} 	
\end{widetext} 
together with the doping constraint 
\begin{align}\label{restr_rho}  
	\nu = \frac12 \int_{[-\pi,\pi]^n} \frac{d^n\bk}{(2\pi)^n} \sum_{r,r'=\pm}\fT\big( E_{r,r'}(\bk)\big).
\end{align} 
The first equation in \eqref{restr_HFeqs} together with \eqref{restr_rho} shows that $d_0 = \nu$ at a solution of the Hartree--Fock equations.
As discussed in Section \ref{sec:summary}, this does {\em not} imply that the variational parameter $d_0$ can be eliminated from the problem by absorbing it into the chemical potential: this is a common but dangerous shortcut leading to incorrect results.

\section{Mathematical derivation of Hartree--Fock theory}\label{sec:BLS}  
Our derivation in Section~\ref{sec:HF} of Hartree--Fock theory in terms of the Hartree--Fock function $\FHF$ is in part heuristic. In this section, we add the parts needed to make our derivation mathematically complete, following Bach, Lieb and Solovej \cite{BLS1994}; see also \cite{BP1996}.

In Section~\ref{sec:MHF}, we sketch the mathematical proof of the min-max principle in \eqref{minmax} for unrestricted Hartree--Fock theory, and in Section~\ref{sec:MrHF} we show how to adapt this to restricted Hartree--Fock theory.   

\subsection{Unrestricted Hartree--Fock theory}\label{sec:MHF}

We recall the definition of the grand canonical potential, 
\begin{align*} 
	\Omega(H;\rho) = \Tr(\rho H) + \frac1\beta\Tr(\rho\ln(\rho)), 
\end{align*} 	 
indicating now that $\Omega$ depends not only on the state $\rho$ but also on the Hamiltonian $H$. 
We derive the Hartree--Fock state $\rhoHF^{*}$ from the following fundamental principle discussed in Section~\ref{sec:foundation},
\begin{align}\label{principle}
	\Omega(H;\rhoHF^{*}) = \min_{\rhoqf}\Omega(H;\rhoqf) . 
\end{align} 	 

We computed $\Tr(\rhoqf H)=\langle H\rangle$ in \eqref{expectationH}. This gives 
\begin{multline}\label{Omegaqf} 
\Omega(H;\rhoqf) = \langle H_0\rangle  + \frac{U}{4}\sum_{\bx}
\big( \langle n_{\bx}-1\rangle^2 - \langle\vec{s}_\bx\rangle^2 \big) \\ +  \frac1\beta\Tr(\rhoqf\ln(\rhoqf)).
\end{multline} 	
We linearize the interaction terms by the following trick: since, trivially, 
	\begin{multline*} 
		0 = \min_{d(\bx)}\big(\langle n_{\bx}-1\rangle - d(\bx)\big)^2 
		\\= \langle n_{\bx}-1\rangle^2 + \min_{d(\bx)}\big(d(\bx)^2 - 2d(\bx)\langle n_{\bx}-1\rangle\big), 
	\end{multline*} 
we can write 
	\begin{align}\label{rhosquare}  
		\langle n_{\bx}-1\rangle^2 =
		\max_{d(\bx)}\big(-d(\bx)^2 + 2d(\bx)\langle n_{\bx}-1\rangle\big), 
\end{align} 
and similarly, 
\begin{align}\label{Ssquare}  
		-\langle \vec{s}_{\bx}\rangle^2 = \min_{\vec{m}(\bx)}(\vec{m}(\bx)^2-2\vec{m}(\bx)\cdot\langle\vec{s}_{\bx}\rangle).
\end{align} 
Thus, 
	\begin{multline*}
	\Omega(H;\rhoqf)
		= \min_{\boldsymbol{\vec{m}}}\max_{\boldsymbol{d}}\Big(\langle H_0\rangle 
		+ \frac{U}{4}\sum_{\bx} \big(\vec{m}(\bx)^2\\ -2\vec{m}(\bx)\cdot\langle\vec{s}_{\bx}\rangle-d(\bx)^2+2d(\bx)\langle n_{\bx}-1\rangle\big) \\
		+ \frac1\beta\Tr(\rhoqf\ln(\rhoqf)) \Big) . 
	\end{multline*}
We observe that 
	\begin{multline*} 
		\langle H_0\rangle + \frac{U}{4}\sum_{\bx} \big(-2\vec{m}(\bx)\cdot\langle\vec{s}_{\bx}\rangle+2d(\bx)\langle n_{\bx}-1\rangle\big)\\
	\end{multline*} 
is equal to $\langle \HHF\rangle$ with $\HHF$ in \eqref{HHF} and, by definition,  
\begin{align*} 
\langle \HHF\rangle + \frac1\beta\Tr(\rhoqf\ln(\rhoqf))
= \Omega(\HHF;\rhoqf) .
\end{align*} 	
Thus
	\begin{multline*}
	\Omega(H;\rhoqf)
		 = \min_{\boldsymbol{\vec{m}}}\max_{\boldsymbol{d}}\Big(\frac{U}{4}\sum_{\bx} \big(\vec{m}(\bx)^2-d(\bx)^2 \big) \\ + \Omega(\HHF,\rhoqf) \Big) .
\end{multline*} 
We insert this in \eqref{principle} and use that the order of $\min_{\boldsymbol{\vec{m}}}\max_{\boldsymbol{d}}$ and $\min_{\rhoqf}$ can be exchanged without changing the result; as discussed in Appendix \ref{app:exchange}, the latter is a non-trivial fact which can be proved by using results in the mathematics literature going back to a famous paper by von Neumann~\cite{vN1928}.
This gives 
\begin{multline}\label{Omegaminmax0}
\Omega(H;\rhoHF^{*})
	= \min_{\boldsymbol{d}}\max_{\boldsymbol{\vec{m}}}\Big(\frac{U}{4}\sum_{\bx} \big(\vec{m}(\bx)^2-d(\bx)^2 \big) \\ + \min_{\rhoqf}\Omega(\HHF;\rhoqf) \Big) .
\end{multline} 
We know that the minimum of $\Omega(\HHF,\rho)$ over all states $\rho$ is $\Omega(\HHF;\rhoHF)$ where $\rhoHF$ is the Gibbs state corresponding to $\HHF$ given in \eqref{rhoHF} (this is proved in Appendix~\ref{app:Gibbs}); since $\rhoHF$ is quasi-free, we can restrict this minimization to quasi-free states without changing the result: 
\begin{align*} 
\min_{\rhoqf}\Omega(\HHF;\rhoqf) = \Omega(\HHF;\rhoHF).
\end{align*} 
Moreover, by using \eqref{rhoqf} and the definition of $\Omega$, 
\begin{multline*} 
\Omega(\HHF;\rhoHF) = \Tr(\rhoHF\HHF) \\+ \frac1\beta\Tr(\rhoHF(-\beta\HHF-\ln(\ZHF)))
= -\frac1\beta\ln(\ZHF). 
\end{multline*} 	  
This yields 
\begin{align}\label{Omegaminmax}
	\Omega(H;\rhoHF^{*})
	= \min_{\boldsymbol{\vec{m}}}\max_{\boldsymbol{d}}
	\GG(\boldsymbol{d},\boldsymbol{\vec{m}};\mu) 
\end{align}
with $\GG$ defined in \eqref{cGdef}; recall that $\FHF=\GG/L^n$. 

To summarize: {\em We proved that the quasi-free state $\rhoHF^{*}$ providing the best approximation of the exact Gibbs state of the Hamiltonian in \eqref{Hdef}--\eqref{H0def} is the Gibbs state for the Hartree--Fock Hamiltonian $\HHF=\HHF(\boldsymbol{d},\boldsymbol{\vec{m}};\mu)$ in \eqref{HHF} with the mean-fields $\boldsymbol{d}=\boldsymbol{d}^{*}$, $\boldsymbol{\vec{m}}=\boldsymbol{\vec{m}}^{*}$ determined by the min-max principle in \eqref{minmax}, and the free energy of this state is $\FHF(\boldsymbol{d}^{*},\boldsymbol{\vec{m}}^{*};\mu)$, with $\FHF=\FHF(\boldsymbol{d},\boldsymbol{\vec{m}};\mu)$ in \eqref{FHF}. 
} 
\subsection{Restricted Hartree--Fock theory}\label{sec:MrHF} 
We now show that the min-max principle also applies to restricted Hartree--Fock theory, taking the restriction to AF states as a representative example. For this, we need some notation. 

Let 
\begin{align}\label{HAF}  
	H_{\mathrm{AF}} = H_0 + \frac{U}{2}\sum_{\bx}\big( d_0(n_{\bx}-1)-(-1)^{\bx}m_1\vec{e}\cdot\vec{s}_{\bx} 
	\big) 	
\end{align} 	 
be the Hartree--Fock Hamiltonian for AF states (with real parameters $d_0$ and $m_1$ and a unit vector $\vec{e}\in\R^3$), and 
\begin{align}\label{rhoAF} 
	\rho_{\mathrm{AF}} = \frac{\ee^{-\beta H_{\mathrm{AF}} }}{\Tr\big( \ee^{-\beta H_{\mathrm{AF}} }\big)}
\end{align} 	
the corresponding Gibbs state. Below, we sometimes write $\rho_{\mathrm{AF}}(d_0,m_1\vec{e};\mu)$, and similarly for $H_{\mathrm{AF}}$, if we want to make clear what  parameters we mean.

As discussed in Section~\ref{sec:foundation}, Hartree--Fock theory restricted to AF states amounts to the following variational problem, 
\begin{align*} 
	\Omega(H;\rho_{\mathrm{AF}}^{*})
	= \min_{\rho_{\mathrm{AF}}} \Omega(H;\rho_{\mathrm{AF}}), 
\end{align*} 	
where the right-hand side is short for 
\begin{align*} 
	 \min_{d_0,m_1\vec{e}}  \Omega(H,\rho_{\mathrm{AF}}(d_0,m_1\vec{e};\mu)). \end{align*} 	 
Since $\rho_{\mathrm{AF}}$ is quasi-free, we can compute  $\Omega(H;\rho_{\mathrm{AF}})$ using \eqref{Omegaqf}. To adapt the tricks in \eqref{rhosquare} and \eqref{Ssquare} to the present case, we note that,  in the AF state $\rho_{\mathrm{AF}}$,  $\langle n_{\bx}-1\rangle=a$ and $\langle\vec{s}_{\bx}\rangle=(-1)^{\bx}\vec{e}\, b$ for some constants $a$ and $b$ (for the convenience of the reader, we prove this fact in Appendix~\ref{app:symm}). For this reason, we can write 
\begin{multline*}
\sum_{\bx}\langle n_{\bx}-1\rangle^2 = L^n a^2
= L^n\max_{\tilde d_0}(-\tilde{d_0}^2+2\tilde d_0 a)\\
= \max_{\tilde d_0}\Big(-L^n\tilde{d_0}^2+2\sum_{\bx} \tilde d_0\langle n_{\bx}-1\rangle\Big)
\end{multline*} 
and
\begin{multline*}
		-\sum_{\bx}\langle \vec{s}_{\bx}\rangle^2 = -L^nb^2 = L^n\min_{\tilde m_1}(\tilde m_1^2-2\tilde m_1b)\\
	= \min_{\tilde m_1} \Big( L^n \tilde m_1^2 - 2 \sum_{\bx} (-1)^{\bx}\tilde m_1\vec{e}\cdot \langle \vec{s}_{\bx}\rangle\Big).    
\end{multline*} 
Thus, in view of \eqref{Omegaqf} and \eqref{HAF}, we get 
\begin{multline*}
	\Omega(H;\rho_{\mathrm{AF}}) = \min_{\tilde m_1}\max_{\tilde d_0}\Big( L^n \frac{U}{4}(\tilde m_1^2-\tilde d_0^2) \\ + \langle \tilde H_{\mathrm{AF}}\rangle +  \frac1\beta\Tr(\rho_{\mathrm{AF}}\ln(\rho_{\mathrm{AF}}))\Big) \\
= \min_{\tilde m_1}\max_{\tilde d_0}\Big( L^n \frac{U}{4}(\tilde{m}_1^2- \tilde{d}_0^2)
+ \Omega(\tilde H_{\mathrm{AF}};\rho_{\mathrm{AF}})\Big) 
\end{multline*} 	
with $\tilde H_{\mathrm{AF}}=H_{\mathrm{AF}}(\tilde d_0,\tilde m_1\vec{e};\mu)$. 
This yields --- again, we change orders of min-max and min; we prove in Appendix \ref{app:exchange} that this does not change the result --- 
\begin{multline}\label{OmegaAFminmax} 
	\Omega(H;\rho_{\mathrm{AF}}^{*})	
	=  \min_{\tilde m_1}\max_{\tilde d_0}\Big( L^n \frac{U}{4}(\tilde m_1^2-\tilde d_0^2)
	\\
	+ \min_{\rho_{\mathrm{AF}}} \Omega(\tilde H_{\mathrm{AF}};\rho_{\mathrm{AF}})\Big).   
\end{multline} 
The remaining argument is very similar to the one in Section~\ref{sec:MHF}: 
Since the minimum of $\Omega(\tilde H_{\mathrm{AF}};\rho)$ over all states $\rho$ is attained for $\rho=\tilde\rho_{\mathrm{AF}}=\rho_{\mathrm{AF}}(\tilde d_0,\tilde m_1\vec{e};\mu)$ contained in the set of all AF states, we have
\begin{align*} 
	\min_{\rho_{\mathrm{AF}}} \Omega(\tilde H_{\mathrm{AF}};\rho_{\mathrm{AF}})
	= \Omega(\tilde H_{\mathrm{AF}};\tilde \rho_{\mathrm{AF}}) 
	= -\frac1\beta\ln(\Tr(\ee^{-\beta\tilde H_{\mathrm{AF}}})).  
\end{align*} 	
Thus, 
\begin{align*} 
	\Omega(H;\rho_{\mathrm{AF}}^{*})	
	=  \min_{m_1}\max_{d_0}\GG_{\mathrm{AF}}(d_0,m_1;\mu)
\end{align*} 	
where 
\begin{align*} 
\GG_{\mathrm{AF}}(d_0,m_1;\mu) 
= L^n \frac{U}{4}(m_1^2-d_0^2) - \frac1\beta\ln(\Tr(\ee^{-\beta H_{\mathrm{AF}}}))
\end{align*} 	
with $H_{\mathrm{AF}}=H_{\mathrm{AF}}(d_0,m_1;\mu)$ (we dropped the tildes). 
Since 
\begin{align*}  
\lim_{L\to\infty} L^{-n}\GG_{\mathrm{AF}}(d_0,m_1;\mu) 
= \FHF(d_0,m_1;\mu) 
\end{align*} 
where $\FHF(d_0,m_1;\mu)$ is given in \eqref{FHF_AF}--\eqref{E_AF}, we obtain that $\rho_{\mathrm{AF}}^{*}$ can be determined by the  min-max principle in \eqref{minmaxAF}. 

To summarize: {\em We proved that the AF state providing the best approximation to the exact Gibbs state of the Hamiltonian \eqref{Hdef}--\eqref{H0def} is the Gibbs state for the Hartree--Fock Hamiltonian in \eqref{HAF}  with parameters $d_0=d_0^{*}$, $m_1=m_1^{*}$ determined by the min-max principle in \eqref{minmaxAF}.}

It is easy to adapt the argument above to F states, and to extend it to restricted Hartree--Fock theory allowing for AF, F, and P states. For a state where a fraction $w$ of the system is in an AF state and a fraction $1-w$ is in an F state (for example), one finds by an argument similar to the one in Section~\ref{sec:details} that 
\begin{align*} 
\Omega(H;\rho_{\mathrm{mixed}}) = w\Omega(H;\rho_{\mathrm{AF}})+(1-w)	\Omega(H;\rho_{\mathrm{AF}})+\Delta\Omega
\end{align*}	
where $\Delta\Omega/L^n\to 0$ in the thermodynamic limit $L\to\infty$; using this, one can further substantiate our treatment in Section~\ref{sec:details}. 

\section{Conclusions}\label{sec:conclusions} 
We showed that Hartree--Fock theory is a more powerful method for Hubbard-like models than is generally believed: if one is careful about stability, Hartree--Fock theory restricted to simple states describing pure phases predicts, in many cases, a mixed phases where exotic physics is to be expected. We gave three example: (i) For the Hubbard model at zero temperature, Hartree--Fock theory predicts an anti-ferromagnetic phase only strictly at half-filling, but away from half-filling there is a significant doping regime (depending on $U/t$) where a mixed phase occurs, and this is true in any dimension; see Figs.~\ref{fig:3D}--\ref{fig:inftyD}. (ii) For a recently introduced extension of the 2D Hubbard model introduced in \cite{DLKK2024} to describe altermagnetism that can be experimentally explored in cold atom systems, Hartree--Fock theory not only predicts altermagnetism but also mixed and ferromagnetic phases at half-filling (depending on parameters); see Fig.~\ref{fig:DLKK}.
 (iii) For the extended Hubbard model with next-nearest-neighbor hopping of strength $t'$ at zero temperature, Hartree--Fock theory predicts a non-trivial Mott transition; this transition is not sharp but rather a crossover, with a mixed phase in a range $U_{c1}<U<U_{c2}$ (depending on $t'/t$) interpolating between a paramagnetic phase for $0<U<U_{c1}$ and an antiferromagnetic phase for $U>U_{c2}$ (Fig.~\ref{fig8} proves this in one example, and we plan to give further details on this elsewhere).  

We presented an update of Hartree--Fock theory for Hubbard-like models which protects against a subtle mistake going back to a seminal paper by Penn \cite{P1966}. Unfortunately, this mistake became well-established, which has made the literature on Hartree--Fock theory for Hubbard-like models unreliable. To convince readers that our updated method is the correct one, we also included a self-contained derivation of Hartree--Fock theory, starting from fundamental principles and including all technical details. Our updated  mean-field phase diagrams of the Hubbard model differ from the established ones, and they are  more interesting in that they also reveal mixed phases in addition to conventional AF, F, and P phases; see Figs.~\ref{fig:3D}--\ref{fig:inftyD}. To demonstrate the power of the method, we also presented similar results for extended Hubbard models; see Fig.~\ref{fig:DLKK}. 

We hope that this paper can re-establish Hartree--Fock theory as a basic, simple, and reliable method for Hubbard-like models in the physics literature.
We also hope that it will cause other Hartree--Fock results on Hubbard-like models in the literature to be checked and, if needed, corrected: we feel that it is important that such checks and corrections are published.

For readers preferring the functional integral derivation of Hartree--Fock theory, we mention that such a derivation leading to the min-max principle in \eqref{minmaxAF} was given in Ref.~\cite{LW1997}; the functional integral derivation is interesting since it provides a complimentary physics interpretation of mixed states.

While this paper is written for physicists, we mention that our results can be made mathematically precise: our updated Hartree--Fock method is based on mathematical theorems proved by Bach, Lieb, and Solovej \cite{BLS1994}. Furthermore,  while the phase diagrams we present were obtained numerically, we recently succeeded to extend methods in \cite{LL2024} to analytic expansions that allow us to compute some of these phase diagrams analytically; these expansions are more accurate than numerical computations, and they can be used to prove the occurrence of mixed phases according to the standards of mathematics  \cite{Comment:expansions}. 

\acknowledgements We thank Christophe Charlier, Joscha Henheik, Asbj{\o}rn Lauritsen, and Valdemar Melin for helpful discussions and collaborations on related projects. E.L. acknowledges support from the Swedish Research Council, Grant No. 2023-04726. 
J.L. acknowledges support from the Swedish Research Council, Grant No.\ 2021-03877.

\appendix 

\section{Definitions and background}\label{app:Gen}
For the convenience of the reader, we collect some well-known definitions and background material. 

\subsection{General fermion models}\label{app:CAR}
We collect definitions and basic facts about a large class of lattice fermion models that include the Hubbard-like models discussed in the main text. 

In general, a lattice fermion model is defined on a fermion Fock space generated by fermion creation and annihilation operators $c_I^\dag$ and $c_I$ labeled by some index $I=1,2,\ldots,\cN$, where $\cN$ is the number of one-particle degrees of freedom. 
The fermion operators are fully characterized by the canonical anticommutator relations 
\begin{equation}\label{CAR}  
\{ c_I^\pdag,c_J^\dag\}=\delta_{I,J},\quad  	
\{c_I,c_J\}=0 
\end{equation} 
for $I,J=1,\ldots,\cN$ (we write $\{A,B\}$ for the anti-commutator $AB+BA$ and $\delta_{I,J}$ is the Kronecker delta), and the requirement that $c_I^\dag$ is the hermitian conjugate of $c_I$. If $|0\rangle$ is the vacuum in the fermion Fock space such that $c_I|0\rangle=0$ for all $I$, then a complete orthonomal set of states in the Fermion Fock space is given by 
$$
|n_1,\ldots,n_{\cN}\rangle := 
(c_1^\dag)^{n_1}\cdots (c_{\cN}^\dag)^{n_{\cN}}|0\rangle 
$$
where $n_I$ is 0 or 1; since there are $2^{\cN}$ distinct such states, the fermion Fock space is isomorphic to $\C^{2^{\cN}}$. 

For example, for the $n$-dimensional Hubbard-like model discussed in the main text, $I=(\bx,\sigma)$ where $\bx=(x_1,\ldots,x_n)$, $x_i=1\ldots,L$ for $i=1,\ldots,n$,   and $\sigma = \uparrow,\downarrow$; thus, $\cN=2L^n$. 

A generic Hamiltonian is defined as 
\begin{equation} 
H = H_0 + \sum_{I,J,K,L} v_{I,J,K,L}^\pdag c_I^\dag c_J^\pdag   c_K^\dag c_L^\pdag  
\end{equation} 
where 
\begin{equation} 
H_0 = \sum_{I,J} t_{I,J}^\pdag c_I^\dag c_J^\pdag 
\end{equation} 
is the non-interacting part (all sums are over $1,\ldots,\cN$). The model parameters are such that $H$ is hermitian, i.e., 
\begin{equation} 
t_{I,J}=\overline{t_{J,I}},\quad  
v_{I,J,K,L}=\overline{v_{L,K,J,I}}
\end{equation} 
for all $I,J,K,L$, where the bar denotes complex conjugation. In particular, $(t_{I,J})_{I,J=1}^{\cN}$ defines a hermitian $\cN\times\cN$ matrix. 

\subsection{Lattice Fourier transform}\label{app:FT} 
For a finite even integer $L>0$, we denote as $\Lambda_L$ the set of all lattice points $\bx=(x_1,\ldots,x_n)$ with $x_i=1,\ldots,L$ for $i=1,\ldots,n$, where $x_i+L$ is identified with $x_i$. We denote as $\Lambda_L^*$ the set of corresponding Fourier modes, i.e., 
$\Lambda_L^*$ is the set of all $\bk=(k_1,\ldots,k_n)$ such that 
\begin{align}\label{bk} 
k_i\in (2\pi/L)\Z,\quad -\pi\leq k_i< \pi\quad (i=1,\ldots,n). 
\end{align} 

The Fourier transform of the fermion creation and annihilation operators is defined as 
\begin{equation} 
c_\sigma(\bk)= \frac{1}{L^{n/2}}\sum_{\bx} c_{\bx,\sigma}\ee^{-\ii \bk\cdot\bx}
\end{equation} 
for $\bk\in\Lambda_L^*$, with $\sum_{\bx}$ short for $\sum_{\bx\in\Lambda_L}$ 
Using
\begin{equation*} 
\sum_{\bx} \ee^{-\ii \bk\cdot\bx} = L^n \delta_{\bk,\mathbf{0}} 
\end{equation*} 
where  $\delta_{\bk,\mathbf{0}}= \prod_{i=1}^n\delta_{k_i,0}$, one finds that the inverse is given by 
\begin{equation} 
c_{\bx,\sigma} = \frac1{L^{n/2}} \sum_{\bk} c_\sigma(\bk)
\ee^{\ii\bk\cdot\bx} 	
\end{equation} 
with $\sum_{\bk}$ short for $\sum_{\bk\in\Lambda_L^*}$
To obtain results in the thermodynamic limit, one uses that, for sufficiently well-behaved functions $g$ of $\bk$, 
\begin{equation}\label{Riemannsum}  
\lim_{L\to\infty} \frac1{L^{n}}\sum_{\bk}g(\bk)
= \int_{[-\pi,\pi]^n} \frac{d^n\bk}{(2\pi)^n}\, g(\bk)
\end{equation} 
(the sum on the left-hand side is a Riemann sum converging in the limit to the integral on the right-hand side). Moreover, if $g$ depends on $\mathbf{k}$ only via the function $\varepsilon_0(\bk)$ defined in \eqref{veps0}, then the $n$-dimensional integral on the right-hand side in \eqref{Riemannsum} can be converted to a one-dimensional integral using the density of states $N_n(\eps)$ in \eqref{DOS}, 
\begin{equation}\label{kintegralwithDOS} 
 \int_{[-\pi,\pi]^n} \frac{d^n\bk}{(2\pi)^n}\, g(\varepsilon_0(\bk)) 
 = \int N_n(\eps)g(\eps)d\eps 
\end{equation} 
(the integral on the right-hand side can be taken over $\R$ or over the interval  $[-2nt,2nt]$ where $N_n(\eps)$ is non-zero).

\subsection{Particle-hole transformation}\label{app:PH}
The particle-hole transformation leaving the Hamiltonian in \eqref{Hdef} invariant is defined as 
\begin{equation}\label{PHdef} 
	c^\pdag_{\bx,\sigma}\to  (-1)^{\bx}c^\dag_{\bx,\sigma},
	\quad c^\dag_{\bx,\sigma}\to  (-1)^{\bx}c^\pdag_{\bx,\sigma}
\end{equation} 
together with $\mu\to -\mu$ and $t'\to -t'$, and this transformation changes $\nu\to -\nu$ (since, under \eqref{PHdef}, $n_{\mathbf{x}, \sigma}-\tfrac12 \to -(n_{\mathbf{x}, \sigma}-\tfrac12)$ and $c_{\bx,\sigma}^\dag c_{\by,\sigma}^\pdag \to \pm c_{\bx,\sigma}^\dag c_{\by,\sigma}^\pdag$, with $+$ and $-$  if the sites $\bx,\by$ are nearest-neighbor and next-nearest-neighbor, respectively). Thus, for $t'=0$ (but not otherwise), half-filling $\nu=0$ corresponds to $\mu=0$.

\subsection{Density of states}\label{app:DOS} 
\begin{figure}
	\vspace{0.4cm}
	\begin{center}
		\hspace{-.6cm}
		\begin{overpic}[width=.46\textwidth]{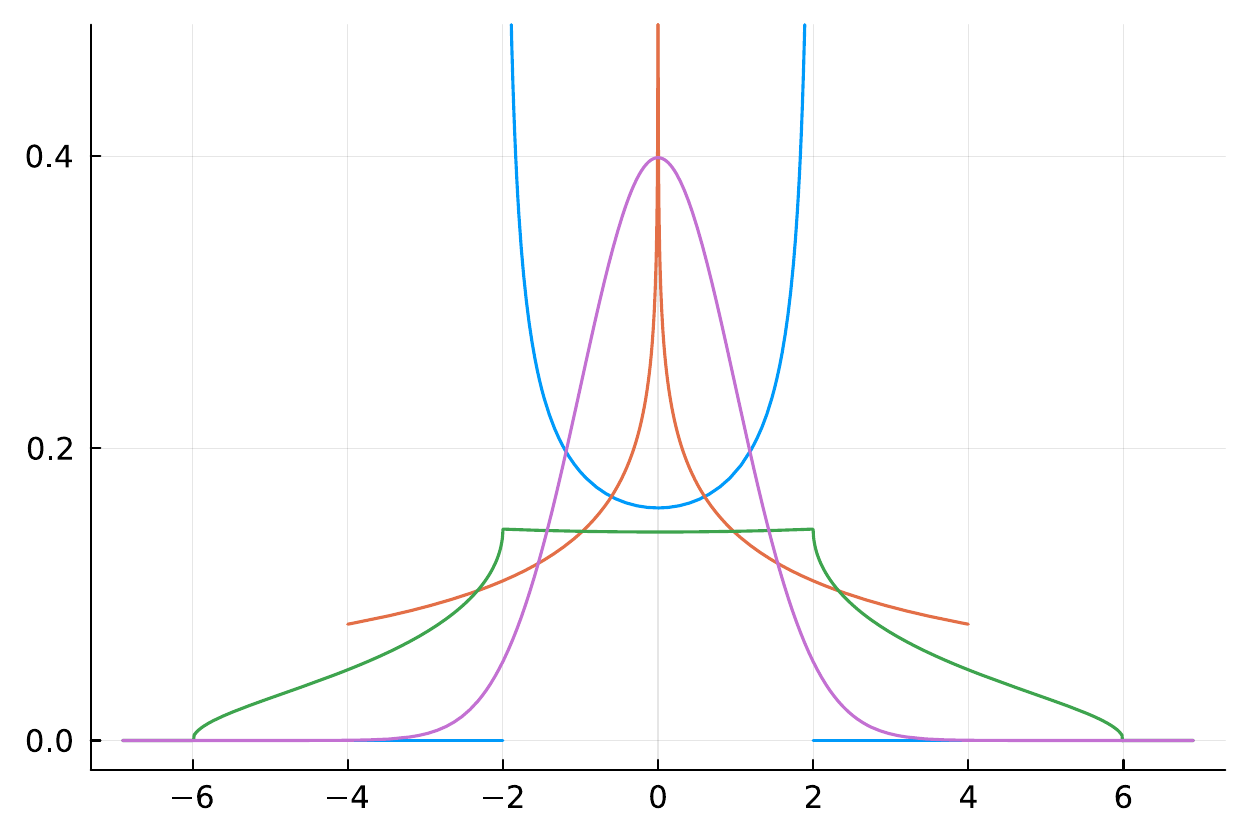 }
			\put(5.5,66.5){\footnotesize $N_n(\eps)$}
			\put(100,4){\footnotesize $\eps$}
		\end{overpic}				
		\caption{\label{fig:DOS} 
Plots of the density of states $N_n(\eps)$ in \eqref{DOS} of the tight-binding band relation in \eqref{veps0} for different dimensions $n=1$ (blue), $n=2$ (red), $n=3$ (green) and $n=\infty$ (magenta); we set $t=1$ for $n=1,2,3$ and $t=1/\sqrt{2n}$ for $n\to \infty$.}
	\end{center}
	\vspace{-0.4cm}
\end{figure}
We summarize known facts about the density of states $N_n(\eps)$ in \eqref{DOS} for the Hubbard model band relation in \eqref{veps0} for dimensions $n=1,2,3$, and $\infty$; see Fig.~\ref{fig:DOS} for a plot of these functions. 

In 1D, it is known that   
\begin{equation} 
N_1(\eps) = \frac{\theta(1-|\eps/2t|)}{2t\pi\sqrt{1-(\eps/2t)^2}}
\end{equation} 
where $\theta$ is the Heaviside function. 
In 2D and 3D, $N_n(\eps)$ are transcendental functions which  can be computed accurrately to any desired precision using efficient expansions; see e.g.\ \cite{LL2024} and references therein. Finally, $N_n(\eps)$ with $t=t^*/\sqrt{2n}$ and fixed $t^*>0$ has a well-defined non-trivial limit as $n\to\infty$ given by the Gaussian distribution \cite{W1983,MV1989} 
\begin{equation} 
N_\infty(\eps) = \frac1{\sqrt{2\pi}t^*} \ee^{-\frac12(\eps/t^*)^2}.
\end{equation} 

\subsection{Gibbs state from grand canonical potential}\label{app:Gibbs}  
We give a proof of the well-known fact that, for fixed Hamiltonian $H$ and $0<\beta<\infty$, the Gibbs state is the unique absolute minimizer of the grand canonical potential among all states.  

We denote as $N$ the (finite) dimension of the pertinent Hilbert space, i.e., the Hamiltonian $H$ is a self-adjoint $N\times N$ matrix and the set $\cS$ of all states contains all self-adjoint $N\times N$ matrices $\rho$ with non-negative eigenvalues satisfying $\Tr(\rho)=1$ (in the main text, $N=2^{\cN}$). 
We also recall the well-known fact that, for fixed $H$, the grand canonical potential $\Omega(\rho)$ in \eqref{Omegadef} defines a strictly convex function  of $\rho\in \cS$,  i.e., for arbitrary distinct  $\rho_1,\rho_2\in\cS$ and $\alpha\in (0,1)$, $\alpha\rho_1+(1-\alpha)\rho_2\in\cS$ and 
\begin{align*}
\Omega(\alpha\rho_1+(1-\alpha)\rho_2) < \alpha\Omega(\rho_1)+(1-\alpha)\Omega(\rho_2)
\end{align*} 
(this is a simple consequence of the well-known strict concavity of entropy $-\Tr(\rho\ln(\rho))$; see \cite[Sec.~II.B]{W1978} for a proof).
We first determine all stationary points of $\Omega(\rho)$ away from the boundary of $\cS$, i.e, we restrict ourselves to states $\rho$ which only have eigenvalues in the open interval $(0,1)$. Pick an arbitrary such $\rho\in\cS$, together with a hermitian $N\times N$ matrix $v$ such that $\Tr(v)=0$; then $\rho+\alpha v\in\cS$ for sufficiently small real $\alpha$. 
 We compute the variation 
$$
\delta_v \Omega(\rho) = \left. 
\frac{d}{d\alpha}\Omega(\rho+\alpha v)\right|_{\alpha=0} 
$$
and determine stationary points $\rho^*$ of $\Omega$ by the following condition, $\delta_v\Omega(\rho)=0$ for all $v$ at $\rho=\rho^*$. We use $\delta_v \Tr(f(\rho))=\Tr(f'(\rho)v)$ for the function $f(x)=x\ln(x)$ of $x\in (0,1)$
(to see this, write $\rho=\sum_k \rho_k|k\rangle\langle k|$ with $\langle k|k'\rangle=\delta_{k,k'}$ and $\rho_k=\langle k|\rho|k\rangle$ for $k,k'=1,\ldots,N$; then the eigenvalues of $\rho+\alpha v$ are $\approx \rho_k + \alpha\langle k|v|k\rangle$ by first-order perturbation theory (where the approximation, $\approx$, amounts to ignoring irrelevant terms of order $\alpha^2$), and thus, if $|\alpha|$ is sufficiently small, 
\begin{multline*}
\Tr(f(\rho+\alpha v)-f(\rho)) \approx  \sum_k [f(\rho_k + \alpha \langle k|v|k\rangle)-f(\rho_k)] \\
\approx  \sum_k \alpha f'(\rho_k) \langle k|v|k\rangle
= \alpha \sum_k \langle k|f'(\rho)v|k\rangle= \alpha \Tr(f'(\rho)v); \end{multline*}
dividing by $\alpha$ and taking the limit $\alpha\to 0$ implies the result); since  
$f'(x)=\ln(x)+1$, 
$$
\delta_v\Omega(\rho) = \Tr([H  + (1/\beta)(\ln(\rho)+ I)]v)
$$
where $I$ is the $N\times N$ unit matrix. Since $\Tr(v)=0$ for allowed variations $v$, the condition $\delta_v\Omega(\rho)=0$ for all $v$ is true if and only if 
$$
H  + (1/\beta)(\ln(\rho)+ I) =c I
$$
for some constant $c$. This gives $
\rho = \ee^{\beta c-1}\ee^{-\beta H}$, 
and the condition $\Tr(\rho)=1$ determines $c$: $\ee^{\beta c-1}=1/Z$ with $Z=\Tr(\ee^{-\beta H})$ the partition function. 
Thus, there is a stationary point of the function $\Omega(\rho)$ of $\rho\in\cS$ given by
\begin{align*} 
\rho^*=\ee^{-\beta H}/\Tr(\ee^{-\beta H}), 
\end{align*} 
which is the Gibbs state. 

To conclude, we recall the well-known argument proving that a stationary point of a strictly convex function is the unique  absolute minimum of this function. 
For arbitrary $\rho\in\cS$ different from $\rho^*$ and for $\alpha\in(0,1)$,
\begin{multline*}
	\Omega(\rho^*+\alpha(\rho-\rho^*))-\Omega(\rho^*) \\
	= 
		\Omega(\alpha\rho+(1-\alpha)\rho^*)-\Omega(\rho^*)\\
		<
		\alpha\Omega(\rho)+(1-\alpha)\Omega(\rho^*) -\Omega(\rho^*)\\
		= \alpha[\Omega(\rho)-\Omega(\rho^*)], 
\end{multline*}
using strict convexity in the third step. Dividing by $\alpha$ and taking the limit $\alpha\downarrow 0$ yields 
$$
\delta_v \Omega(\rho^*) <  \Omega(\rho)-\Omega(\rho^*) 
$$
for $v=\rho-\rho^*$ (which clearly satisfies $\Tr(v)=0$). Since $\delta_v \Omega(\rho^*)=0$, this implies  $\Omega(\rho^*)< \Omega(\rho)$ for all $\rho\in\cS$ different from $\rho^*$. This completes the proof.

\subsection{Remarks on convexity and mixed states}\label{app:remarks} 
The proof in Section~\ref{app:Gibbs} gives an example of the power of abstract mathematics in theoretical physics.
In this Appendix, we give a few loosely  connected remarks illustrating how such mathematics can shed light on restricted Hartree--Fock theory and mixed states.

\medskip
 
\noindent {\bf R1:} In Appendix~\ref{app:Gibbs} we proved that the grand canonical potential $\Omega$ is strictly convex for finite system size $L$. As discussed in Remark~R3 below, the mixed states constructed in Section~\ref{sec:details} are examples proving that, in the thermodynamic limit $L\to\infty$, the grand canonical potential $\Omega(\rho)$ is only convex (and not strictly convex), i.e.,
\begin{align*}
	\Omega(\alpha\rho_1+(1-\alpha)\rho_2) \leq  \alpha\Omega(\rho_1)+(1-\alpha)\Omega(\rho_2)
\end{align*} 
for distinct $\rho_1,\rho_2\in\cS$ and $\alpha\in(0,1)$. 

In the limit $L\to\infty$, if we restrict $\cS$ to a convex subset $\cS_c$ (i.e.,   $\alpha\rho_1+(1-\alpha)\rho_2\in\cS_c$ for all $\rho_1,\rho_2\in\cS_c$ and $\alpha\in(0,1)$), the function $\Omega(\rho)$ of $\rho\in\cS_c$ inherits convexity from the convexity of $\Omega(\rho)$ as a function of $\rho\in\cS$. Thus, the corresponding approximation method 
\begin{align} 
\Omega(\rho^*) = \min_{\rho\in\cS_c}\Omega(\rho) 
\end{align} 
can profit from general results about convex function; e.g., one knows without further work that the stationary points  of the function $\Omega(\rho)$ of $\rho\in\cS_c$ are degenerate minima which form a convex set (this follows from general properties of convex functions).
 
\smallskip

\noindent {\bf R2}: Remark~R1 highlights the significance of the set $\cS_{\mathrm{qf}}$ of quasi-free states not being convex: Hartree--Fock equations may have many distinct solution, $\rho^{(i)}$, $i=1,\ldots,M$, for some $M \geq 2$, and these solutions can have different free energies $\Omega(\rho^{(i)})$ which not even need to be local minima  (we discussed examples in the main text); this would not be possible if the set $\cS_{\mathrm{qf}}$ was convex. A similar remark applies to restricted Hartree--Fock theory. 

\smallskip

\noindent {\bf R3:} The mixed states constructed in Section~\ref{sec:details} can be regarded as affine linear combinations of the AF and P states $\rho_{\mathrm{AF}}$ and $\rho_{\mathrm{P}}$:  
\begin{align*}
	\rho=w\rho_{\mathrm{AF}} + (1-w)\rho_{\mathrm{P}}
\end{align*}
for $w\in(0,1)$ \cite{LW1997}. Since the free energy of the mixed state $\rho$ in the limit $L\to\infty$ is  
\begin{align*}
	\Omega(\rho) =w\Omega(\rho_{\mathrm{AF}})  + (1-w)\Omega(\rho_{\mathrm{P}}), 
\end{align*}
the strictness of the concavity of the grand canonical potential is lost in the thermodynamic limit. 
More generally, if we define a set $\cS_{\mathrm{qf,rst}}$ of quasi-free states by some ansatz on the Hartree--Fock fields (one example would be  \eqref{Penn_ansatz}), one can construct corresponding mixed states 
\begin{align}\label{rho_hull} 
	\rho = \sum_{i=1}^K \alpha_i \rho_i
\end{align} 	
where macroscopically large regions of the system coincide with the state $\rho_i\in\cS_{\mathrm{qf,rst}}$, with $\alpha_i\in (0,1)$ the fraction of lattice sites in the state $\rho_i$ ($\sum_{i=1}^K\alpha_i=1$ and $K=2,3,\ldots$; the special case $K=1$ are the quasi-free states and, for fixed $K=2,3,\ldots$, we exclude the cases where $\alpha_i=0$ or  $\alpha_i=1$ for any $i$ since these cases are already included in the set of such states for some smaller value of $K$). By arguments generalizing the one in Section~\ref{sec:details}, the grand canonical potential of such a state in the limit $L\to\infty$ is 
\begin{align}\label{Om_hull}
	\Omega(\rho) = \sum_{i=1}^K \alpha_i \Omega(\rho_i)
\end{align} 
(for finite $L$, there are corrections $\Delta\Omega$ coming from phase boundaries, but these corrections become negligible in the limit $L\to\infty$). 	
By setting $\alpha_K=1-\sum_{i=1}^{K-1}\alpha_i$, the $\alpha_i$ for $i=1,\ldots,K-1$ can be regarded as variational parameters; a necessary condition for $\Omega$ to be minimized at a state for which all $\alpha_i\in(0,1)$ is that
$\frac{\partial\Omega}{\partial\alpha_i}=0$ for $i=1,\ldots,K-1$, which implies $\Omega(\rho_i)=\Omega(\rho_K)$ for $i=1,\ldots,K-1$, i.e., 
\begin{align}\label{condition_hull} 
	\Omega(\rho_i) = \Omega(\rho_j)\quad (1\leq i<j\leq K). 
\end{align} 	
The case $K=2$ correspond to the mixed states discussed in the main text. Our arguments above suggest that there should exist generalizations \eqref{rho_hull} of these mixed states to $K=3,4,\ldots$ and, for such a state to exist, one has to have distinct quasi-free states $\rho_i$ minimizing the grand canonical potential and such that \eqref{condition_hull} holds. In Remark~R4 below we give examples of such states, and we explain that  the existence of these states are interesting for the physics interpretation of restricted Hartree--Fock theory. 

Thus, if $\cS_{\mathrm{qf,rst}}$ is a set of quasi-free states defining some restricted Hartree--Fock theory, it is natural from a physics point of view to add states as in \eqref{rho_hull},  and to extend the definition of $\Omega$ to these states using \eqref{Om_hull}. From a mathematical point of view, the resulting set of states is the convex hull, $\cS^{\mathrm{cv.h.}}_{\mathrm{qf,rst}}$, of $\cS_{\mathrm{qf,rst}}$, i.e., it is the smallest convex subset of $\cS$ containing $\cS_{\mathrm{qf,rst}}$. 
Moreover, \eqref{Om_hull} extends the definition of $\Omega$ in a simple way from $\cS_{\mathrm{qf,rst}}$ to $\cS^{\mathrm{cv.h.}}_{\mathrm{qf,rst}}$. As discussed in Remark~R1, this convex generalization of the restricted Hartree--Fock theory has nice mathematical properties. 

\smallskip 

\noindent {\bf R4:} As an example of our discussion in Remark~R3, consider the set $\cS_{\mathrm{F}}$ of all F-states defined in \eqref{ansatz_F} and its convex hull $\cS^{\mathrm{cv.h.}}_{\mathrm{F}}$. If $\vec{e}_i$, $i=1,\ldots,K$, are different unit vectors in $\R^3$ and $\rho_i$ the corresponding F-states with the same value of $d_0=d_0^*$ and $m_0=m_0^*>0$ given by a stable solution of the Hartree--Fock equations, then \eqref{rho_hull} can be interpreted physically as a state where macroscopically large F regions with different magnetization directions $\vec{e}_i$ coexist such that, in the thermodynamic limit, the free energy of this mixed state is the same as for an F-state where $\vec{e}$ is constant. This is interesting for the physical interpretation of F solutions in Hartree--Fock theory: if one finds an F state in Hartree--Fock theory minimizing Hartree--Fock theory restricted to AF, F, and P states, one cannot conclude from this that the system has a non-zero magnetization; to know this, one would have to go beyond Hartree--Fock theory (since the F-solution with fixed $\vec{e}$, which indeed has a non-zero magnetization, is only one representative in a large set of mixed states with the same free energy and where many of them have zero magnetization).
 
\bigskip

\section{Computation details}\label{app:details}  
To keep this paper self-contained, we give derivations of well-known mathematical results used in the main text.

\subsection{Hartree--Fock partition function}\label{app:QF}
We compute the partition function $\Zqf$ in \eqref{rhoqf}--\eqref{Hqf} to derive the result in \eqref{TlnZqf}.

The spectral theory of self-adjoint matrices guarantees the existence of a unitary matrix $U=(U_{I,J})$ and a diagonal real matrix $E=\diag(E_1,\ldots,E_{\cN})$ such that $h = U^\dag E U$, or equivalently 
\begin{align}\label{h_diag} 
	h_{I,J} = \sum_{K} (U^\dag)_{I,K} E_K U_{K,J},\quad (U^\dag)_{I,K}=\overline{U_{K,I}} 
\end{align} 
with $E_K$ the eigenvalues of the one-particle Hamiltonian $h$. Using this, we can construct a Bogoliubov transformation diagonalizing the Hartree--Fock Hamiltonian: the unitarity of $U$ implies that 
\begin{align}\label{BT}  
	\tilde c_I=\sum_J U_{I,J}c_J
\end{align} 
are fermion operators satisfying the canonical anti-commutator relations in \eqref{CAR}, and the formulas above imply that $\Hqf$ in \eqref{Hqf} can be expressed as  
\begin{multline}\label{HHFdiag}  
	\Hqf = \sum_{I,J,K} \overline{U^\pdag_{K,I}}E^\pdag_K U^\pdag_{K,J} \big( c_I^\dag c_J^\pdag-\tfrac12\delta_{I,J}\big) 
	\\ = \sum_{K} E^\pdag_K \big( \tilde c_K^\dag \tilde c_K^\pdag-\tfrac12\big) .     	
\end{multline} 
To compute the trace of $\exp(-\beta\Hqf)$, 
we use the orthonormal Fock space basis 
$$
(\tilde c_1^\dag)^{n_1}\cdots (\tilde c_\cN^\dag)^{n_{\cN}}|0\rangle 
$$
where $n_K=0,1$ for $K=1,\ldots,\cN$:  
\begin{multline}\label{ZHFcomputation} 
	\Tr(\ee^{-\beta\Hqf}) = 
	\sum_{n_1=0,1}\cdots \sum_{n_{\cN}=0,1}
	\langle 0|  (\tilde{c}_{\cN})^{n_{\mathcal{N}}}\cdots (\tilde c_1)^{n_{1}}
	\\ \times 
	\prod_K \ee^{-\beta E_K(\tilde c_K^\dag \tilde c_K^\pdag-\tfrac12)}
	(\tilde c_1^\dag)^{n_1}\cdots (\tilde c_\cN^\dag)^{n_{\cN}}|0\rangle \\
	= \prod_K \sum_{n_K=0,1} \ee^{-\beta E_K(n_K-\tfrac12)} 
	\\
	= \prod_K(\ee^{\beta E_K/2} + \ee^{-\beta E_K/2}), 
\end{multline}
which implies 
\begin{multline*} 
	\frac1\beta\ln\Zqf =
	\frac1\beta \ln\prod_K(\ee^{\beta E_K/2}+\ee^{-\beta E_K/2}) =\sum_K\LnT(E_K)
\end{multline*} 
with $\LnT(E)$ defined in \eqref{LT}. This proves \eqref{TlnZqf}.

\subsection{A useful identity}\label{app:Id}  
The computation in \eqref{ni-1} is non-trivial since the operators $n_{\bx}-1$ and $\HHF$ do not commute. We thus give a mathematical result justifying this computation:   

{\em If $A$ and $B$ are linear operators on the fermion Fock space (for $L<\infty$) and $\alpha$ is a real parameter, then} 
\begin{equation}\label{Id}  
	\frac{\Tr\big(\ee^{(A+\alpha B)}B\big)}{\Tr\big(\ee^{(A+\alpha B)}\big)}	
	=  \frac{\partial}{\partial \alpha }\ln\Tr\big(\ee^{(A+\alpha B)}\big)
\end{equation} 
(while this is obviously true if $A$ and $B$ commute, it is non-trivial in general). Note that \eqref{ni-1} is the following special case of this result: $A+\alpha B=-\beta\HHF$, $B=n_{\bx}-1$,  $\alpha=-\beta(U/2)d(\bx)$. 

This result can be verified by the following computations, 
$$
\frac{\partial}{\partial \alpha}\ln\Tr\big(\ee^{(A+\alpha B)}\big)
= \frac{1}{\Tr\big(\ee^{(A+\alpha B)}\big)}  
\frac{\partial}{\partial\alpha}\Tr\big(\ee^{(A+\alpha B)}\big) 	
$$
and 
\begin{align*}
	\frac{\partial}{\partial \alpha } \Tr(&\ee^{(A + \alpha B)}) 
	 = \sum_{k=0}^\infty \frac{ \Tr(\frac{\partial}{\partial \alpha} (A+\alpha B)^k)}{k!}
	\\
	& = \sum_{k=1}^\infty \frac{ \Tr((A+\alpha B)^{k-1}B)}{(k-1)!}
	\\
	& = \Tr\bigg(\sum_{k=0}^\infty \frac{ (A+\alpha B)^{k}}{k!} B\bigg)
	= \Tr(\ee^{(A + \alpha B)}B), 
\end{align*}
where we used the cyclicity of the trace to obtain the second equality.

\subsection{Mathematical facts about quasi-free states}\label{app:qf} 
Due to the Wick theorem in \eqref{Wick}, one can completely characterize a quasi-free state by the expectation values 
\begin{align}\label{gammaIJ} 
	\gamma_{I,J} = \langle c_J^\dag c_I^\pdag \rangle
\end{align} 
where, as before, we use the short-hand notation $I=(\bx,\sigma)$ and similarly for $J$. 
These expectation values are the matrix elements of a hermitian matrix $\gamma=(\gamma_{I,J})$ of size $\cN\times\cN$, where $\cN=2L^n$. 
Thus, $\gamma$ can be regarded as an operator on the Hilbert space $\C^{\cN}$, which  has the natural physics interpretation as the {\em one-particle Hilbert space}: as elaborated in Appendix~\ref{app:CAR}, the fermion Fock space is obtained by second quantization from this one-particle Hilbert space $\C^{\cN}$. 

Any such $\gamma$ defines a self-adjoint operator such that $0\leq \gamma\leq 1$ (this is a convenient way of saying that all eigenvalues, $\gamma_I$, of $\gamma$ are in the range $0 \leq \gamma_I\leq 1$). Thus, there is a one-to-one correspondence between quasi-free states in the fermion Fock space and self-adjoint one-particle operators $\gamma$ such that $0\leq\gamma\leq 1$.

In  \eqref{HHFgen1}, we used another one-particle operator to characterize quasi-free states, namely the one-particle Hamiltonian $h=(h_{I,J})$ which, in general, can be any self-adjoint operator on $\C^{\cN}$. Both characterizations are equivalent: indeed, as shown below, the pertinent operators $\gamma$ and $h$ are related to each other in a simple way as follows, 
\begin{align}\label{hgamma}  
	\gamma = \frac{1}{\ee^{\beta h}+1}  
	\Leftrightarrow  h = \frac1\beta\ln(\gamma^{-1}-1),  
\end{align}
i.e., the same unitary  operator $U=(U_{I,J})$ on $\C^{\cN}$ diagonalizing $h$ as in \eqref{h_diag} diagonalizes $\gamma$, 
\begin{align*} 
 \gamma_{I,J} = \sum_K (U^\dag)_{I,K}\gamma_K U_{K,J} , 
\end{align*} 	
and the eigenvalues $E_{K}$ of $h$ and $\gamma_{K}$ of $\gamma$ are related as in \eqref{hgamma}, i.e., $\gamma_{K} =1/(\ee^{\beta E_{K}}+1)$.  

\begin{proof}[Proof of \eqref{hgamma}] This relation can be proved as in Section \ref{app:QF} (we use the same notation as there): by \eqref{HHFdiag}, $\Hqf=\sum_{K} E_{K}^\pdag (\tilde c_{K}^\dag \tilde c_{K}^\pdag - 1/2)$		
where $\tilde c_{K}$ are Bogoliubov transformed fermion fields, and, by \eqref{ZHFcomputation},
\begin{align*} 
	\Tr(\ee^{-\beta\Hqf})= 
	\prod_{K} \sum_{n_K=0,1} \ee^{-\beta E_K(n_K-\tfrac12)}.
\end{align*}
Since
\begin{multline*} 
\Tr\big(\ee^{-\beta\Hqf} \tilde c^\dag_I \tilde c_J^\pdag\big) 
	 = \sum_{n_1=0,1}\cdots \sum_{n_{\cN}=0,1}
	\langle 0| (\tilde{c}_{\cN})^{n_{\mathcal{N}}}\cdots (\tilde c_1)^{n_{1}}
	\\ \times 
	\prod_K \ee^{-\beta E_K(\tilde c_K^\dag \tilde c_K^\pdag-\tfrac12)}\cdot \tilde c_I^\dag \tilde c_J^\pdag \cdot  
	(\tilde c_1^\dag)^{n_1}\cdots (\tilde c_\cN^\dag)^{n_{\cN}}|0\rangle \\
	= \delta_{I,J}\prod_{K\neq I} \sum_{n_K=0,1} \ee^{-\beta E_K(n_K-\tfrac12)}\cdot \sum_{n_I=0,1} n_I \ee^{-\beta E_I(n_I-\tfrac12)} \\
	= \delta_{I,J}\Tr(\ee^{-\beta\HHF}) \frac{\sum_{n_I=0,1} n_I \ee^{-\beta E_I(n_I-\tfrac12)}}{\sum_{n_I=0,1} \ee^{-\beta E_I(n_I-\tfrac12)}},  
\end{multline*} 	
we obtain
\begin{multline*} 
	\langle \tilde c^\dag_I \tilde c_J^\pdag\rangle = 
	\frac{\Tr\big(\ee^{-\beta\Hqf}\tilde c^\dag_{I} \tilde c_{J}^\pdag\big)}{\Tr\big(\ee^{-\beta\Hqf}\big)} \\ =
	\delta_{I,J}\frac{\ee^{-\beta E_I/2} } {\ee^{\beta E_I/2}+ \ee^{-\beta E_I/2}}  =  \delta_{I,J}\frac{1}{\ee^{\beta E_I}+1}.
\end{multline*} 
The inverse of the Bogoliubov transformation in \eqref{BT} is 
\begin{align}\label{invBT} 
 c_I = \sum_{J} (U^\dag)_{I,J}\tilde c_J, 	
\end{align} 	
which implies 
\begin{multline*} 
\gamma_{I,J} = \langle c_J^\dag c_I^\pdag\rangle = 
\sum_{K,L}\underbrace{\overline{U^\dag_{J,K}}}_{U_{K,J}}(U^\dag)_{I,L}\underbrace{\langle\tilde c^\dag _K \tilde c_L^\pdag\rangle}_{\delta_{K,L}/(\ee^{\beta E_K}+1)} \\
= \sum_K (U^\dag)_{I,K} \frac1{\ee^{\beta E_K}+1} U_{K,J}.
\end{multline*} 	
This proves \eqref{hgamma}.
\end{proof} 

In what follows, we prove the folklore mentioned at the beginning of Section~\ref{sec:formulas}, namely, that a state is quasi-free if and only if it can be represented as in \eqref{rhoqf} for some quadratic Hamiltonian $\Hqf$, or can be obtained as a limit of such states.

Assume first that $\Hqf$ is a quadratic Hamiltonian and that $\rhoqf$ is given by \eqref{rhoqf}.
To show that $\rhoqf$ is quasi-free, we need to verify identities of the form \eqref{Wick}.
In the proof above, we computed $\langle c_I^\dag c_J^\pdag\rangle$. In a similar way, one can compute expectation values of products of arbitrary numbers of creation and annihilation operators. For example, it is straightforward to generalize the computation above to obtain 
	\begin{multline*} 
	\langle \tilde c^\dag_I  \tilde c^\dag_K \tilde c_L^\pdag \tilde c_J^\pdag\rangle = 
	\frac{\Tr\big(\ee^{-\beta\Hqf}\tilde c^\dag_{I}  \tilde c^\dag_K \tilde c_{J}^\pdag\tilde c_L^\pdag\big)}{\Tr\big(\ee^{-\beta\Hqf}\big)} 
	\\ 
	= \left(\delta_{I,J}\delta_{K,L} - \delta_{I,L}\delta_{K,J} \right) \frac{1}{\ee^{\beta E_I}+1}\frac{1}{\ee^{\beta E_K}+1}, 
\end{multline*} 		
and by using \eqref{invBT} and its adjoint one finds  \eqref{Wick} (the details of the computations are similar to the ones in the proof above). Moreover, we show in Appendix~\ref{app:exchange} that the set of quasi-free states is closed. Hence, by taking limits, we conclude that any limit of states of the form \eqref{rhoqf} is quasi-free.

Assume now that $\rho$ is a quasi-free state. Then, as mentioned above, $\rho$ is completely characterized by the expectation values $\gamma_{I,J} = \Tr(\rho c_J^\dag c_I^\pdag)$. Assuming that $\gamma$ is invertible, define $h$ in terms of $\gamma$ by \eqref{hgamma}, 
define the quadratic Hamiltonian $\Hqf$ in terms of $h$ by \eqref{Hqf}, and define $\rhoqf$ in terms of $\Hqf$ by \eqref{rhoqf}.
The proof of \eqref{hgamma} shows that
$$\Tr(\rhoqf c_J^\dag c_I^\pdag) = \gamma_{I,J} = \Tr(\rho c_J^\dag c_I^\pdag).$$
Since $\rho$ is fully determined by these expectation values, we conclude that $\rho = \rhoqf$.	
This shows that $\rho$ can be represented as in \eqref{rhoqf} for some quadratic Hamiltonian $\Hqf$ whenever $\gamma$ is invertible. 
If $\gamma$ is not invertible, then write $\gamma= \sum_k \gamma_k|k\rangle\langle k|$ with $\langle k|k'\rangle=\delta_{k,k'}$ and $\gamma_k =\langle k|\gamma|k\rangle$ for $k,k'=1,\ldots,\mathcal{N}$. Applying the above argument to the quasi-free state $\rho^{(\epsilon)}$ ($0 < \epsilon < 1$) corresponding to the invertible one-particle operator $\gamma^{(\epsilon)} = \gamma + \epsilon \sum_{k:\gamma_k = 0} |k\rangle\langle k|$ and taking the limit $\epsilon \downarrow 0$, we deduce that any quasi-particle state can be obtained as a limit of states of the form \eqref{rhoqf}. This completes the proof.

To conclude, we note that the reversed order of the indices in \eqref{gammaIJ} is natural since it allows us to write the expectation value of the Hartree--Fock Hamiltonian in \eqref{HHFgen1} as  
\begin{align*}  
	\langle \Hqf\rangle = \sum_{I,J} h_{I,J} \big( \gamma_{J,I} -\tfrac12\delta_{I,J}\big) = \tr\big( h(\gamma-\tfrac12)\big) , 	
\end{align*} 	
where $\tr$ denotes the trace in the one-particle Hilbert space (which is the usual trace of $\cN\times\cN$ matrices).

\subsection{Rotation invariance of Hubbard interaction}\label{app:nunu}	
In this section, we derive \eqref{HH0Usumnunu} by showing that 
	$$
	(n_\uparrow -\tfrac12)(n_\downarrow -\tfrac12)
	= \tfrac14(n + \vec{e}\cdot\vec{s}-1)(n - \vec{e}\cdot\vec{s}-1) 
	$$
	for an arbitrary unit vector $\vec{e}\in\R^3$, suppressing the common subscript $\bx$ of $n_{\uparrow,\downarrow}$, $n$, $\vec{s}$, and $\vec{e}$. 
	
	Since $n_{\uparrow,\downarrow} =\tfrac12(n\pm s^z)$, this identity is equivalent to 
	\begin{align*} 
		(n + s^z-1)(n - s^z-1) = (n + \vec{e}\cdot\vec{s}-1)(n - \vec{e}\cdot\vec{s}-1), 
	\end{align*} 
	which by straightforward computations, using $[n,\vec{s}\,]=0$, can be simplified to 
	\begin{align}\label{Sid1} 
		(\vec{e}\cdot\vec{s}\,)^2 = (s^z)^2.
	\end{align} 
The relation \eqref{Sid1} follows from the identities 
	\begin{equation} 
		\begin{split} 
			(s^x)^2 = (s^z)^2,\quad (s^y)^2 = (s^z)^2,\\
			\{s^x, s^y\} = \{s^y, s^z\} =  \{s^z, s^x\} = 0.
		\end{split} 
		\label{Sid2}
	\end{equation} 	
Indeed, by expanding $(\vec{e}\cdot\vec{s}\,)^2 = (e^x s^x+e^ys^y+e^zs^z)^2$ and using \eqref{Sid2} and $\vec{e}^{\,2}=1$, one obtains \eqref{Sid1}. 
	
To complete the argument, one has to verify all the identities in \eqref{Sid2}. This can be done by straightforward but tedious computations, inserting definitions and using the canonical anticommutator relations in \eqref{CAR}.

\subsection{Structure of AF states}\label{app:symm}  
In the main text we used that, for the Hamiltonian $H_{\mathrm{AF}}$ in \eqref{HAF}, $\langle n_{\bx}-1\rangle=a$ and $\langle\vec{s}_{\bx}\rangle=(-1)^{\bx}\vec{e}\,b$ for some (known) constants $a$ and $b$ (this is usually only shown for $\vec{e}=(0,0,1)$; the general result follows from this by rotation invariance of the Hubbard Hamiltonian). 
For the convenience of the reader, we give a proof of this fact based on results already obtained. 

By Fourier transformation, we can write the Hartree--Fock Hamiltonian in \eqref{HHF} as
$$
\HHF = H_0 + \frac{U}{2}\sum_{\bk}\big( 
\tilde{d}_0 (-\bk)(\tilde{n}_{\bk}-L^{n/2}\delta_{\bk,\mathbf{0}}) - \tilde{\vec{m}}(-\bk)\cdot \tilde{\vec{s}}_{\bk}\big), 
$$ 
where $\tilde{n}_{\bk}$ and $\tilde{d}(\bk)$ are defined for $\bk\in \Lambda_L^*$ by
$$
\tilde n_{\bk} = \frac1{L^{n/2}}\sum_{\bx} n_{\bx}\ee^{-\ii\bk\cdot\bx},\quad 
 \tilde{d}(\bk) = \frac1{L^{n/2}}\sum_{\bx} d(\bx)\ee^{-\ii\bk\cdot\bx}, 
$$
and similarly for $\tilde{\vec{s}}_{\bk}$ and  $\tilde{\vec{m}}(\bk)$. By the result stated and proved in Appendix~\ref{app:Id}, it follows that
\begin{align*} 
\frac{U}{2}\langle \tilde{n}_{\bk}\rangle &= -\frac1\beta\frac{\partial}{\partial \tilde{d}(-\bk)}\ln\ZHF\quad (\bk\neq\mathbf{0}), \\ 
\frac{U}{2}\langle \tilde{\vec{s}}_{\bk}\rangle &= \frac1\beta\frac{\partial}{\partial\tilde{\vec{m}}(-\bk)}\ln\ZHF, 
\end{align*} 
with $\ZHF=\Tr(\ee^{-\beta\HHF})$. For the special case of $\HHF$ in \eqref{HAF}, we have
$\tilde d(\bk)=d_0 L^{n/2} \delta_{\mathbf{k}, \mathbf{0}}$ and $\tilde{\vec{m}}(\mathbf{k}) = m_1\vec{e}\, L^{n/2} \delta_{\mathbf{k}, \mathbf{Q}}$.
In particular, $\tilde{d}(\bk)$ is non-zero only if $\bk=\mathbf{0}$ and $\tilde{\vec{m}}(\bk)$ is non-zero only if $\bk=\bQ$; thus, in this case, $\ZHF$ only depends on $\tilde{d}(\mathbf{0})$ and $\tilde{\vec{m}}(\bQ)^2$ (the latter is true due to rotation invariance).  
Thus, for $\HHF$ in \eqref{HAF},  
$$
\langle \tilde{n}_{\bk}\rangle = 
\delta_{\bk,\mathbf{0}}\langle \tilde{n}_{\mathbf{0}}\rangle,\quad 
\langle \tilde{\vec{s}}_{\bk}\rangle = 
\delta_{\bk,\bQ}\vec{e}\langle \vec{e}\cdot\tilde{\vec{s}}_{\bQ}\rangle, 
$$
which implies that $\langle n_{\bx}\rangle=a$ and $\langle \vec{s}_{\bx}\rangle=(-1)^{\bx}\vec{e}\, b$ for some constants $a$ and $b$.

\section{Exchanging max and min}\label{app:exchange}
In general, for functions $f(x,y)$ of two sets of variables $x$ and $y$, the orders of minimization, $\min_x$,  and maximization, $\max_y$, cannot be exchanged without changing the result, as illustrated by the example $f(x,y)=\sin(x+y)$: $\min_{x\in\R}\max_{y\in\R}\sin(x+y)=1$ is different from $\max_{y\in\R}\min_{x\in\R}\sin(x+y)=-1$ (there are many other such examples). 
Thus, there is a need to justify the interchanges of orders of $\min$ and $\min\max$	in Section~\ref{sec:BLS}. 
Fortunately, there exist non-trivial results in the mathematics literature which can be used to justify these interchanges. In this appendix, we give details on how this can be done. 
 
In Section \ref{sec:BLS}, we used the following identity to derive equation \eqref{Omegaminmax}:
$$\min_{\rhoqf} \min_{\boldsymbol{\vec{m}}}\max_{\boldsymbol{d}}A(\boldsymbol{d}, \boldsymbol{\vec{m}}, \rhoqf)
= \min_{\boldsymbol{\vec{m}}}\max_{\boldsymbol{d}} \min_{\rhoqf} A(\boldsymbol{d}, \boldsymbol{\vec{m}}, \rhoqf)$$ 
where 
\begin{multline}\label{calAdef}
A(\boldsymbol{d}, \boldsymbol{\vec{m}}, \rhoqf)
= \frac{U}{4}\sum_{\bx} \big(\vec{m}(\bx)^2-d(\bx)^2 \big) + \Omega(\HHF,\rhoqf)
	\\
= \frac{U}{4}\sum_{\bx} \big(\vec{m}(\bx)^2-d(\bx)^2 \big)  
	\\
+ \Tr\bigg(\rhoqf \bigg(H_0 + \frac{U}{2}\sum_{\bx}\big( d(\bx)(n_{\bx}-1)-\vec{m}(\bx)\cdot\vec{s}_{\bx} 
\big)\bigg)\bigg)
	\\
+ \frac1\beta\Tr(\rhoqf\ln(\rhoqf)), 
\end{multline} 
with $\HHF=\HHF(\boldsymbol{d},\boldsymbol{\vec{m}})$ in \eqref{HHF} with $\boldsymbol{d}=(d(\bx))\in \R^{L^n}$ and  $\boldsymbol{\vec{m}}=(\vec{m}(\bx))\in \R^{3L^n}$ ($\min\min\max A$ is short for $\min(\min(\max(A)))$, and similarly for $\min\max\min A$). 
We now give a proof of this identity.

It is obvious that $\min_{\rhoqf}$ can be exchanged with $\min_{\boldsymbol{\vec{m}}}$, and it is not hard to show that
\begin{equation}\label{minmaxA}
\min_{\rhoqf} \max_{\boldsymbol{d}} A(\boldsymbol{d}, \boldsymbol{\vec{m}}, \rhoqf) \geq 
\max_{\boldsymbol{d}} \min_{\rhoqf}  A(\boldsymbol{d}, \boldsymbol{\vec{m}}, \rhoqf), 
\end{equation}
regardless of the structure of the function $A$ (the latter is known in mathematics as the {\em max-min inequality}).
Thus, what is non-trivial is to prove that equality holds in \eqref{minmaxA}. 
This follows essentially from an application of Sion's minimax theorem \cite{S1958, K1988} generalizing a result due to von Neumann \cite{vN1928}. However, since the set of quasi-free states, $\cS_{\mathrm{qf}}$,
over which the minimum over $\rhoqf$ is taken is not convex in general, the standard versions of Sion's minimax theorem do not apply. Instead we will appeal to a more recent generalization due to Mosconi \cite{M2012}. 

We recall that states $\rho$ are self-adjoint $N\times N$ matrices with eigenvalues $\geq 0$ such that $\Tr(\rho)=1$, that $\cS$ denotes the set of all states $\rho$, and that
$\cS_{\mathrm{qf}}$ is the subset of $\cS$ containing all quasi-free states, i.e., all limits of states of the form \eqref{rhoqf}--\eqref{Hqf} with $h=(h_{\bx,\sigma;\by,\sigma'})$ an $\cN\times\cN$ self-adjoint matrix where $N=2^{\cN}$ and $\cN=2L^n$.
For simplicity, we assume that $\beta$ is finite; the case $\beta=\infty$ is included as a limit $\beta\to\infty$.

To show that equality holds in \eqref{minmaxA}, note first that, for fixed $\boldsymbol{\vec{m}}$ and $\rhoqf$, $A(\boldsymbol{d}, \boldsymbol{\vec{m}}, \rhoqf)$ is a strictly concave function of $\boldsymbol{d} \in \R^{L^n}$ (this is easy to verify).
On the other hand, the second to last term in \eqref{calAdef} is linear in $\rhoqf$, and it is well-known that $\tr(\rho \ln \rho)$ is a strictly convex function of $\rho \in \mathcal{S}$ (see \cite[Sec.~II.B]{W1978}). 
The set $\cS$ is compact. (To see this, note that $\cS$ is bounded in $\C^{N \times N}$ because the norm of $\rho \in \cS$ equals the largest eigenvalue which is $\leq 1$. Moreover, $\cS$ is closed because it is an intersection of closed sets: $\cS = \cap_{\{v \in \C^N : \langle v|v\rangle=1\}} \cS_v$, where $\cS_v$ is the set of all self-adjoint $N\times N$ matrices $\rho$ such that $0\leq \langle v|\rho| v\rangle\leq 1$ and $\Tr(\rho)=1$. This shows that $\cS$ is a closed and bounded subset of $\C^{N \times N}$, hence compact.)
The subset $\mathcal{S}_{\mathrm{qf}} \subset \mathcal{S}$ of all quasi-free states is closed in $\mathcal{S}$ (and hence compact) because it is obtained from $\mathcal{S}$ by imposing a finite number of conditions of the form $f_i(\rho) = 0$ where each $f_i$ is continuous. Therefore, we may apply \cite[Theorem 4.2]{M2012} (see also \cite[Theorem 1.B]{R2016}) to conclude that either equality holds in \eqref{minmaxA} or else there exist $\boldsymbol{d}$ and $\boldsymbol{\vec{m}}$ such that the function $A(\boldsymbol{d}, \boldsymbol{\vec{m}},\rho_{\mathrm{qf}})$ of $\rho_{\mathrm{qf}}\in \cS_{\mathrm{qf}}$ has at least two global minima.
Since we proved in Appendix \ref{app:Gibbs} that the Gibbs state $\rhoHF=\ee^{-\beta\HHF}/\Tr(\ee^{-\beta\HHF}) \in \mathcal{S}_{\mathrm{qf}}$ is the unique global minimizer of $\Omega(\HHF,\rho)$ over all states $\rho \in \mathcal{S}$, we know that the function $A(\boldsymbol{d}, \boldsymbol{\vec{m}}, \rho)$ of $\rho\in\cS_{\mathrm{qf}}$ does not have two global minima. This shows that equality holds in \eqref{minmaxA} and completes the derivation of \eqref{Omegaminmax}. 

The identity \eqref{OmegaAFminmax} follows from the minimax result \cite[Theorem 4.2]{M2012} in a similar way as \eqref{Omegaminmax}. Indeed, since the set $\mathcal{S}_{\mathrm{AF}}$ of states $\rho_{\mathrm{AF}}$ of the form \eqref{rhoAF} is closed in $\mathcal{S}$, we may apply \cite[Theorem 4.2]{M2012} in a similar way as above to interchange $\min_{\rho_{\mathrm{AF}}}$ and $\max_{\tilde{d}_0}$. In fact, the less general result \cite[Theorem 1]{R2010} can be utilized instead of \cite[Theorem 4.2]{M2012} in the present situation because $\tilde{d}_0 \in \R$.

%\bibliographystyle{ieeetr.bst}
%\bibliographystyle{apsrev4-1}
%\bibliography{Hubbard}

\begin{thebibliography}{10}
\bibitem{Comment:qualitative}
	For example, an anti-ferromagnetic and a paramagnetic solution would be
	qualitatively different, but two ferromagnetic solutions with different
	magnetization directions would not.
\bibitem{Comment:expansions}
	Up to now, our analytical results are restricted to the Hubbard model at zero
	temperature in 2D (on-going work with Christophe Charlier) and 1D (on-going
	work with Valdemar Melin), but we are confident that these methods can be
	generalized to other phase diagrams as well (we plan to present these results
	elsewhere).
\end{thebibliography}
		
\end{document}